\documentclass[12pt]{article}
\usepackage{amsmath}
\usepackage{graphicx,caption}
\usepackage{enumerate}
\usepackage[round,sort]{natbib}
\usepackage{url}
\usepackage{amsthm}
\usepackage{amssymb}
\usepackage{xargs}[2008/03/08]
\usepackage{xcolor}
\usepackage{enumitem}
\usepackage{authblk}
\usepackage{algorithm}
\usepackage{algorithmic}
\usepackage{multirow}
\usepackage{tikz}
\usepackage{makecell}
\usepackage{pdfpages}
\usepackage{microtype}

\newcommand{\blind}{0}

\usepackage{soul}
\usepackage{cancel}

\theoremstyle{plain}

\newtheorem{theorem}{Theorem}

\newtheorem{proposition}{Proposition}
\newtheorem{corollary}{Corollary}

\theoremstyle{remark}
\newtheorem{remark}{Remark}

\newtheorem*{example}{Example}

\newtheorem{assumption}{Assumption}
\newtheorem*{assumption*}{Assumption}
\newtheorem*{condition*}{Condition}

\usepackage[hidelinks]{hyperref}
\hypersetup{
	colorlinks,
	linkcolor={red!50!black},
	citecolor={blue!50!black},
	urlcolor={blue!80!black}
}

\global\long\def\expect{\mathbb{E}}%
\global\long\def\prob{\mathbb{P}}%
\global\long\def\real{\mathbb{R}}%
\global\long\def\metricsp{\mathcal{M}}%
\global\long\def\metricfn{d}%
\global\long\def\manifold{\mathcal{M}}%
\global\long\def\asymplt{\lesssim}%
\global\long\def\asympeq{\asymp}%
\newcommandx\tangentspace[2][usedefault, addprefix=\global, 1=\manifold]{T_{#2}#1}%
\global\long\def\ball#1#2{B(#1,#2)}%
\newcommandx\lpnorm[3][usedefault, addprefix=\global, 1=r, 2=]{\|#3\|_{\mathcal{L}^{#1}}^{#2}}%
\newcommandx\lp[1][usedefault, addprefix=\global, 1=p]{\mathcal{L}^{#1}}%
\global\long\def\innerprodM#1#2#3{\langle#1,#2\rangle_{#3}}%
\global\long\def\innerprodE#1#2{\langle#1,#2\rangle_{E}}%
\global\long\def\tanbundle#1#2{T_{#1}#2}%
\newcommandx\normE[1]{\|#1\|_{E}}
\global\long\def\metricE#1#2{d_{E}(#1,#2)}%
\global\long\def\metric#1#2{d(#1,#2)}%
\global\long\def\metricpoly#1#2#3{d^{#3}(#1,#2)}%
\global\long\def\metricsq#1#2{\metricpoly{#1}{#2}{2}}%
\global\long\def\geod#1#2{\gamma_{#1}^{#2}}%

\global\long\def\Exp#1#2{\mathrm{Exp}_{#1}#2}%
\newcommandx\vfnorm[3][usedefault, addprefix=\global, 1=\mu, 2=]{\|#3\|_{#1}^{#2}}%
\newcommandx\vfinnerprod[2][usedefault, addprefix=\global, 1=\mu]{\llangle#2\rrangle_{#1}}%
\global\long\def\define{:=}%
\global\long\def\trainset{\mathcal{D}_{\mathrm{train}}}%
\global\long\def\testset{\mathcal{D}_{\mathrm{test}}}%

\global\long\def\indicator#1{\mathbb{I}_{\{#1\}}}
\newcommandx\opnorm[3][usedefault, addprefix=\global, 1=\mu, 2=]{\vertiii{#3}_{#1}^{#2}}%
\newcommandx\fronorm[2][usedefault, addprefix=\global, 1=]{|#2|_{F}^{#1}}%
\global\long\def\spdmfd#1{{\mathcal{S}_{#1}^{+}}}%
\global\long\def\alexanlge#1#2#3{\angle_{#1}(#2, #3)}%
\global\long\def\comanlge#1#2#3{\bar{\angle}_{#1}(#2, #3)}%
\DeclareMathOperator*{\argmin}{argmin}
\DeclareMathOperator*{\argmax}{argmax}
\global\long\def\estbeta{\hat{\beta}_n}%
\global\long\def\truebeta{\beta^*}%
\global\long\def\truemu{\mu^*}%
\global\long\def\estmu{\hat{\mu}_n}%
\global\long\def\covernum#1#2#3{N\left(#1, #2, #3\right)}%
\global\long\def\excessrisk#1#2{\mathcal{E}(#1, #2)}%
\global\long\def\classloss#1{\mathcal{L}(#1)}%
\global\long\def\emplik#1{L_n(#1)}%
\global\long\def\poplik#1{L(#1)}%
\global\long\def\emplikfn{L_n}%
\global\long\def\poplikfn{L}%

\global\long\def\compoint#1{\bar{#1}}%

\def\hess{\text{Hess}}
\def\munbh{r}

\usepackage{xr}
\makeatletter
\newcommand*{\addFileDependency}[1]{
	\typeout{(#1)}
	\@addtofilelist{#1}
	\IfFileExists{#1}{}{\typeout{No file #1.}}
}
\makeatother
\newcommand*{\myexternaldocument}[1]{%
	\externaldocument{#1}%
	\addFileDependency{#1.tex}%
	\addFileDependency{#1.aux}%
}
\myexternaldocument{CoM-supp}

\def\xref#1{\ref*{#1}}

\begin{document}
	
	
	
	\def\spacingset#1{\renewcommand{\baselinestretch}%
		{#1}\small\normalsize} \spacingset{1}

	
	\title{\bf Binary Regression and Classification with Covariates in Metric Spaces}
	\author{Yinan Lin}
	\author{Zhenhua Lin\thanks{linz@nus.edu.sg. Research partially supported by NUS start-up grant A-0004816-00-00.}}
	\affil{Department of Statistics and Data Science, National University of Singapore}
	\date{}
	\maketitle
	
	\bigskip
	\begin{abstract}
		Inspired by logistic regression, we introduce a  regression model for data tuples consisting of a binary response and a set of covariates  residing in a metric space without vector structures. Based on the proposed  model we also develop a binary classifier for metric-space valued data. We propose a maximum likelihood estimator for the metric-space valued regression coefficient in the model, and provide upper bounds on the estimation error under various metric entropy conditions that quantify complexity of the underlying  metric space. Matching lower bounds are derived for the important metric spaces commonly seen in statistics, establishing optimality of the proposed estimator in such spaces. Similarly, an upper bound on the excess risk of the developed classifier is provided for general metric spaces. A finer upper bound and a matching lower bound, and thus optimality of the proposed classifier,  are established for Riemannian manifolds. To the best of our knowledge, the proposed regression model and the above minimax bounds are the first of their kind for analyzing a binary response with covariates residing in general metric spaces. 
		We also investigate the numerical performance of the proposed estimator and classifier via simulation studies, and illustrate their practical merits via an application to task-related fMRI data.
	\end{abstract}

	\noindent%
	
	\vspace{0.5cm}    
	\noindent    
	\textit{MSC2020 subject classification}: 62R20, 62J12, 62H30.
	
	\vfill
	
	\newpage
	\spacingset{1.5}

	\section{Introduction}\label{sec:intro}
	
	Object data sampled from general metric spaces emerge recently in many domains, posing fundamental challenges to traditional statistical methodologies geared to vector structures.  Examples of such data include phylogenetic trees arising from  evolutionary biology \citep{Billera2001}, distribution functions sampled from the nonlinear Wasserstein geometry for modeling differentiating cells in developmental biology \citep{Schiebinger2019}, symmetric positive-definite matrices from a nonlinear Riemannian manifold for representing diffusion tensors and brain connectivities in neuroscience  \citep[e.g.,][]{Arsigny2007,Dryden2009, yuan2012}, graph Laplacians for studying networks \citep{Dubey2022}, shapes of tumors \citep{Bharath2020} in medical sciences, and more in \cite{Marron2021}. Although there is a body of literature on regression analysis for this type of data \cite[e.g.,][among many others]{hein2009robust, faraway2014regression,petersen2019frechet, Qiu2022, Ying2022, Tucker2023,Zhang2023}, very few studies explore parametric regression models specifically designed for binary response variables with metric-space valued covariates.

	In the analysis of a binary response variable linked to a set of covariates, the crucial concepts of odds and log odds emerge across various fields, with logistic regression serving as a widely used parametric model to examine the impact of covariates on the odds. Motivated by this background, this paper presents a new model designed to analyze a binary response affected by covariates that reside in a metric space 
	$\metricsp$ \emph{without a vector structure}. 
	Specifically, for a binary variable $Y$ and a covariate $X\in\metricsp$, we propose the model
	\begin{equation}
		\text{log}\bigg(\frac{P_X}{1-P_X}\bigg) = \metric{\truemu}{X} \metric{\truemu}{\truebeta} \cos(\alexanlge{\truemu}{X}{\truebeta}),
		\label{eq:intro-model}
	\end{equation}
	where $P_{X} \define \prob(Y=1|X)$ is success probability, $\metricfn$ is the distance on $\metricsp$, $\truemu\in \metricsp$ is the (Fr{\'e}chet) mean of $X$, $\truebeta$ is the coefficient object to be estimated, and $\alexanlge{\truemu}{X}{\truebeta}$ is the angle between the geodesics emanating from $\truemu$ to $X$ and $\truebeta$. The left-hand side of \eqref{eq:intro-model} represents the log odds. The right-hand side of \eqref{eq:intro-model}, referred to as the Alexandrov inner product between the two geodesics, reduces to $(X-\truemu)^\top(\truebeta-\truemu)$ when $\metricsp=\real^D$, thus mimicking and extending the linear covariate effect on the log odds. Similarly, $\metric{\truemu}{X} \cos(\alexanlge{\truemu}{X}{\truebeta})$ mimics the projection of the geodesic connecting $\truemu$ to $X$ onto the geodesic connecting $\truemu$ to $\truebeta$; see Figure \ref{fig:alex-projection} for an illustration. Intuitively, the parameter $\truebeta$ points to a ``direction'' in which the log odds changes at the fastest rate, and the influence of $X$ on the log odds is determined by its projection onto this direction. This direction could therefore offer insights into the association between the response and the covariates, highlighting the practical merit of the proposed model \eqref{eq:intro-model}; see Section \ref{sec:application} for a demonstration.
	In Section \ref{sec:meth}, we provide further details about \eqref{eq:intro-model}  and develop an efficient estimator $\estbeta$ for $\truebeta$ based on the principle of maximum likelihood.  	To the best of our knowledge, this represents the first parametric model designed for the odds and its association with covariates from general metric spaces. 
	
	\begin{figure}
		\centering
		\begin{tikzpicture}
			\coordinate (O) at (0,0);
			\coordinate (B) at (4,0);
			\coordinate (X) at (2,2);
			\coordinate (P) at (2,0);
			\coordinate (N) at (2,-1);
			
			\draw[fill=black] (O) circle (.3ex);
			\node[below] at (O) {$\truemu$};
			
			\draw[fill=black] (X) circle (.3ex);
			\node[above] at (X) {$X$};
			
			\draw[fill=black] (B) circle (.3ex);
			\node[below] at (B) {$\truebeta$};
			
			\draw (O) -- (X);
			\draw (O) -- (B);
			\draw[dashed] (X) -- (P);
			\draw[red, line width=1] (O) -- (P);
			
			\node at (N) {Euclidean Space};

			\coordinate (O) at (7,0);
			\coordinate (B) at (11,0);
			\coordinate (X) at (9,2);
			\coordinate (P) at (10,0.15);
			\coordinate (N) at (9,-1);

			\draw[fill=black] (O) circle (.3ex);
			\node[below] at (O) {$\truemu$};
			
			\draw[fill=black] (X) circle (.3ex);
			\node[above] at (X) {$X$};
			
			\draw[fill=black] (B) circle (.3ex);
			\node[below] at (B) {$\truebeta$};
			
			\draw (O) to[out=30,in=250] (X);
			\draw (O) to[out=10,in=170] (B);
			\draw[dashed] (X) to[out=-80,in=150] (P);
			\draw[red, line width=1] (O) to[out=10,in=175] (P);
			\node at (N) {Curved Metric Space};
			
		\end{tikzpicture}
		\caption{Illustration of geodesic projection: The projection in Euclidean space (left) and its analogy in curved metric space (right), where the segments in red represent the projection $d(\truemu,X)\cos(\angle_{\truemu}(X,\truebeta))$ of the geodesic connecting $\truemu$ to $X$ onto the geodesic connecting $\truemu$ to $\truebeta$.}
		\label{fig:alex-projection}
	\end{figure}
	
	The proposed model \eqref{eq:intro-model} also induces a  classifier for data residing in general metric spaces. Via modeling the log odds in a parametric form, this classifier is less prone to the curse of dimensionality, and thus is preferable when interpretable results are desired or when the sample size is small relative to the complexity of $\metricsp$; see Section \ref{sec:application} for a numeric demonstration. Meanwhile we should remark that, when the association between log odds and covariates is highly nonlinear,  dedicated classifiers \citep[e.g.,][among others]{gottlieb2016nearly, qiao2019rates,huang2017riemannian}, which are generally nonparametric, may produce more accurate classification results at the cost of interpretability; see Section \ref{sec:simulation} for a numeric illustration.
	
	In addition to the aforementioned methodological contributions, we establish upper and matching lower bounds on the  estimation error $\metricsq{\estbeta}{\truebeta}$ under various metric entropy conditions that characterize complexity of a metric space. This demonstrates the optimality of the proposed estimator for metric spaces  commonly encountered in statistics.  For instance,  under certain  regularity conditions, the proposed estimator $\estbeta$ achieves the optimal rate
	$\expect[\metricsq{\estbeta}{\truebeta}]\asympeq	n^{-1/\alpha}$, where $\alpha$ is linked to the geometric characteristics of $\metricsp$; see Section \ref{sec:theory-estimation} for details.
	To the best of our knowledge, this is the first optimality result for regression of logistic type like \eqref{eq:intro-model}  on general metric spaces.
	
	Another major theoretical contribution is the establishment of upper and corresponding lower bounds on the excess risk of the classifier induced by the proposed mode \eqref{eq:intro-model}. Roughly speaking, we show that, when the covariate resides in a $D$-dimensional Riemannian manifold, the excess risk decays at the optimal rate $\sqrt{D/n}$; see Section \ref{sec:theory-classification} for details. To  our knowledge, this represents the first optimality result in classification for manifold-valued objects. Notably, our results encompass the optimal rate  in \cite{abramovich2019high} for Euclidean logistic regression as a special case. Additionally, we  develop an upper bound on the excess risk for general metric spaces. This upper bound, presented in Theorem \ref{thm:excessrisk-upper-simplified-metric} and expressed in terms of the estimation error, characterizes the influence of the geometric features and complexity of the underlying metric space on the excess risk. 
	
	Theoretical analysis of the proposed estimator and classifier is challenged by the absence of a vector structure in a general metric space. For instance, the regression model \eqref{eq:intro-model} is defined geometrically, and our theoretical analysis is significantly complicated by the nonlinear metric geometry of the underlying space, including but not limited to the space curvature. This contrasts with the Euclidean case where the presence of rich vector and algebraic structures makes the analysis much more tractable. In addition, the likelihood functions (see Section \ref{sec:meth}), combining   geometric constructions with the nonlinear logistic function, introduces extra complexity into our theoretical studies; {see Claim \xref{lemma:maximal-ineq-expectation}, Corollary \xref{cor:fmse} and Lemma \xref{lem:mid-seg-comparison-model-space} of Supplementary Material for examples.}     
	
	The rest of the paper is organized as follows. Section \ref{sec:background} provides a brief introduction on metric and Riemannian geometry. In Section \ref{sec:meth}, we introduce the proposed model \eqref{eq:intro-model} in details and develop an estimator for the model parameter. We study theoretical properties of the estimator in Section \ref{sec:theory-estimation} and the excess risk of classification in  Section \ref{sec:theory-classification}. Section \ref{sec:simulation} is devoted to assessing the empirical performance of the proposed method via  simulation studies. An application to task-related functional magnetic resonance imaging (fMRI) data is presented in Section \ref{sec:application}. 	We conclude the article in Section \ref{sec:discussion}. 
	All proofs are relegated to Supplementary Material.

	\section{Background on Metric and Riemannian Geometry}\label{sec:background}
	
	Below we briefly review some geometric concepts that are essential to our developments, while refer readers  
	to \cite{Burago2001} and \cite{lee2018introduction} for a comprehensive treatment on metric and Riemannian geometry; {see also \cite{holopainen2009metric} and Section S1 of \cite{Shao2022} for self-contained notes on basic concepts of metric and Riemannian geometry.}
	
	\vspace{3mm}
	\noindent
	\textbf{Geodesic.} For a metric space $(\metricsp, \metricfn)$ and two points $p, q\in \metricsp$, a geodesic connecting $p$ and $q$ is a map $\gamma$ from a closed interval $[0, T]\subset \real$ to $\metricsp$ with $T=d(p,q)$ such that, $\gamma(0)=p$, $\gamma(T)=q$ and $\metric{\gamma(t)}{\gamma(t^\prime)}=|t-t^\prime|$ for all $t, t^\prime \in [0, T]$.  The image of $\gamma$ on $[0, T]$ is called a geodesic segment. 
	{Intuitively, a geodesic segment is a path/curve with a shortest length (with respect to the underlying metric) that connects two given points.}
	A metric space $(\metricsp, \metricfn)$ is a geodesic space if every two points in $\metricsp$ can be connected by a geodesic, and is a uniquely geodesic space if this geodesic is unique. We use $\geod{p}{q}$ to denote the geodesic connecting $p$ and $q$ in a uniquely geodesic space.

	\vspace{3mm}
	\noindent
	\textbf{Curvature.} Curvature quantifies the deviation of a general metric space from being flat. A standard way to define curvature is to compare geodesic triangles on the underlying space with geodesic triangles on the following reference spaces $(M_{\kappa}^2,d_\kappa)$:
	\begin{itemize}
		\item When $\kappa=0$, $M_{\kappa}^2=\real^2$ with the standard Euclidean distance;
		\item When $\kappa<0$, $M_{\kappa}^2$ is the hyperbolic space $\mathbb{H}^2=\{(x, y, z)\in \real^3: x^2+y^2-z^2=-1 \text{ and } z>0\}$ with the hyperbolic distance function $d_\kappa(p,q)=\cosh^{-1}(z_p z_q - x_p x_q - y_p y_q)/\sqrt{-\kappa}$, where $p=(x_p, y_p, z_p)$ and $q=(x_q, y_q, z_p)$;
		\item When $\kappa>0$, $M_{\kappa}^2$ is the sphere $\mathcal{S}^2=\{(x, y, z)\in \real^3: x^2+y^2+z^2=1\}$ with the angular distance function $d_\kappa(p,q)=\cos^{-1}(x_p x_q + y_p y_q + z_p z_q)/\sqrt{\kappa}$.
	\end{itemize}
	A geodesic triangle $\triangle(p,q,r)$ with vertices $p, q, r$ in a uniquely geodesic space $\metricsp$ is formed by three geodesic segments, termed sides, that connect $p$ to $q$, $p$ to $r$ and $q$ to $r$, respectively. The sum $\metric{p}{q}+\metric{p}{r}+\metric{r}{q}$ is called the perimeter of $\triangle(p,q,r)$.
	A comparison triangle of $\triangle(p,q,r)$ in the reference space $M_{\kappa}^2$ is a geodesic triangle on  $M_{\kappa}^2$ consisting of vertices $\compoint{p}, \compoint{q}, \compoint{r} \in M_{\kappa}^2$ and sides $\geod{\compoint{p}}{\compoint{q}}, \geod{\compoint{p}}{\compoint{r}}, \geod{\compoint{q}}{\compoint{r}} \subset M_{\kappa}^2$ such that $d_{\kappa}(\compoint{p}, \compoint{q})=\metric{p}{q}$, $d_{\kappa}(\compoint{p}, \compoint{r})=\metric{p}{r}$ and $d_{\kappa}(\compoint{q}, \compoint{r})=\metric{q}{r}$. 
	In addition, $\compoint{x}\in \geod{\compoint{p}}{\compoint{q}}$ is a comparison point of $x\in \geod{p}{q}$ if $d_{\kappa}(\compoint{p}, \compoint{x})=\metric{p}{x}$. Let $D_{\kappa}=\pi / \sqrt{\kappa}$ if $\kappa>0$ and $D_{\kappa}=\infty$ otherwise.
	We say the (global) curvature of $\metricsp$ is lower (upper, respectively) bounded by $\kappa$ if every geodesic triangle with perimeter less than $2D_{\kappa}$ satisfies the following property: There exists a comparison triangle $\triangle(\compoint{p}, \compoint{q}, \compoint{r})$ in $M_{\kappa}^2$ such that $\metric{x}{y}\ge d_{\kappa}(\compoint{x}, \compoint{y})$ ($\metric{x}{y} \le d_{\kappa}(\compoint{x}, \compoint{y})$, respectively) for all $x\in \geod{p}{q}$ and $y\in \geod{p}{r}$ and their comparison points $\compoint{x}$ and $\compoint{y}$ on $\triangle(\compoint{p}, \compoint{q}, \compoint{r})$. Metric spaces with lower or upper bounded curvature are called Alexandrov spaces, and complete geodesic spaces with curvature upper bounded by $0$ are called Hadamard spaces. 
	
	\vspace{3mm}
	\noindent
	\textbf{Angle.} For a metric space $(\metricsp, \metricfn)$ and three distinct points $p, q, r\in \metricsp$, the comparison angle between $q$ and $r$ at $p$, denoted by $\comanlge{p}{q}{r}$, is defined by 
	\[
	\comanlge{p}{q}{r} = \arccos \frac{\metricsq{p}{q}+\metricsq{p}{r}-\metricsq{q}{r}}{2\metric{p}{q}\metric{p}{r}}.
	\]
	Geometrically, for $p,q,r\in\metricsp$ that form a geodesic triangle $\triangle (p,q,r)$, the comparison angle $\bar\angle_p(q,r)$ is the Euclidean angle at $\bar p$ of the comparison triangle $\triangle(\bar p,\bar q,\bar r)$ in $\real^2$. 
	The Alexandrov angle between two geodesics $\gamma$ and $\eta$ emanating from $p$ in a uniquely geodesic space, denoted by $\alexanlge{p}{\gamma}{\eta}$, is defined by
	\begin{equation}\label{eq:alex-angle}
		\alexanlge{p}{\gamma}{\eta} = \lim_{r\to 0} \sup_{0<t,s<r} \comanlge{p}{\gamma(t)}{\eta(s)}.
	\end{equation}
	In a curved space, the comparison angle $\comanlge{p}{\gamma(t)}{\eta(s)}$ varies with  $t$ and $s$ even though the directions of $\gamma$ and $\eta$ remain unchanged. In contrast, the Alexandrov angle $\alexanlge{p}{\gamma}{\eta}$ intrinsically depends only on the directions of $\gamma$ and $\eta$ (i.e., invariant to  $t$ and $s$), but not on their lengths. 
	For three distinct points $p, q, r\in\metricsp$, we define the angle $\alexanlge{p}{q}{r}=\alexanlge{p}{\geod{p}{q}}{\geod{p}{r}}$.
	
	\vspace{3mm}
	\noindent
	\textbf{Riemannian Manifolds.} A differentiable manifold is a topological manifold  with a maximal differentiable atlas that induces a differential structure on the manifold. 	When all transition maps in the atlas are $C^{\infty}$ differentiable, the manifold is a smooth manifold. 	For each point $p$ on a smooth manifold $\metricsp$, there is a linear space $\tanbundle{p}{\metricsp}$ attached to $p$, called the tangent space at $p$. The elements in $\tanbundle{p}{\metricsp}$ are called tangent vectors. 	A Riemannian manifold is a smooth manifold $\metricsp$ with a Riemannian metric that defines a positive-definite inner product $\innerprodM{\cdot}{\cdot}{p}$ on the tangent space $\tanbundle{p}{\metricsp}$ for each $p\in\manifold$ and  smoothly varying with $p$. 	An affine connection on a smooth manifold provides a way to compare tangent vectors at distinct points on the manifold via the notion of parallel transport. Each Riemannian manifold has a canonical affine connection, the Levi--Civita connection, which is torsion-free and compatible with the Riemannian metric.
	For a smooth function $f:\manifold\rightarrow\real$,  the Hessian of $f$ at $p\in\manifold$ is defined by 
	\begin{equation}\label{eq:hessian}
		(\hess_p f)(\xi) = \nabla_{\xi} \text{grad }f
	\end{equation}
	for all $\xi \in \tanbundle{p}{\metricsp}$, where $\nabla$ is the Levi--Civita connection  and $\text{grad}$ is the gradient operator \citep[Page 45,][]{lee2018introduction}. It can be seen that is a linear operator on the tangent space $\tanbundle{p}{\metricsp}$. 
	
	\section{Binary Regression in Metric Spaces}\label{sec:meth}
	
	\subsection{Model}
	To motivate our model for a binary response with a covariate residing in a metric space, we observe that the Euclidean logistic regression model with the covariate $X=(X_1,\ldots,X_D)\in\real^D$, when appropriately reparameterized, can be represented as follows:
	\begin{equation}
		\text{logit}(P_{X}) = \innerprodE{X-\mu^\ast}{{\beta}^{*}-\mu^\ast}=\metricE{\truemu}{X}  \metricE{\truemu}{\truebeta} \cos(\alexanlge{\truemu}{X}{\truebeta}),
		\label{LR-model}
	\end{equation}
	where {$P_{X} = \prob(Y=1\mid X)$ is the success probability, $\text{logit}(t)=\log(\frac{t}{1-t})$ is the logit function for $t \in [0,1]$, ${\beta}^{*}=({\beta}_1^{*}, \ldots, {\beta}_D^{*})$ is the true coefficient vector,  $\innerprodE{\cdot}{\cdot}$ is the standard Euclidean inner product, $\truemu=\expect X$ is the mean, $d_E$ is the standard distance on $\real^D$, and $\alexanlge{\truemu}{X}{\truebeta}$ is the Euclidean angle between $X$ and $\truebeta$ at $\truemu$. Without loss of generality, in the above we have assumed that the intercept is zero; an extension to the general case is straightforward (see also Remark \ref{rem:1}). 	
		
		The logistic regression model in the form of \eqref{LR-model} inspires us to model the log odds influenced by a metric-space valued covariate $X$, as follows. Let $(\metricsp,d)$ be a uniquely geodesic metric space and $X\in\metricsp$, where $d$ is the distance on $\metricsp$. First, the Euclidean mean is replaced with its popular extension in general metric spaces, i.e., the  Fr{\'e}chet mean 
		\begin{equation}
			\truemu = \argmin_{\mu \in \metricsp} \expect[\metricpoly{X}{\mu}{2}],
			\label{frechet-mean}
		\end{equation} 
		which is well studied in the literature and is reduced to the Euclidean mean $\expect X$ when $\metricsp=\real^D$. In addition, the angle $\alexanlge{\truemu}{X}{\truebeta}$ in \eqref{LR-model} is naturally generalized to general metric spaces by the Alexandrov angle between the geodesics $\geod{\truemu}{X}$ and $\geod{\truemu}{\truebeta}$ emanating from $\truemu$. 
		Then we propose to model the effect of the metric-space valued covariate $X\in\metricsp$ on the log odds by 
		\begin{equation}
			\text{logit}(P_{X}) = h(\truebeta;X, \truemu) \define \metric{\truemu}{X} \metric{\truemu}{\truebeta} \cos(\alexanlge{\truemu}{X}{\truebeta}),
			\label{LR-model-metric}
		\end{equation}
		where  $\truebeta\in \metricsp$ is a population parameter to be estimated and $\alexanlge{\truemu}{X}{\truebeta} = \alexanlge{\truemu}{\geod{\truemu}{X}}{\geod{\truemu}{\truebeta}}$ is the Alexandrov angle. 
		
		The right-hand side of \eqref{LR-model-metric}, termed the Alexandrov inner product  \citep{lin2021total}, is interpreted as the conditional log odds of $Y=1$ and $Y=0$ given $X$. In addition, $\metric{\truemu}{X} \cos(\alexanlge{\truemu}{X}{\truebeta})$ in \eqref{LR-model-metric} is interpreted as the projection of the geodesic $\geod{\truemu}{X}$ onto the geodesic $\geod{\truemu}{\truebeta}$, and thus $d(\truemu,\truebeta)$ represents the expected change in log odds based on a one-unit change of this projection. Alternatively, the projection $\metric{\truemu}{\truebeta} \cos(\alexanlge{\truemu}{X}{\truebeta})$  represents the expected change in log odds based on a one-unit change of $X$ along the geodesic $\gamma_{\truemu}^X$ emanating from $\truemu$ to $X$; {in Section \xref{sec:odds-ratio} of Supplementary Material, we also show that $\metric{\truemu}{\truebeta} \cos(\alexanlge{\truemu}{X}{\truebeta})$ represents a  log odds ratio.} These indicate that the geodesic $\geod{\truemu}{\truebeta}$ identifies the direction in which the log odds change most rapidly, thereby encapsulating critical information of the covariates for distinguishing the two classes of the response variable; see the applications below for a demonstration.

		The proposed model \eqref{LR-model-metric} has at least two potential applications. First, based on the relationship between the log odds and the geodesic $\geod{\truemu}{\truebeta}$ in the above,  practitioners may use \eqref{LR-model-metric} as a regression model to  gain insights into {distinctive patterns in covariates} between the two classes via examining an estimate of the geodesic $\geod{\truemu}{\truebeta}$. For an illustration, take the example of investigating how the brain functional connectivity is associated with different tasks related to certain human functions \citep{Salvo2021}, such as the motor and language processing tasks. To study this question of scientific interest, one may focus on 8 brain regions associated with these two tasks and utilize a set of $8\times8$ SPD (symmetric positive-definite) matrices derived from fMRI images, each matrix representing connectivity among these regions when a subject performs a task. Applying model \eqref{LR-model-metric} and a maximum likelihood estimator (given below) to analyze this data provides the geodesic estimate $\gamma_{\hat\mu}^{\hat\beta}(t)$, as shown in Figure \ref{fig:geodesic-hat}. As $t$ increases, it is observed that Region 1 (the superior temporal gyrus) becomes more active, while its connectivity with Regions 5–8 (encompassing the primary motor and visual cortex) diminishes. Thus, practitioners may infer that there is a reduced functional interaction between the superior temporal gyrus and the primary motor and visual cortex during the language processing task. Further demonstration of this application is available in Section \ref{sec:application}.
		
		Secondly, the classifier derived from \eqref{LR-model-metric}, which classifies $X$ into the class $Y=1$ if $P_X\geq 0.5$ and into $Y=0$ otherwise, allows practitioners to classify a metric-space valued covariate $X$ based on an estimate of the conditional success probability $P_X=\prob(Y=1\mid X=x)=\{1+\exp(-h(\truebeta;x,\truemu))\}^{-1}$; see Section \ref{sec:theory-classification} for more details. For instance, in the context of analyzing functional connectivity and task-specific brain functions, the classifier can be used to predict the task being executed, utilizing the observed $8\times 8$ SPD matrix that depicts connectivity among the 8 brain regions. As outlined in Section \ref{sec:application}, our findings suggest that this model-based classifier can surpass nonparametric methods like the $k$-nearest-neighbors algorithm and neural networks in performance. This could be attributed to the model's interpretability and its efficiency in handling the relatively high dimensionality inherent to the data of $8\times 8$ SPD matrices, which have an intrinsic dimension of 36, where nonparametric classifiers might struggle due to the curse of dimensionality.

		\begin{remark}\label{rem:1}
			The  model \eqref{LR-model-metric} can naturally accommodate both Euclidean and metric-space valued covariates. For instance, suppose $X_1\in\real^D$ and $X_2\in\manifold$. To include both $X_1$ and $X_2$ into the proposed model, we can treat $X=(X_1,X_2)$ jointly as a covariate on the product metric space $\real^D\times\manifold$, and then apply the model \eqref{LR-model-metric} to $X$. {For instance, with $X_1\equiv 1$, we can include an intercept term in the model \eqref{LR-model-metric}.} A similar strategy can be used to accommodate multiple covariates potentially residing in different metric spaces.
		\end{remark}
		
		\begin{remark}
			The proposed model \eqref{LR-model-metric} encompasses the Euclidean logistic model as a special case when  $\metricsp=\real^D$. 
			In light of the resemblance between our model \eqref{LR-model-metric} and the Euclidean logistic model \eqref{LR-model}, the former may be referred to as the metric-space logistic regression model, although it does not possess all properties of the latter due to the lack of Euclidean geometry and algebraic structures in \eqref{LR-model-metric}. Nonetheless, we opt to use the term ``binary regression'' in this paper rather than ``logistic regression'' to prevent potential misconceptions.
			
			The strength of the proposed model \eqref{LR-model-metric} stems from its resistance to the curse of dimensionality (when compared to nonparametric methods) and  its ability to help practitioners investigate the relationship between the log odds and metric-space valued covariates, as demonstrated in the above and in Section \ref{sec:application}. 
			Alternative models that are inspired by other features of Euclidean logistic regression are beyond the scope of this paper and left for future studies.
		\end{remark}

		\subsection{Maximum Likelihood Estimator}
		To estimate the parameters $\truemu$ and $\truebeta$, let $(X_i, Y_i)$, $i=1,\ldots, n$, be  
		independent and identically distributed (i.i.d.) observations of $(X, Y)\sim F$ for a joint distribution $F$ on $\manifold\times \{0,1\}$. We first estimate the Fr{\'e}chet mean $\truemu$ by the sample Fr{\'e}chet mean
		\begin{equation}
			\estmu = \argmin_{\mu \in \metricsp} \frac 1 n \sum_{i=1}^{n} \metricpoly{X_i}{\mu}{2}.
			\label{sample-frechet-mean}
		\end{equation}
		For $\truebeta$, according to the  proposition below, we observe that it is the maximizer of the population log-likelihood function $\poplik{\beta}$, i.e.,  
		\begin{equation}
			\truebeta = \argmax_{\beta \in \metricsp} \poplik{\beta} \quad\text{with}\quad \poplik{\beta} \define \expect \left[ Y h(\beta;X, \truemu) - \log(1+e^{h(\beta;X, \truemu)}) \right].
			\label{pop-argmax}
		\end{equation}
		\begin{proposition}\label{prop:well-specified}
			If  $(X, Y)$ is a random pair sampled from the model \eqref{LR-model-metric} with the parameters $\truebeta$ and $\truemu$, then $\truebeta \in \argmax_{\beta \in \metricsp} \poplik{\beta}$. Moreover, $\truebeta = \argmax_{\beta \in \metricsp} \poplik{\beta}$ when $\poplik{\beta}$ has a unique maximizer. 
		\end{proposition} 
		
		This  motivates us to estimate $\truebeta$ by maximizing the empirical version of $\poplik{\beta}$, 
		\begin{align}
			\notag
			\emplik{\beta}
			&= n^{-1}\sum_{i=1}^n \left\{ Y_i h(\beta;X_i, \estmu) - \log(1+e^{h(\beta;X_i, \estmu)}) \right\},
		\end{align}
		where $h(\beta;X_i, \estmu) = \metric{\estmu}{X_i} \metric{\estmu}{\beta} \cos(\alexanlge{\estmu}{X_i}{\beta})$. 
		The maximum likelihood estimator (MLE) of $\truebeta$ is then
		\begin{equation}
			\estbeta {\in} \argmax_{\beta \in \metricsp} L_n(\beta).
			\label{empirical-argmax}
		\end{equation}
		Existence and uniqueness of $\estbeta$, as well as $\truemu$, $\truebeta$ and $\estmu$, are discussed in Section \ref{subsec:assump}.
		
		Optimization methods for \eqref{empirical-argmax} in both general metric spaces and Riemannian manifolds, including how to compute the Alexandrov inner product in these spaces, are detailed in Section \xref{sec:RGD} of Supplementary Material.

		\section{Theoretical Properties for Estimation}\label{sec:theory-estimation}
		
		\subsection{Assumptions}\label{subsec:assump}
		
		To study the {theoretical} properties of $\estbeta$ in \eqref{empirical-argmax}, we require the following two geometric assumptions on the metric space $\metricsp$. 
		
		\begin{assumption}\label{as:space}
			$\metricsp$ is a complete uniquely geodesic metric space.
		\end{assumption}
		
		\begin{assumption}\label{as:lip}
			There exist constants $C_U>0$ and {$\alpha_U \in (0, 1]$} such that, for all $p,q,r,u\in \metricsp$, 
			\[
			|\metric{p}{q} \cos(\alexanlge{p}{u}{q}) - \metric{p}{r} \cos(\alexanlge{p}{u}{r})| \le C_U \metricpoly{q}{r}{\alpha_U}.
			\]
		\end{assumption}
		
		Assumption \ref{as:space} is standard in metric-space data analysis. With this assumption, any pair of points in $\metricsp$ can be connected by a unique geodesic. Assumption \ref{as:lip} is a regularity condition on the curvature of $\metricsp$, which plays a similar role of the weak quadruple inequality in \cite{schotz2019convergence}. These two assumptions are satisfied with $\alpha_U=1$ for Hadamard spaces and some subspaces of positively curved Alexandrov spaces, according to Propositions \xref{prop:had-lip} and \xref{prop:alex-lip} in Section \xref{sec:proposition-lip-metric} of Supplementary Material. 
		
		\begin{example}\label{ex:spaces}
			Among many others, below are some concrete examples of metric spaces satisfying  both Assumption \ref{as:space} and Assumption \ref{as:lip} (with $\alpha_U=1$).
			\begin{enumerate}
				\item[(a)] The space of SPD matrices, endowed with either of the Riemannian metrics of \cite{moakher2005differential,Arsigny2007,lin2019riemannian}, for modeling diffusion tensors and brain functional connectivity in brain sciences.
				\item[(b)] The positive quadrant $\{(x_0,\ldots,x_D)\in\mathcal S^D: x_j\geq 0\text{ for }j=0,\ldots,D\}$ of the unit sphere $\mathcal S^D$, endowed with the angular distance, for modeling compositional data of $D+1$ components.
				\item[(c)] The tree space \citep{Billera2001} in evolutionary biology.
				\item[(d)] The Wasserstein space of probability distributions defined on a compact domain of $\real$, endowed with the optimal transport geometry, for analyzing human mortality data \citep{Chen2021}. 
				
			\end{enumerate}
		\end{example}
		
		We also require the following two \emph{distributional} assumptions on $(X,Y)$. Their variants are commonly adopted for metric-space data analysis, and their 
		Euclidean counterparts are commonly seen in the literature of $M$-estimation; see the discussions below.
		
		\begin{assumption}\label{as:existence}
			The distribution of $X\in \metricsp$ has a bounded support, and the objects $\truemu$ and $\estmu$ exist and are unique, the later almost surely. 
			Moreover, $\estbeta$ exists almost surely. 
		\end{assumption}
		
		\begin{assumption}\label{as:margin-global}
			There exist constants $\lambda_X>0$, $C_L>0$ and $\alpha_L > 1+{2/m}$ for some constant $m\geq 4$ with  $\expect[\metricpoly{\truemu}{X}{m}]<\infty$, such that, for any $\beta\in \metricsp$, 
			\[
			\expect\big[|h(\beta;X, \truemu)-h(\truebeta;X, \truemu)|^2\big] \ge C_L \lambda_{X} \metricpoly{\beta}{\truebeta}{\alpha_L}.
			\]
		\end{assumption}

		Existence and uniqueness of the Fr\'echet means $\truemu$ and $\estmu$, often presumed in data analysis within general metric spaces, are influenced by the distribution of $X$. Nonetheless, in certain metric spaces, the requirement of a finite second moment of $X$ is adequate. For example,  for any Hadamard space $\metricsp$, both $\truemu$ and $\estmu$ exist and are unique  when  $\expect[\metricpoly{X}{u}{2}] < \infty$ for some $u\in\metricsp$  \citep[Proposition 4.3,][]{sturm2003probability}, regardless of the distribution of $X$. For other metric spaces,  a convexity condition on the distance function can guarantee existence and uniqueness of both $\truemu$ and $\estmu$ when data reside in a geodesically convex region \citep{lin2021total}. See also \cite{Bhattacharya2003,Afsari2011} for sufficient conditions on $X$ when $\metricsp$ is a Riemannian manifold.
		
		For $\truebeta$,  Proposition \ref{prop:well-specified} establishes its existence. In Section \xref{subsec:uniqueness-MLE} of Supplementary Material, Proposition \xref{prop:unique-MLE} further establishes the uniqueness of $\truebeta$ under additional Assumption \ref{as:margin-global}, with two concrete examples provided in the same section. For $\estbeta$, our theoretical analysis does not require uniqueness of $\estbeta$, since the theoretical results in the sequel hold for \emph{any} measurable choice of the maximizers $\estbeta$ in \eqref{empirical-argmax}; such measurability is commonly adopted in the M-estimation theory \citep[e.g., Pages 285--286 of][]{van1996weak} and object data analysis \citep[e.g., Theorem 2.3 of][]{Bhattacharya2003}.
		
		The existence of the MLE $\estbeta$ appears to be relatively subtle. When $\metricsp$ is complete and totally bounded, $\estbeta$ exists (Proposition \xref{prop:exist-MLE} in Supplementary Materials). For unbounded metric spaces, specific distributional conditions for the pair $(X,Y)$ are required. This necessity applies even within Euclidean spaces $\metricsp=\real^D$, where conditions such as the overlap condition \citep{albert1984existence} and the proper convexity condition \citep{silvapulle1981existence} are pertinent. To illustrate this type of distributional conditions outside of Euclidean realms, in Section \xref{subsec:ex-as-exist} of Supplementary Material we construct an analogous condition through the example of the SPD matrix manifold. A detailed investigation into the distributional conditions on $(X,Y)$ in the context of metric spaces, whether sufficient or necessary for $\estbeta$'s existence, is a subject deserving of its own comprehensive study in future research.

	Assumption \ref{as:margin-global} is common in the  $M$-estimation theory. It regulates the behavior of $\emplikfn-\poplikfn$ near $\truebeta$ and is related to the rate of convergence \citep[see,][Section 3]{ahidar2020convergence}. In addition, if $\manifold=\real^D$, then Assumption \ref{as:margin-global} is reduced to the usual condition that 
	$\expect[(X-\truemu)(X-\truemu)^{\top}]$ is positive-definite with $\lambda_{X}$ being the smallest eigenvalue of $\expect[(X-\truemu)(X-\truemu)^{\top}]$. Therefore, Assumption \ref{as:margin-global} generalizes this basic distributional condition from Euclidean space to general metric spaces. 
	Examples of $X$ satisfying Assumption \ref{as:margin-global} are given in Section \xref{subsec:ex-as3-sphere} of Supplementary Material. 
	
	\begin{remark}
		Although certain conditions of Assumption \ref{as:existence}  may be satisfied for certain metric spaces according to the above discussions, we shall  highlight that both Assumption \ref{as:existence} and Assumption \ref{as:margin-global} generally relate to the distribution of $(X,Y)$ and are not solely dependent on the geometry of $\metricsp$, even in the special case of Euclidean space. Hence, it is generally not possible to specify which spaces meet these two assumptions without considering the underlying distributions.
	\end{remark}
	
	\begin{remark}
		We do not assume the space $\metricsp$ to be bounded.  In this general setting, the bounded support or similar conditions on $X$ are typically assumed to establish the consistency and convergence rates of the MLE $\estbeta$ even for the Euclidean logistic regression model \eqref{LR-model} \citep{fahrmeir1985consistency, fahrmeir1986asymptotic, yin2006asymptotic,van2008high, meier2008group, bach2010self,liang2012maximum, negahban2012unified, ning2017general, guo2021inference}.
		This boundedness condition on the data support, as one might expect, could potentially be relaxed to allow the data space to expand at a certain rate as the sample size grows; a similar relaxation in Euclidean spaces has been explored, for example, by \cite{fahrmeir1986asymptotic}. Alternatively, it may be replaced with certain conditions (e.g., sub-Gaussianity) on the tail probability of the distribution of $X$. Since such relaxations involve much heavier technicalities without providing additional insights, we have chosen to omit them in this paper to maintain a more focused presentation.
	\end{remark}
	
	\begin{remark}
		The constraint $m \ge 4$ in Assumption \ref{as:margin-global} is imposed to obtain the rate of convergence for the expected {squared} estimation error $\expect\big[\metricpoly{\truebeta}{\estbeta}{2}\big]$; for the expected estimation error $\expect\big[\metricpoly{\truebeta}{\estbeta}{}\big]$, $m \ge 2$ is sufficient. 
		Especially, if $\metricsp$ is bounded, the constraint $\alpha_L>1+2/m$ in Assumption \ref{as:margin-global} can be reduced to $\alpha_L>1$, since now $\expect [d^m(\truemu,X)]<\infty$ for any arbitrarily large but fixed $m$.
		In addition,  $\alpha_L > 1$ is sufficient for establishing the convergence rate of  $\metricpoly{\truebeta}{\estbeta}{}$ in probability.
	\end{remark}
	
	In the following subsections we present convergence rates of $\estbeta$  under two types of metric entropy conditions that are widely used in the literature to  quantify  space complexity.
	
	\vspace{3mm}
	\noindent
	\textbf{Notation.}
	For two non-negative sequences $\{a_n\}$ and $\{b_n\}$, we write $a_n \lesssim b_n$ (respectively, $ a_n\gtrsim b_n$) if there is a constant $c>0$ not depending on $n$, such that $a_n \le c b_n$ (respectively, $a_n\ge c b_n$) for all sufficiently large $n$. We write $a_n \asymp b_n$ if and only if both $a_n \lesssim b_n$ and $a_n \gtrsim b_n$ are true.
	
	\subsection{Rates of Convergence and Optimality with Local Metric Entropy Condition}
	
	The first type of the condition on metric entropy, termed the {(local)} metric entropy condition in this paper, concerns the metric entropy in a local ball of $\metricsp$. Let $\ball{\truebeta}{\delta} \subset \metricsp$ be the (closed) ball of the radius $\delta>0$ centered at $\truebeta$ and $\covernum{\epsilon}{\ball{\truebeta}{\delta}}{\metricfn}$ be its covering number \citep[][Section 5.1]{wainwright2019high} using balls of the radius $\epsilon>0$. Metric entropy then refers to $\log \covernum{\epsilon}{\ball{\truebeta}{\delta}}{\metricfn}$.  We consider the following two  metric entropy conditions of different scales.
	
	\begin{assumption}\label{as:entropy-global}
		There is a constant $D>0$, such that one of the following holds:
		\begin{enumerate}
			\item[(a)](Log-polynomial metric entropy) $ \covernum{\epsilon}{\ball{\truebeta}{\delta}}{\metricfn} \lesssim   \left(\frac{\delta}{\epsilon}\right)^{D}$, 
			\item[(b)](Polynomial metric entropy)  $\log \covernum{\epsilon}{\ball{\truebeta}{\delta}}{\metricfn} \lesssim \left(\frac{\delta}{\epsilon}\right)^{D}$.
		\end{enumerate}
	\end{assumption}
	
	Assumption \ref{as:entropy-global}(a), in which the metric entropy $\log \covernum{\epsilon}{\ball{\truebeta}{\delta}}{\metricfn}$ is of the order $D\log(\delta/\epsilon)$, is often satisfied by \emph{finite-dimensional} spaces, such as the spaces (a)--(c) in Example \ref{ex:spaces}.  
	More generally, Assumption \ref{as:entropy-global}(a) holds for $D$-dimensional Riemannian manifolds of bounded curvature and  {metric spaces} that admit an Ahlfors--David regular measure \citep[see Example 2.3,][]{ahidar2020convergence}.
	
	Assumption \ref{as:entropy-global}(b), when compared to  \ref{as:entropy-global}(a), allows the covering number to grow in a much faster rate and thus accommodates spaces of much higher complexity, such as the Wasserstein space in Example \ref{ex:spaces}(d). In fact, for this Wasserstein space, 
	one can show  $\log \covernum{\epsilon}{\ball{\truebeta}{\delta}}{\metricfn_W} \asympeq {\delta}/{\epsilon}$, where $d_W$ is the $2$-Wasserstein distance, based on the metric entropy of the set of all non-decreasingly bounded functions on $[0,1]$ \citep{yang1999minimax} and by following the arguments in the proof of Proposition 1 in \cite{petersen2019frechet}. Thus, this Wasserstein space satisfies Assumption \ref{as:entropy-global}(b) with $D=1$, but not  Assumption \ref{as:entropy-global}(a).
	
	Define 
	\[
	v_{L,n} = \left(\frac{D}{n}\right)^{\frac{1}{2(\alpha_L - \alpha_U)}}
	\]
	when $\manifold$ satisfies Assumption \ref{as:entropy-global}(a), and
	
	\[
	v_{L,n} =
	\begin{cases}
		\big(\frac{1}{n}\big)^{\frac{1}{2(\alpha_L - \alpha_U)}} & if~ D<2\alpha_U, \\
		\big(\frac{\log^2 n}{n}\big)^{\frac{1}{2(\alpha_L - \alpha_U)}}  & if~ D=2\alpha_U, \\
		{\big(\frac{1}{n}\big)^{\frac{\alpha_U}{D(\alpha_L-\alpha_U)}}} & if~ D>2\alpha_U
	\end{cases}
	\]
	when $\manifold$ satisfies Assumption \ref{as:entropy-global}(b). 
	With this definition, the following theorems provide upper and lower bounds on the rate of convergence for the estimator $\estbeta$ under the  metric entropy assumption. 
	
	\begin{theorem}\label{thm:upperbound-global-expect-entropy-simplified}
		Suppose that Assumptions \ref{as:space}--\ref{as:entropy-global} hold. Then, there is a constant $c>0$ depending on the constants in the assumptions, such that for any measurable sequence $\estbeta$ indexed by $n$ and determined by \eqref{empirical-argmax}, 
		\[
		\expect	 \big[\metricfn^2(\truebeta,\estbeta)\big] \le c v_{L,n}^2.
		\]
	\end{theorem}

	\begin{theorem}\label{thm:lowerbound-global-entropy-simplified}
		Suppose $\sigma_{X}^2= \expect[\metricsq{\truemu}{X}]<\infty$ and $\metricsp$ satisfies Assumptions \ref{as:space} and \ref{as:lip}. 
		If $(X_i, Y_i), i=1,\ldots, n$, are i.i.d. sampled from the model \eqref{LR-model-metric}, then 
		\[
		\inf_{\hat{\beta}_n} \sup_{\beta \in \metricsp} {\expect_{\beta}[\metricfn^2(\beta,\estbeta)]}
		\gtrsim 
		\begin{cases}
			\left(\frac{D}{\sigma_X^2 n}\right)^{\frac 1{\alpha_U}} & 
			\text{ when } \covernum{\epsilon}{\ball{\truebeta}{\delta}}{\metricfn} \asymp  \left(\frac{\delta}{\epsilon}\right)^{D}, \\
			\left(\frac{1}{\sigma_X^2 n}\right)^{\frac{1}{\alpha_U}} & \text{ when }\log \covernum{\epsilon}{\ball{\truebeta}{\delta}}{\metricfn} \asymp \left(\frac{\delta}{\epsilon}\right)^{D}.
		\end{cases}
		\]
		In the above, the infimum is taken over all estimators $\estbeta$, and 
		$\expect_{\beta}$ is the expectation with respect to the distribution induced by the model \eqref{LR-model-metric} with $\truebeta=\beta$.
	\end{theorem}
	
	According to Theorem \ref{thm:upperbound-global-expect-entropy-simplified} and Theorem \ref{thm:lowerbound-global-entropy-simplified}, for the commonly encountered scenario where $\alpha_L=2\alpha_U$,  
	the proposed estimator $\estbeta$ is rate-optimal under Assumption \ref{as:entropy-global}(a) of the log-polynomial  metric entropy condition. It is also optimal under Assumption \ref{as:entropy-global}(b) of the polynomial metric entropy condition with $D \le 2\alpha_U$ (up to a logarithmic factor when $D=2\alpha_U$). 
	To the best of our knowledge, this is the first optimality result in parametric binary regression with covariates residing in a general metric space.
	
	Among nonlinear metric spaces, Riemannian manifolds, which satisfy Assumption \ref{as:entropy-global}(a),  are  most commonly seen in statistics and have a wide range of applications \citep[e.g.,][]{pennec2006riemannian, Dryden2009}. 
	As an application of Theorems \ref{thm:upperbound-global-expect-entropy-simplified} and \ref{thm:lowerbound-global-entropy-simplified}, the following corollary demonstrates that, for certain Riemannian manifolds, the proposed estimator is rate-optimal in estimation. Recall that a Hadamard manifold is a simply connected and complete Riemannian manifold  that is also a Hadamard space. Let $h(\truebeta)=h(\truebeta;X,\truemu)$ be viewed as a function of $\truebeta$ and let $\Lambda_{\min}(\cdot)$ denote the smallest eigenvalue of a linear operator. Recall the Hessian defined in \eqref{eq:hessian}.
	\begin{corollary}
		\label{cor:opt-estimation-manifold}
		For a $D$-dimensional Hadamard manifold $\metricsp$ of bounded sectional curvature, if the distribution of $X$ has a bounded support, the smallest eigenvalue $\lambda_{X}$ of $\expect[\hess_{\truebeta}h]$ satisfies $\lambda_{X}>0$, and {$\estbeta$ exists almost surely}, then
		\[
		\expect[\metricfn^2(\beta,\estbeta)] \asymp D/n.
		\]
		The same conclusion holds for a complete and simply connected $D$-dimensional Riemannian manifold  $\manifold$ with non-negative sectional curvature upper bounded by $\kappa>0$ and of diameter no larger than $D_\kappa/2$ such that $\inf_{p,q\in\manifold}\Lambda_{\min}(\hess_q d_p^2)>0$, where $d_p^2(\cdot)=d^2(\cdot,p)$.
	\end{corollary}
	A proof of this corollary follows directly from verifying Assumptions \ref{as:space}--\ref{as:entropy-global}, which is presented in the proof of Theorem \ref{thm:excessrisk-upper-simplified}. Examples of  bounded and unbounded Riemannian manifolds that satisfy the requirement for the Hessian of $h$ are provided in Section \xref{subsec:ex-as3-sphere} of Supplementary Material.

	\subsection{Rates of Convergence and Optimality with Global Metric Entropy Assumption}
	In the case that $\mathcal M$ is totally bounded, the global metric entropy condition   \citep{yang1999information}, concerning the minimal number of balls to cover the \emph{entire} metric space $\manifold$, is  another widely adopted  condition on metric entropy.
	\begin{assumption}\label{as:global-entropy-global}
		There is a constant $D>0$, such that one of the following holds:
		\begin{enumerate}
			\item[(a)](Log-polynomial global metric entropy) $\covernum{\epsilon}{\metricsp}{\metricfn} \lesssim   \left(\frac{1}{\epsilon}\right)^{D}$, 
			\item[(b)](Polynomial global metric entropy)  $\log \covernum{\epsilon}{\metricsp}{\metricfn} \lesssim \left(\frac{1}{\epsilon}\right)^{D}$.
		\end{enumerate}
	\end{assumption}
	
	Assumption \ref{as:global-entropy-global} can be deduced from Assumption \ref{as:entropy-global} when $\metricsp$ is bounded, and hence is weaker than the latter. This may make Assumption \ref{as:global-entropy-global} relatively easier to check. All finite-dimensional compact Riemannian manifolds satisfy Assumption \ref{as:global-entropy-global}(a). A collection of $D$-dimensional Riemannian manifolds embedded in a common compact subspace of $\real^D$, when endowed with the Hausdorff distance,  satisfies Assumption \ref{as:global-entropy-global}(b) \citep[Theorem 9,][]{genovese2012minimax}. Assumption \ref{as:global-entropy-global} is widely adopted in the literature of classification and/or metric-space data analysis \citep[][among  others]{ahidar2020convergence, yang1999minimax, genovese2012minimax, lin2021total}.  For instance,  \cite{ahidar2020convergence} derived upper bounds on the rate of convergence of the generalized Fr{\'e}chet mean in metric spaces under Assumption \ref{as:global-entropy-global}(b). 
	
	Let $\gamma_0=W_0(1/2)\approx 0.35173$, where $W_0$ is the Lambert W function. With Assumption \ref{as:global-entropy-global}, define 
	\[
	v_{G,n} = \left(\frac{D}{n}\right)^{\frac{1}{2(\alpha_L - \alpha_U+\gamma_0)}}
	\]
	when $\manifold$ satisfies Assumption \ref{as:global-entropy-global}(a), and
	\[
	v_{G,n} = 
	\begin{cases}
		\big(\frac{1}{n}\big)^{\frac{1}{2(\alpha_L - \alpha_U + D/2)}} & if~ D<2\alpha_U, \\
		\big(\frac{\log^2 n}{n}\big)^{\frac{1}{2\alpha_L}}  & if~ D=2\alpha_U, \\
		{\big(\frac{1}{n}\big)^{\frac{\alpha_U}{D\alpha_L}}} & if~ D>2\alpha_U
	\end{cases}
	\]
	when $\manifold$ satisfies Assumption \ref{as:global-entropy-global}(b).
	The following theorems provide upper and lower bounds on $\expect	 \big[\metricfn^2(\truebeta,\estbeta)\big]$ under each condition of Assumption \ref{as:global-entropy-global}.
	
	\begin{theorem}\label{thm:upperbound-global-expect-global-entropy-simplified}
		Suppose that Assumptions \ref{as:space}--\ref{as:margin-global} and \ref{as:global-entropy-global} hold. Then, there is a constant $c>0$ depending on the constants in the assumptions, for any measurable sequence $\estbeta$ indexed by $n$ and determined by \eqref{empirical-argmax}, such that
		\[
		\expect	 \big[\metricfn^2(\truebeta,\estbeta)\big] \le c v_{G,n}^2.
		\]
	\end{theorem}
	
	\begin{theorem}\label{thm:lowerbound-global-global-entropy-simplified}
		Suppose $\sigma_{X}^2= \expect[\metricsq{\truemu}{X}]<\infty$ and $\metricsp$ satisfies Assumptions \ref{as:space} and \ref{as:lip}.  
		If $(X_i, Y_i), i=1,\ldots, n$, are i.i.d. sampled from the model \eqref{LR-model-metric}, then 
		\[
		\inf_{\hat{\beta}_n} \sup_{\beta \in \metricsp} {\expect_{\beta}[\metricfn^2(\beta,\estbeta)]} 
		\gtrsim
		\begin{cases}
			\left(\frac{D}{\sigma_X^2 n}\right)^{\frac {1}{\alpha_U}} & \text{ when }\covernum{\epsilon}{\metricsp}{\metricfn} \asymp \left(\frac{1}{\epsilon}\right)^{D}, \\
			\left(\frac{1}{\sigma_{X}^2 n}\right)^{\frac{1}{\alpha_U+D/2}} & \text{ when }\log \covernum{\epsilon}{\metricsp}{\metricfn} \asymp \left(\frac{1}{\epsilon}\right)^{D}.
		\end{cases}
		\]
		In the above, the infimum is taken over all estimators $\estbeta$ and 
		$\expect_{\beta}$ is the expectation with respect to the distribution induced by the model \eqref{LR-model-metric} with $\truebeta=\beta$.
	\end{theorem}
	
	Theorems \ref{thm:upperbound-global-expect-global-entropy-simplified} and \ref{thm:lowerbound-global-global-entropy-simplified} show that, if $\alpha_L=2\alpha_U$, then the  estimator $\estbeta$ is rate-optimal under the polynomial global metric entropy condition with $D \le 2\alpha_U$ (up to a logarithmic factor when $D=2\alpha_U$). Under the log-polynomial global metric entropy condition, there is a small gap $\gamma_0$  between the upper and lower bounds. {Similar gaps are also observed in other estimation problems \citep{yang1999information,ma2015volume}. Closing such a gap is challenging especially for nonlinear metric spaces and  left for future study.}

	\section{Logistic Classifiers}\label{sec:theory-classification}
	
	Plugging the estimators $\estbeta$ and $\estmu$ into the model \eqref{LR-model-metric}, we can estimate the success probability $\prob(Y=1\mid X=x)$ and obtain a classifier by classifying the instance $x\in\manifold$ accordingly. 
	Specifically, with $\hat{P}_{X} = \{1+\exp(-h(\estbeta; X, \estmu))\}^{-1}$ denoting the estimated success probability, we propose the classifier
	\begin{equation}
		\hat s(X) = \left\{
		\begin{aligned}
			1 &\quad \text{if}~\hat P_{X} \ge 1/2~ \text{or, equivalently, } h(\estbeta;X, \estmu)\ge 0, \\
			0 &\quad \text{otherwise}.
		\end{aligned}
		\right.
	\end{equation}

	Of theoretical interest is the rate of the excess risk of $\hat s$ relative to the  Bayes classifier.  For a random pair $(X, Y) \in \metricsp\times \{0,1\}$ {sampled from the model \eqref{LR-model-metric}}, the Bayes classifier   $s^*$,  minimizing the error probability, is  
	\begin{equation}\label{bayes-classifier}
		s^*(X) = \left\{
		\begin{aligned}
			1 &\quad \text{if}~P_{X} \ge 1/2~ \text{or, equivalently, } h(\truebeta;X, \truemu)\ge 0, \\
			0 &\quad \text{otherwise}.
		\end{aligned}
		\right.
	\end{equation}
	The excess risk of $\hat{s}$ is defined by $\excessrisk{s^*}{\hat{s}} = \expect\classloss{\hat{s}} - \expect\classloss{s^*}$, where  $\classloss{s}=\indicator{s(X) \ne Y}$ signifies whether an error is made by the classifier $s$ and $\indicator{\cdot}$ is the indicator function. 
	The excess risk, commonly investigated in the literature of classification \citep{massart2006risk, audibert2007fast}, quantifies the classification accuracy of $\hat s$ (relative to the Bayes classifier $s^\ast$), since
	$\mathbb{P}\{\hat{s}(X) = Y\}=1-\expect\classloss{\hat{s}} = (1 - \expect\classloss{s^*}) - \excessrisk{\hat{s}}{s^*}$,
	where $1 - \expect\classloss{s^*}$ is the classification accuracy of the Bayes classifier $s^*$. 
	
	We first provide an upper bound on the excess risk in terms of the estimation errors of $\estbeta$ and $\estmu$ for Alexandrov spaces. This bound requires some regularity on the Alexandrov inner product $h(\truebeta;x,\mu)$ defined in \eqref{LR-model-metric}.
	We say $h(\truebeta;x,\mu)$ is regular near $\truemu$ if there exist  constants $C_{1},C_{2}>0$ and \emph{arbitrarily small but fixed} constants $0<\munbh_1\leq \munbh_2$ such that, for all  $\mu\in\ball{\truemu}{\munbh_2}$, the following two Lipschitz continuity conditions are satisfied: 
	\begin{itemize}[leftmargin=5ex]
		\item[1)] $|h(\truebeta;x,\mu)-h(\truebeta;x,\truemu)|\leq C_{2}d(\mu,\truemu)$ for all $x\in\ball{\truemu}{\munbh_2}$, and
		\item[2)]   $|\metric{\truebeta}{\mu}\cos(\alexanlge{\mu}{x}{\truebeta})-\metric{\truebeta}{\truemu}\cos(\alexanlge{\truemu}{x}{\truebeta})|\leq C_{1} d(\mu,\truemu)$ for all $x$ with $\metric{x}{\truemu}=\munbh_1$.
	\end{itemize}
	This mild regularity condition, guaranteeing adequate (Lipschitz) continuity of the Alexandrov inner product $h(\truebeta;x,\mu)$ near $\truemu$, holds for all $\truemu$ in Riemannian manifolds (e.g., the SPD matrix manifold and the unit sphere in Example \ref{ex:spaces}) since  $h(\truebeta;x,\mu)$ is smooth in both $x$ and $\mu$, and for the Wasserstein space in Example \ref{ex:spaces}(d). It also generally holds for other metric spaces when $\truemu$ is  a ``nonsingular'' point, e.g.,  any interior point of an orthant in Example \ref{ex:spaces}(c) of the Billera--Holmes--Vogtmann tree space \citep{Billera2001}. 
	
	\begin{remark} The constants $C_1$ and $C_2$ in the above typically depend on the constants $r_1$ and $r_2$, respectively. For instance, in the special case of Euclidean space (i.e., when $\metricsp=\real^D$), $C_1$ diverges as $r_1$ approaches 0, while $C_2$ grows as $r_2$ approaches $\infty$. 
		Therefore, the above regularity of $h(\truebeta;x,\mu)$ can only be assumed locally \emph{near} $\truemu$ (but not globally in the entire space $\metricsp$); this  significantly intensifies the complexity of the proofs (e.g., see  Corollary \xref{cor:fmse} and Lemma \xref{lem:mid-seg-comparison-model-space} of Supplementary Material).   
	\end{remark}
	
	The following theorem upper bounds the excess risk in terms of the estimation error of $\estbeta$ and $\estmu$, where we recall $D_k=\pi/\sqrt{\kappa}$ if $\kappa>0$ and $D_\kappa=\infty$ otherwise. 
	
	\begin{theorem}\label{thm:excessrisk-upper-simplified-metric}
		Suppose $\metricsp$ is a complete Alexandrov space with curvature upper bounded by $\kappa$ and of diameter no larger than $D_\kappa/2$. Assume that the Alexandrov inner product $h(\truebeta;x,\mu)$ is regular near $\truemu$. Then, under Assumptions \ref{as:existence}, there exists a constant $c>0$ such that
		\[
		\excessrisk{s^*}{\hat{s}} \le c  \left(
		\expect\left[\metricsq{\truemu}{\estmu}\right] 
		+
		\expect\big[ \metricsq{\truebeta}{\estbeta}\big] 
		\right)^{1/2}.
		\]
	\end{theorem}
	The assumed space in the above theorem includes 
	Hadamard spaces with $\kappa=0$ ($D_\kappa=\infty$ in this case) and positively curved spaces such as spheres with $\kappa>0$.
	The convergence rate of $\estmu$ established in \cite{schotz2019convergence, ahidar2020convergence} and the  rate of $\estbeta$ in  Theorems \ref{thm:upperbound-global-expect-entropy-simplified} and \ref{thm:upperbound-global-expect-global-entropy-simplified} imply  $\expect\left[\metricsq{\truemu}{\estmu}\right] \asymplt\expect[ \metricsq{\truebeta}{\estbeta}]$. Consequently, the above theorem, together with  Theorems \ref{thm:upperbound-global-expect-entropy-simplified} and \ref{thm:upperbound-global-expect-global-entropy-simplified},  provides upper bounds on the rate of the excess risk under the log-polynomial and polynomial metric entropy conditions.
	
	For Riemannian manifolds, we can obtain a finer upper bound on the excess risk by exploiting the Riemannian structure. Let $h(\truebeta)=h(\truebeta;X,\truemu)$ be viewed as a function of $\truebeta$ and let $\Lambda_{\min}(\cdot)$ denote the smallest eigenvalue of a linear operator. Recall the Hessian defined in \eqref{eq:hessian}.
	
	\begin{theorem}
		\label{thm:excessrisk-upper-simplified}
		For a $D$-dimensional Hadamard manifold of bounded sectional curvature, suppose that the distribution of $X$ has a bounded support, the smallest eigenvalue $\lambda_{X}$ of $\expect[\hess_{\truebeta}h]$ satisfies $\lambda_{X}>0$, and {$\estbeta$ exists almost surely}. Then, there is a constant $c>0$ depending on $\lambda_{X}$ and $\sigma_X^2=\expect[\metricsq{\truemu}{X}]$, such that
		\[
		\excessrisk{s^*}{\hat{s}} \le c \sqrt{{D}/{n}}.
		\]
		The same conclusion holds for a complete and simply connected $D$-dimensional Riemannian manifold  $\manifold$ with non-negative sectional curvature upper bounded by $\kappa>0$ and of diameter no larger than $D_\kappa/2$ such that $\inf_{p,q\in\manifold}\Lambda_{\min}(\hess_q d_p^2)>0$, where $d_p^2(\cdot)=d^2(\cdot,p)$.
	\end{theorem}
	
	Below we provide a matching lower bound  for the class $\mathcal{C}=\{s(X)=\indicator{h(\beta; X, \truemu)\ge 0}: \beta\in \metricsp\}$ for  Riemannian manifolds, establishing the asymptotic minimax optimality of the proposed classifier in terms of both $n$ and the dimension $D$;  a similar class of classifiers was considered by \cite{abramovich2019high} in the Euclidean setting. 
	{To the best of our knowledge, this is the first optimality result for classifiers on covariates without a vector structure.} 
	
	\begin{theorem}\label{thm:excessrisk-lower}
		For a $D$-dimensional Riemannian manifold $(\metricsp, \metricfn)$, 
		if $(X_i, Y_i), i=1,\ldots, n$, are i.i.d. drawn from the model \eqref{LR-model-metric}, then
		\[
		\inf_{\hat{s}} \sup_{s^* \in \mathcal{C}} \excessrisk{s^*}{\hat{s}}  
		\gtrsim
		\sqrt{{D}/{n}},
		\]
		where the infimum is taken over all classifiers.
	\end{theorem}
	
	\begin{remark}
		For general metric spaces, \cite{gottlieb2016nearly} provides a lower bound of the rate $n^{-1}$ (without a matching upper bound) under a restrictive condition that the Bayes classifier has zero error probability. This lower bound, when applied to Riemannian manifolds, is sub-optimal, as it  requires the restrictive condition and does not include the dimension $D$.  Finding matching lower bounds and establishing optimality for the excess risk in general metric spaces are much more challenging and  left for future exploration.	
	\end{remark}

	\section{Simulation Studies}\label{sec:simulation}
	We illustrate the numeric performance of the proposed  regression from the perspectives of both estimation and classification by using the manifold $\manifold=\spdmfd{3}$ of $3\times 3$ symmetric positive-definite (SPD) matrices endowed with the Log-Cholesky metric \citep{lin2019riemannian}. For estimation, we consider
	\begin{itemize}
		\item Case 1: data are sampled from the proposed model \eqref{LR-model-metric}.
	\end{itemize}
	For the Fr\'echet mean $\truemu$, we consider two settings, namely, the $3\times 3$ identity matrix $I_{3}$ and the  matrix  $\Sigma_{AR}=(0.5^{|i-j|})_{1\le i,j \le 3}$. For the coefficient $\truebeta$, we also consider two settings, namely,  $3I_{3}$ and $3\Sigma_{AR}$. For classification, we consider the following additional data models:
	\begin{itemize}
		\item Case 2: {$\text{logit}(P_{X}) = 2\sin(\pi h(3\Sigma_{AR};X, I_3))+h(3\Sigma_{AR};X, I_3)$},
		\item Case 3: {$\text{logit}(P_{X}) = (\log 3) \sin(\pi \vec{X}_{1}) + \frac{\sqrt 3}{2}\vec{X}_{2}^2 + \frac{\sqrt 3}{4} \exp(\vec{X}_{3}) + \frac{3}{4}\vec X_5 + 2(\log \frac{3}{2})(\vec X_4 + \vec X_6)$},
	\end{itemize} 
	where $P_{X}$ is the conditional success probability, $h$ is defined in \eqref{LR-model-metric} and $\vec X=(\vec X_1,\ldots,\vec X_6)$ is the vector representation of  $X\in \spdmfd{3}$, {i.e., the $6$-dimensional vector obtained by stacking elements in the lower triangular part of $X$.}
	{The model in Case 2 includes a nonlinear term of $h$, while the model in Case 3 is  nonparametric and highly nonlinear in the covariate; both are designed to challenge the proposed classifier.}
	
	For each of the above models, we synthesize  data under different settings. Specifically, we consider two sample sizes, $n=100$ and $n= 500$.  The data are synthesized, as follows. 
	For each $i=1, \ldots, n$, we draw the covariate $X_i=\Exp{\truemu}{S_i}$ with ${\vec{S_i}} \sim N(0, r I_{6})$, where $r=1, 4$ determines the variance $\sigma_{X}^2=\expect[\metricsq{\truemu}{X_i}]$ and $\Exp{\truemu}$ denotes the Riemannian exponential map at $\truemu$.   
	Finally, we compute $P_{X_i}$ according to the models and sample $Y_i$ via Bernoulli trial with the success probability $P_{X_i}$. 
	For each model and each setting, we perform $500$ independent Monte Carlo replicates. 
	In each replicate, we split the dataset into a training set $\trainset$ (80\%) and a test set $\testset$ (20\%). 
	
	\vspace{3mm}
	\noindent
	\textbf{Estimation quality.} 
	As mentioned in the above, we assess the quality of the proposed estimators $\estmu$ and $\estbeta$ computed from  $\trainset$ in Case 1, and quantify estimation quality by $\metric{\truemu}{\estmu}$ and $\metric{\truebeta}{\estbeta}$, respectively. In addition, based on $\estmu$ and $\estbeta$,  for each data point $i\in \testset$ in the test set, we compute the estimated success probability $\hat{P}_{X_i}$ which is of interest in practice. To quantify the estimation quality of the success probability, we consider the root-mean-squared error (RMSE)
	\[
	\mathrm{RMSE} = \sqrt{\frac{1}{|\testset|} \sum_{i \in \testset} \left(\text{logit}(\hat{P}_{X_i}) - \text{logit}(P_{X_i})\right)^2}.
	\]
	Tables \ref{tab:simu-est-I3AR} summarizes the results of the simulation studies with $\truemu=I_{3}$ and $\truebeta=3\Sigma_{AR}$. 
	We see that, the estimation errors and RMSE decrease as the sample size grows, as indicated by our theoretical results in Section \ref{sec:theory-estimation}.
	We also observe that the error $\metric{\truebeta}{\estbeta}$ decreases  as $\sigma_{X}^2$ grows. Intuitively, this is because, from the perspective of estimating $\truebeta$, a larger $\sigma_{X}^2$ indicates a larger signal-to-noise ratio, making estimation  of the coefficient $\truebeta$ relatively easier; {see Example 8.4 in \cite{duchi2019information} for a similar observation in the Euclidean space.} This phenomenon is also dictated by Theorems \ref{thm:lowerbound-global-entropy-simplified} and \ref{thm:lowerbound-global-global-entropy-simplified}, where the lower bounds decrease with respect to  $\sigma_X^2$. In contrast,  the error $\metric{\truemu}{\estmu}$ increases as $\sigma_X^2$ increases. This is because, like estimating the mean of a Gaussian sample in the Euclidean space, the error of estimating the Fr{\'e}chet mean $\truemu$ is scaled up according to the variance $\sigma_X^2$ of $X$. 
	Similar results are observed for the other settings presented in Section \xref{sec:further-numerical} of Supplementary Material. 
	
	\begin{table}[t]
		\centering
		\caption{Empirical estimation performance of the proposed method in Case 1 with $\truemu=I_{3}$ and $\truebeta=3\Sigma_{AR}$. The standard deviations based on the $500$ independent Monte Carlo replicates are given in the parentheses. The value of  $\sigma_{X}^2$ is approximated numerically.}
		\begin{tabular}{ccccc}
			\hline
			$\sigma_{X}^2$ & $n$   & $\metric{\truemu}{\estmu}$ & $\metric{\truebeta}{\estbeta}$ & RMSE \\
			\hline
			\multirow{2}{*}{2.248} & 100   & 0.145(0.047) & 1.137(0.442) & 0.155(0.062) \\
			& 500   & 0.063(0.020) & 0.438(0.138) & 0.027(0.008) \\
			\hline
			\multirow{2}{*}{4.496} & 100   & 0.205(0.066) & 0.921(0.403) & 0.173(0.079) \\
			& 500   & 0.089(0.028) & 0.348(0.112) & 0.030(0.009) \\
			\hline
		\end{tabular}%
		\label{tab:simu-est-I3AR}%
	\end{table}%
	
	\vspace{3mm}
	\noindent
	\textbf{Classification performance.}
	We compare the proposed classifier with the SPDNet \citep{huang2017riemannian}, $k$-NN \citep{chaudhuri2014rates}, kernel support vector machine \citep[kSVM,][]{jayasumana2013kernel} and kernel density classifier \citep[Section 6.6.2,][]{hastie2009elements}. {The details of these methods on SPD matrices are given in Section \xref{sec:further-numerical} of Supplementary Material}. For evaluation, we consider all models in the above, and measure classification performance by accuracy, area under the receiver operating characteristic curve (AUC), sensitivity, specificity and empirical excess risk (EER) defined by
	\[
	\mathrm{EER} = \frac{1}{|\testset|} \sum_{i \in \testset} \{\classloss{\hat{s}(X_i)} - \classloss{s^*(X_i)} \},
	\]
	where $\hat s$ is a classifier,  $s^*$ is the Bayes classifier, and $\classloss{\cdot}$ is defined in Section \ref{sec:theory-classification}. 
	
	Table \ref{tab:simu-class-I3AR} summarizes the results with $\sigma_{X}^2=2.248$ (corresponding to $r=1$) and with $\truemu=I_{3}$, $\truebeta=3\Sigma_{AR}$ for Case 1; similar results for other settings are presented in Section \xref{sec:further-numerical} of Supplementary Material.
	First, in Case 1 where the data are sampled from the proposed model, the EER of the proposed classifier decreases as the sample size grows, which numerically illustrates the theoretical results in Section \ref{sec:theory-classification}. Also, as expected, compared with other methods, the proposed classifier has superior performance in Case 1. For Case 2, although the data model contains a nonlinear term of the Alexandrov inner product, the proposed classifier still outperforms the other methods.  In Case 3 that strongly favors nonparametric methods,  the proposed classifier can produce reasonable but suboptimal results relative to the other classifiers. 
	In summary, there is no universally best method, and the proposed classifier has excellent performance when the data model is linear or moderately nonlinear while the other classifiers generally produce better results when the data model is highly nonlinear and the manifold has relatively low dimensions.

	\begin{table}[htbp]
		\centering
		\caption{Empirical classification performance of classifiers when $\sigma_{X}^2=2.248$. The standard deviations based on the $500$ independent Monte Carlo replicates are in the parentheses.}
		\scalebox{0.7}{%
			\begin{tabular}{cccccccc}
				\hline
				Model & $n$   & Method & Accuracy & AUC   & Sensitivity & Specificity & EER \\
				\hline
				\multirow{10}{*}{\shortstack{Case 1 \\  ($\truemu=I_3$) \\ ($\truebeta=3\Sigma_{AR}$)}} & \multirow{5}{*}{100} & SPDNet & 0.609(0.112) & 0.657(0.132) & 0.603(0.186) & 0.615(0.194) & 0.040(0.111) \\
				&       & $k$-NN & 0.583(0.118) & 0.596(0.100) & 0.571(0.227) & 0.589(0.234) & 0.067(0.133) \\
				&       & kSVM  & 0.584(0.121) & 0.592(0.101) & 0.568(0.287) & 0.595(0.283) & 0.066(0.135) \\
				&       & KDC   & 0.561(0.117) & 0.571(0.093) & 0.583(0.311) & 0.538(0.321) & 0.088(0.138) \\
				&       & Proposed & 0.622(0.114) & 0.685(0.118) & 0.612(0.169) & 0.634(0.163) & 0.028(0.106) \\
				\cline{2-8}          & \multirow{5}{*}{500} & SPDNet & 0.638(0.048) & 0.694(0.054) & 0.624(0.086) & 0.653(0.090) & 0.022(0.037) \\
				&       & $k$-NN & 0.621(0.052) & 0.621(0.050) & 0.624(0.115) & 0.619(0.102) & 0.039(0.048) \\
				&       & kSVM  & 0.642(0.048) & 0.642(0.047) & 0.647(0.107) & 0.637(0.103) & 0.018(0.038) \\
				&       & KDC   & 0.627(0.054) & 0.628(0.049) & 0.639(0.192) & 0.615(0.186) & 0.033(0.052) \\
				&       & Proposed & 0.651(0.049) & 0.709(0.053) & 0.652(0.073) & 0.649(0.071) & 0.009(0.030) \\
				\hline
				\multirow{10}{*}{Case 2} & \multirow{5}{*}{100} & SPDNet & 0.636(0.115) & 0.670(0.135) & 0.631(0.191) & 0.637(0.200) & 0.147(0.126) \\
				&       & $k$-NN & 0.614(0.115) & 0.618(0.107) & 0.606(0.228) & 0.611(0.218) & 0.168(0.132) \\
				&       & kSVM  & 0.634(0.116) & 0.631(0.107) & 0.618(0.266) & 0.638(0.254) & 0.149(0.133) \\
				&       & KDC   & 0.600(0.115) & 0.600(0.100) & 0.608(0.280) & 0.579(0.284) & 0.183(0.133) \\
				&       & Proposed & 0.664(0.115) & 0.693(0.129) & 0.659(0.169) & 0.672(0.170) & 0.118(0.125) \\
				\cline{2-8}          & \multirow{5}{*}{500} & SPDNet & 0.689(0.048) & 0.706(0.055) & 0.676(0.084) & 0.703(0.086) & 0.094(0.053) \\
				&       & $k$-NN & 0.680(0.050) & 0.681(0.050) & 0.685(0.089) & 0.676(0.091) & 0.102(0.055) \\
				&       & kSVM  & 0.717(0.046) & 0.717(0.046) & 0.724(0.080) & 0.710(0.082) & 0.066(0.048) \\
				&       & KDC   & 0.665(0.065) & 0.665(0.064) & 0.678(0.145) & 0.652(0.151) & 0.118(0.067) \\
				&       & Proposed & 0.723(0.047) & 0.720(0.054) & 0.726(0.069) & 0.720(0.070) & 0.060(0.048) \\
				\hline
				\multirow{10}{*}{Case 3} & \multirow{5}{*}{100} & SPDNet & 0.707(0.105) & 0.724(0.131) & 0.422(0.216) & 0.852(0.110) & 0.060(0.099) \\
				&       & $k$-NN & 0.673(0.111) & 0.585(0.105) & 0.867(0.136) & 0.300(0.240) & 0.093(0.118) \\
				&       & kSVM  & 0.673(0.106) & 0.550(0.091) & 0.933(0.102) & 0.166(0.228) & 0.094(0.114) \\
				&       & KDC   & 0.655(0.112) & 0.534(0.077) & 0.929(0.147) & 0.129(0.232) & 0.112(0.122) \\
				&       & Proposed & 0.638(0.109) & 0.727(0.123) & 0.599(0.141) & 0.717(0.189) & 0.128(0.120) \\
				\cline{2-8}          & \multirow{5}{*}{500} & SPDNet & 0.726(0.047) & 0.762(0.054) & 0.454(0.100) & 0.867(0.051) & 0.048(0.044) \\
				&       & $k$-NN & 0.704(0.047) & 0.618(0.051) & 0.890(0.062) & 0.346(0.121) & 0.069(0.048) \\
				&       & kSVM  & 0.717(0.048) & 0.633(0.056) & 0.901(0.057) & 0.365(0.128) & 0.056(0.046) \\
				&       & KDC   & 0.660(0.048) & 0.527(0.047) & 0.948(0.105) & 0.105(0.182) & 0.113(0.054) \\
				&       & Proposed & 0.667(0.051) & 0.755(0.054) & 0.627(0.064) & 0.746(0.080) & 0.106(0.052) \\
				\hline
			\end{tabular}%
		}
		\label{tab:simu-class-I3AR}%
	\end{table}%

	\section{Data Application}\label{sec:application}
	
	We apply the proposed method to the task-based fMRI data from the Human Connectome Project \citep{van2013wu}. In the project, each subject was asked to complete several tasks that are related to certain human functions. For instance, in the motor task, subjects were asked to tap their fingers, squeeze their toes, or move their tongue with the given visual cues, while in the language processing task, subjects were asked to answer related questions after brief auditory stories, or complete addition and subtraction math problems which are presented auditorily. 
	During each task, the temporal blood-oxygen-level dependent (BOLD) signals from each brain region were recorded. Based on the BOLD signals, functional connectivity of the brain, represented by covariance matrices of the BOLD signals, can be derived. 	
	Of interest is  the effect of functional connectivity  on the motor  and language processing tasks. 
	
	We focus on $201$ subjects performing only the motor task and $193$ subject performing only the language processing task; data\footnote{\scriptsize Available at \url{https://www.humanconnectome.org/study/hcp-young-adult/document/500-subjects-data-release}.} collected from these subjects form a dataset of the sample size $n=394$.  
	We consider $m=8$ brain regions related to motor skills and language processing, and apply a preprocessing pipeline described in Section \xref{sec:preprocess-application} of Supplementary Material to obtain the observations $(X_i, Y_i) \in \spdmfd{8}\times \{0,1\}$ for $i=1,\ldots, n$, where $X_i$ is the functional connectivity, $Y_i=0$ for the subject completing the motor task and $Y_i=1$ for the subject completing the language processing task. As the Euclidean distance on $\spdmfd{8}$ severely suffers from the ``swelling effect'' that artificially inflates the variation in statistical analysis of diffusion tensors and functional connectivities \citep{Arsigny2007}, to avoid this undesirable effect, we adopt the Log-Cholesky distance \citep{lin2019riemannian} on $\spdmfd{8}$ specifically designed for eliminating the swelling effect;  the Log-Euclidean \citep{Arsigny2007} and affine-invariant \citep{moakher2005differential} distances yield similar results that are thus not presented.

	\begin{figure}[t]
		\begin{center}
			\scalebox{0.85}{%
				\begin{tikzpicture}
					\def\y{1.4}
					\def\x{-7.2}
					\def\d{1.8}
					\node at (0,0) {\includegraphics[scale=0.42]{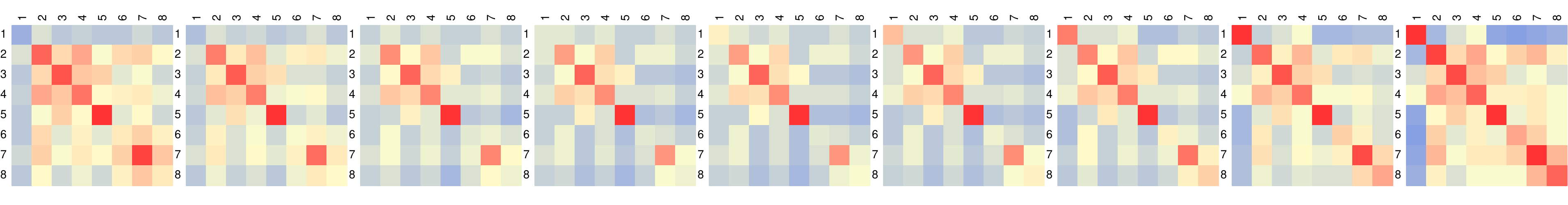} };
					\node at (0,-2) {\includegraphics[scale=0.42]{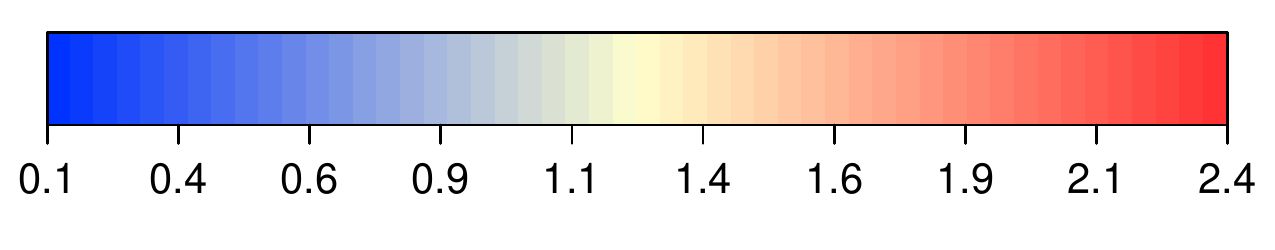} };
					\node at (\x,\y) {\tiny $\geod{\estmu}{\estbeta}(-0.015)$};
					
					\node at (\x+4*\d,\y) {\tiny $\geod{\estmu}{\estbeta}(0)=\estmu$};
					
					\node at (\x+8*\d,\y) {\tiny $\geod{\estmu}{\estbeta}(0.015)$};
					
					\def\oddsy{-1.05}
					\def\oddsx{0.05}
					\def\oddsmove{1.815}
					\node at (\oddsx+\oddsmove*4, \oddsy){\tiny $\boldsymbol{382}$};
					\node at (\oddsx+\oddsmove*3, \oddsy){\tiny $\boldsymbol{86.4}$};
					\node at (\oddsx+\oddsmove*2, \oddsy){\tiny $\boldsymbol{19.5}$};
					\node at (\oddsx+\oddsmove, \oddsy){\tiny $\boldsymbol{4.42}$};
					\node at (\oddsx, \oddsy){\tiny $\boldsymbol{1.0}$};
					\node at (\oddsx-\oddsmove, \oddsy){\tiny $\boldsymbol{0.226}$};
					\node at (\oddsx-\oddsmove*2, \oddsy){\tiny $\boldsymbol{0.051}$};
					\node at (\oddsx-\oddsmove*3, \oddsy){\tiny $\boldsymbol{0.012}$};
					\node at (\oddsx-\oddsmove*4, \oddsy){\tiny $\boldsymbol{0.003}$};
				\end{tikzpicture}
			}
		\end{center}
		\caption{The geodesic $\geod{\estmu}{\estbeta}(t)$ on $\spdmfd{8}$ with $t=-0.015+k\delta$ for $\delta=0.00375$ and $k=0,1,\ldots,8$. The number below each matrix is the odds associated with $X=\geod{\estmu}{\estbeta}(t)$ for the corresponding $t$. The log odds is increased by $1.486$ when the functional connectivity advances along the positive direction of $\geod{\estmu}{\estbeta}$ by $\delta$ unit.}
		\label{fig:geodesic-hat}
	\end{figure}
	
	We fit the proposed  model \eqref{LR-model-metric} by using the above data and depict the estimated parameters in Figure \ref{fig:geodesic-hat}, where we  plot the geodesic $\geod{\estmu}{\estbeta}$ emanating from $\estmu$ in both directions.   To assess the covariate effect, we test the null hypothesis $H_0:\truebeta=\truemu$, which corresponds to no different effect of functional connectivity on the two tasks, by using the  permutation test outlined in Section \xref{subsec:perm-test} of Supplementary Material. The empirical $p$-value is less than 0.001, strongly suggesting that functional connectivity has significantly different effect on motor skills and language processing.

	The geodesic $\geod{\estmu}{\estbeta}$  in Figure \ref{fig:geodesic-hat} is normal to the empirical  decision boundary between the two tasks, with the positive direction (larger odds) extending into the language processing task and the negative direction (smaller odds) extending into the motor task.  From the figure we also find that, for any $t$, if the functional connectivity is changed from $\geod{\estmu}{\estbeta}(t)$ to $\geod{\estmu}{\estbeta}(t+\delta)$ with $\delta=0.00375$ (i.e., the functional connectivity advances along the direction $\geod{\estmu}{\estbeta}$ by $\delta$ unit), then the change in the log odds is $1.486$. Moreover,  we observe that the odds becomes much larger than 1 as Region 1 (the superior temporal gyrus) becomes more active, which suggests that Region 1 is substantially more involved in auditory and language  processing; the finding matches the previous study on this brain region \citep{bigler2007superior}.
	
	In addition, we observe that the correlation between Region 1 and other regions exhibits different patterns in the two tasks. For example, the correlation  between Region 1 and Region 4 (which is part of the inferior frontal gyrus related to  speech and language processing) in the direction of larger odds is visibly stronger than that in the opposite direction, suggesting a stronger functional connectivity  between Region 1 and Region 4  in the language processing task. In contrast, the correlation between Region 1 and Regions 5--8 in the direction of larger odds is weaker than that in the opposite direction. This may indicate that, in the language processing task which does not invoke motor  and visual processing, the functional connectivity between Region 1 and Regions 5--8 which include the primary motor cortex and visual cortex may be suppressed to certain extent. 
	
	\begin{table}[!t]
		\centering
		\caption{Comparison of classifiers.}%
		\begin{tabular}{ccccc}
			\hline
			Method & Accuracy & AUC   & Sensitivity & Specificity \\
			\hline
			SPDNet & 0.8667 & 0.9205 & 0.8613 & 0.8690 \\
			$k$-NN  & 0.8333 & 0.8349 & 0.8401 & 0.8297 \\
			kSVM & 0.8436 & 0.8444 & 0.8667 & 0.8221 \\
			KDC   & 0.7744 & 0.7811 & 0.8313 & 0.7309 \\
			Logistic-BOLD & 0.7641 & 0.8120 & 0.8490 & 0.6984 \\
			Proposed & 0.8718 & 0.9060 & 0.9158 & 0.8376 \\
			\hline
		\end{tabular}
		\label{tab:real-data-comparisons}%
	\end{table}
	
	\begin{figure}[!t]%
		\centering
		{\includegraphics[scale=0.25]{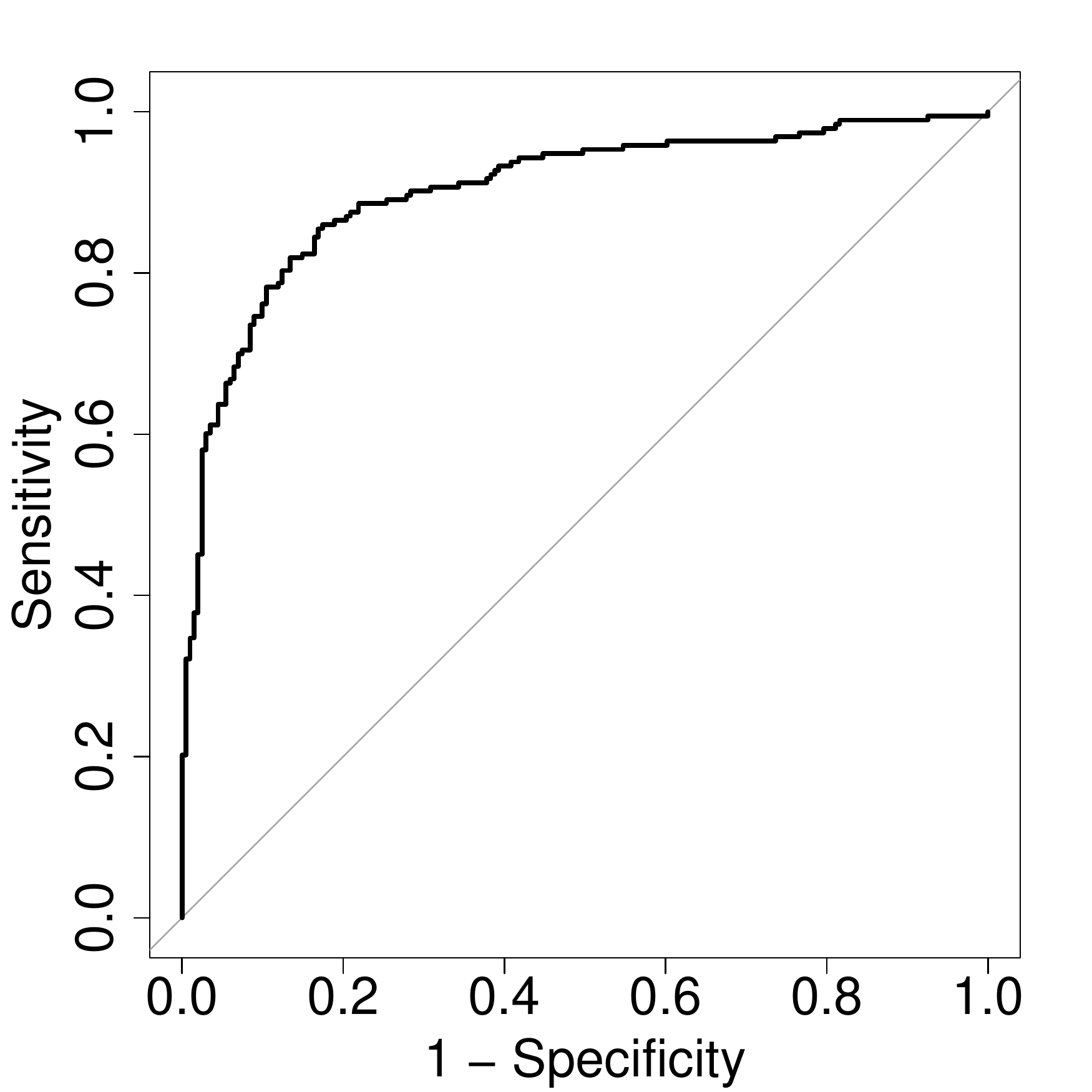}}
		\caption{The in-sample ROC curve of the proposed classifier.}
		\label{fig:real-data-fitted}
	\end{figure}
	
	The in-sample ROC (receiver operating characteristic) curve in {Figure \ref{fig:real-data-fitted}} shows  that the proposed  model is adequate to classify the two tasks based on the functional connectivity. 
	In addition, as in Section \ref{sec:simulation}, we compare the proposed classifier with SPDNet, $k$-NN, kernel SVM (kSVM) and kernel density classifier (KDC). The $10$-fold cross-validated accuracy, AUC, sensitivity and specificity  are presented in {Table \ref{tab:real-data-comparisons}}.
	From the table, it can be seen that, even compared with SPDNet that is based on neural networks, the proposed  classifier  has slightly higher accuracy, and outperforms the other methods, perhaps because some nonparametric classifiers start to suffer from the curse of dimensionality as the manifold $\spdmfd{8}$ has 36 dimensions. Moreover, the proposed  classifier is  more interpretable. For example, as discussed in the above, the geodesic $\geod{\estmu}{\estbeta}$ in Figure \ref{fig:geodesic-hat}, which is the  direction normal to the empirical decision boundary between the two tasks, reveals  interesting patterns of functional connectivity in these two tasks. 
	
	The aforementioned different patterns of inter-connectivities (the off-diagonal entries of each $X_i\in \spdmfd{8}$) also play a significant role in classifying the two tasks. To illustrate this, we fit a {Euclidean} logistic regression model (denoted by Logistic-BOLD in Table \ref{tab:real-data-comparisons}) that includes only the BOLD signals (diagonal entries of each $X_i\in \spdmfd{8}$) as covariates. Compared with the model \eqref{LR-model-metric}, this na\"ive logistic model, which does not consider inter-connectivities, has substantially lower  accuracy and other deteriorated performance measures. 
	
	In summary, the proposed model can uncover interesting and interpretable effects of functional connectivity on the motor and language processing tasks, while offering near state-of-the-art performance in classifying  the two tasks.

	\section{Concluding Remarks}\label{sec:discussion}
	The proposed model \eqref{LR-model-metric}, which incorporates covariates located in a metric space without a vector structure, offers an intuitive approach to modeling the effect of metric-space valued covariates on the log odds, and the model parameters enable practitioners to gain insights into the fundamental distinctions between the two classes under consideration. 
	While  we have established the first set of optimality results for  regression with a binary response and covariates on general metric spaces, several open problems remain for future studies. For example, optimality in the case of $D>2\alpha_U$ remains largely unknown. In addition, under the log-polynomial global metric entropy condition,  there is a small gap between the upper  and lower bounds  as presented  in Theorems \ref{thm:upperbound-global-expect-global-entropy-simplified} and \ref{thm:lowerbound-global-global-entropy-simplified}, respectively. Closing this gap is challenging and, to the best of our knowledge, remains unresolved even in the Euclidean setting. Furthermore, finding the minimax lower bound on the excess risk for general metric spaces awaits future exploration.
		
		\if0\blind
		\section*{Acknowledgments}
		Z. Lin's research is partly supported by the NUS startup grant A-0004816-01-00. 
		Data were provided in part by the Human Connectome Project, WU-Minn Consortium (PI: David Van Essen and Kamil Ugurbil; 1U54MH091657) funded by the 16 NIH Institutes and Centers that support the NIH Blueprint for Neuroscience Research; and by the McDonnell Center for Systems Neuroscience at Washington University.
		\fi
		
		\bibliographystyle{asa}
		\bibliography{ref-CoM}
		
		

\section*{Supplementary material}

# Supplementary Material for "Binary Regression and Classification with non-Euclidean Covariates"

Yinan Lin and Zhenhua Lin*

*Department of Statistics and Data Science, National University of Singapore*

# Contents

**Organization.** In Section A, we provide optimization methods to compute the proposed maximum likelihood estimator. We describe data preprocessing for the fMRI data used in the application and outline a permutation test for the covariate effect in Section B. In Section C we provides a discussion on how the proposed model is linked to log odds ratio. In Section D we present the proof of Propositions 3.1. The proofs of Theorems 4.6 and 4.10 about the upper bounds for estimation are given in Section E, followed by the proofs of Theorems 4.7 and 4.11 about the lower bounds in Section F. Section G is dedicated to the proofs of the theoretical results for classifications in Section 5. Section H provides further discussions and remarks on Assumptions. Technical lemmas are collected in Section I. Finally, in Section J we detail the competing classifiers referenced in simulations and data applications, along with additional simulation results.

**Notation.** We fix additional notations used in this supplementary file. Denote by $\log a = \log_e a$ the logarithm of $a$ with basis $e$ and by $\mathbb{N}$ the set of natural numbers. For a pseudo-metric space $(Q, b)$, denote by $N(\delta, Q, b)$ and $M(\delta, Q, b)$ the covering and packing number

for $Q$ under the pseudo-metric $b$ with the radius $\delta > 0$. For $n \in \mathbb{N}$, define

$$\mathsf{entr}_n(Q, b) := \inf_{\epsilon > 0} \left( \epsilon \sqrt{n} + \int_\epsilon^\infty \sqrt{\log N(r, Q, b)} dr \right). \tag{S1}$$

For two distributions $P$ and $Q$, denote by $D_{\mathrm{kl}}(P\|Q)$ the Kullback–Leibler divergence (KL-divergence) between them. For $p \in [0, 1]$, let $\mathsf{Bernoulli}(p)$ be the Bernoulli distribution with success probability $p$. For $p, q \in [0, 1]$, let $D_{\mathrm{kl}}(p\|q)$ be the KL-divergence between $\mathsf{Bernoulli}(p)$ and $\mathsf{Bernoulli}(q)$. Similarly, let $D_{\chi^2}(p\|q)$ be the $\chi^2$-divergence between $\mathsf{Bernoulli}(p)$ and $\mathsf{Bernoulli}(q)$. In the sequel, unless otherwise stated, $c, C, c_1, C_1, \ldots$ are positive constants not depending on $n$ but may vary from place to place.

# A  Computational Details

We provide optimization methods to find the estimator $\hat{\beta}_n$. For the general metric space, we extend the LIPO algorithm proposed in Malherbe and Vayatis (2017). When the underlying metric is a Riemannian manifold, the Riemannian manifold gradient descent (Absil et al., 2009) can be adopted.

## A.1  Optimization for General Metric Spaces

Linearity and convexity of the objective function $L_n$ are absent from general metric spaces. Consequently, it is generally infeasible to apply (sub)gradient based methods, such as the (sub)gradient descent method, or optimization methods relying on convexity, such as the proximal point algorithm. To address this computational challenge, observing that the objective function $L_n$ is Hölder continuous with a Hölder constant $K > 0$ and an exponent $\alpha_U \in (0, 1]$ (i.e., $|L_n(\beta_1) - L_n(\beta_2)| \leq K d^{\alpha_U}(\beta_1, \beta_2)$ for all $\beta_1, \beta_2 \in \mathcal{M}$) under Assumption 4.2, we consider extending the Lipschitz optimization (Hansen and Jaumard, 1995). Specifically,

we extend the LIPO algorithm (Malherbe and Vayatis, 2017) to Hölder continuous functions on the general metric space in Algorithm 1.

In the algorithm, we need to provide a Hölder constant $K$ and an exponent $\alpha_U$. It can be seen that $K = 2C_U n^{-1} \sum_{i=1}^n d(X_i, \hat{\mu}_n)$ is a Hölder constant for $L_n$, where $C_U$ is the constant defined in Assumption 4.2. Both constants $C_U$ and $\alpha_U$ are determined by the underlying space. For Hadamard spaces, one can take $C_U = 5$ and $\alpha_U = 1$. We also need to uniformly sample points from the underlying metric space $\mathcal{M}$, which can be accomplished by the method in Kovács (2014).

---

**Algorithm 1** The Generalized LIPO Algorithm on Metric Spaces

---

**Require:** An integer $T > 0$, a Hölder constant $K > 0$ and an exponent $\alpha_U \in (0, 1]$.

Let $\beta_{(0)}$ be sampled uniformly from $\mathcal{M}$

$t \leftarrow 0$

**while** $t < T$ **do**

    Sample $\beta$ uniformly from $\mathcal{M}$

    **if** $\min_{j=0,\ldots,t} \left( L_n(\beta_{(j)}) + K \cdot d^{\alpha_U}(\beta, \beta_{(j)}) \right) \geq \max_{j=0,\ldots,t} L_n(\beta_{(j)})$ **then**

        Set $\beta_{(t+1)} = \beta$ and update $t \leftarrow t + 1$

    **end if**

**end while**

**return** $\beta_{(\tau)}$ where $\tau \in \mathrm{argmax}_{t=0,\ldots,T} L_n(\beta_{(t)})$

---

A straightforward extension of the proof of Proposition 11 in Malherbe and Vayatis (2017) leads to the following consistency property of Algorithm 1.

**Proposition S1.** *For a metric space $(\mathcal{M}, d)$ satisfying Assumption 4.2, if $K > 0$ is a Hölder constant with an exponent $\alpha_U \in (0, 1]$ for $L_n$, then Algorithm 1 is consistent for $L_n$,*

4

*i.e.*

$$\max_{t=1,\ldots,T} L_n(\beta_{(t)}) \to \max_{\beta \in \mathcal{M}} L_n(\beta)$$

*in probability, as $T \to \infty$.*

## A.2 Optimization for Riemannian Manifolds

When $\mathcal{M}$ is a Riemannian manifold, the Riemannian structure can be exploited to design a Riemannian gradient descent algorithm (Absil et al., 2009) which is similar to the gradient descent algorithm in Euclidean spaces. Suppose $(\mathcal{M}, g)$ is a $D$-dimensional geodesically complete Riemannian manifold and $\beta \in \mathcal{M}$. Let $\ell : \mathcal{M} \to \mathbb{R}$ be a real-valued function defined on $\mathcal{M}$. Denote by $\text{Exp}_\beta$ the Riemannian exponential map at $\beta$, and by $\text{grad}\ell|_\beta$ the gradient of $\ell$ at $\beta$. Let $\gamma_{(t)}$, for each step $t$, be a sequence of step sizes. In practice, $\gamma_{(t)}$ can be the Armijo step size (Absil et al., 2009, Definition 4.2.2) or a fixed number. We stop the iteration once $\|\text{grad}\ell|_{\beta_{(t)}}\|_{\beta_{(t)}} < \varepsilon$ for some small number $\varepsilon > 0$, where $\|\cdot\|_{\beta_{(t)}}$ is the norm induced by the Riemannian metric on $\mathcal{M}$ at $\beta_{(t)}$. The optimization procedure is summarized in Algorithm 2.

To apply Algorithm 2, one needs to find the gradient of $\ell$ at $\beta$. To this end, recall that the gradient of $\ell$ at $\beta$, $\text{grad}\ell|_\beta$, is the unique tangent vector in tangent space $T_\beta \mathcal{M}$ satisfying

$$\langle \text{grad}\ell|_\beta, \xi \rangle_\beta = D\ell(\beta)[\xi], \quad \forall \xi \in T_\beta \mathcal{M}, \tag{S2}$$

where $\langle \cdot, \cdot \rangle_\beta$ is the Riemannian metric on $T_\beta \mathcal{M}$, and

$$D\ell(\beta)[\xi] = \left.\frac{\mathrm{d}}{\mathrm{d}t}\ell(\gamma(t))\right|_{t=0} \tag{S3}$$

is the directional derivative of the scalar field $\ell$ on $\mathcal{M}$ at $\beta$ in the direction $\xi$, with $\gamma$ being a smooth curve on $\mathcal{M}$ such that $\gamma(0) = \beta$ and $\gamma'(0) = \xi$. Therefore, for any given $\xi \in T_\beta \mathcal{M}$, one may evaluate the gradient $\text{grad}\ell|_\beta$ numerically via $D\ell(\beta)[\xi]$, as follows:

5

**Algorithm 2** Riemannian Gradient Descent Algorithm

**Require:** An initial point $\beta_{(0)} \in \mathcal{M}$, a sequence of step sizes $\{\gamma_{(t)}\} \subset \mathbb{R}$ and a tolerance level $\varepsilon > 0$.

$t \leftarrow 0$

$\eta_{(0)} = -\text{grad}\ell|_{\beta_{(0)}}$

**while** $\|\eta_{(t)}\|_{\beta_{(t)}} \geq \varepsilon$ **do**

$\quad t \leftarrow t + 1$

$\quad$ Move in the direction $\eta_{(t-1)}$, i.e., $\beta_{(t)} = \text{Exp}_{\beta_{(t-1)}}(\gamma_{(t-1)}\eta_{(t-1)})$

$\quad$ Compute the negative gradient of $\ell$ at $\beta_{(t)}$, i.e., $\eta_{(t)} = -\text{grad}\ell|_{\beta_{(t)}}$

**end while**

**return** $\beta_{(t)}$

---

1. Choose $D = \dim(\mathcal{M})$ tangent vectors $\xi^1, \ldots, \xi^D \subset T_\beta\mathcal{M}$ that are orthonormal with respect to the inner product $\langle \cdot, \cdot \rangle_\beta$.

2. Calculate the generalized Fourier coefficient $C_j = D\ell(\beta)[\xi^j]$ according to (S3) for each $1 \leq j \leq D$.

3. Calculate the gradient $\text{grad}\ell|_\beta = \sum_{j=1}^{D} C_j \xi^j$.

### A.3 Compute the Alexandrov Inner Product

To evaluate the empirical likelihood $L_n$ at $\beta$, given $\hat{\mu}_n$ and $X_i$, one needs to calculate the Alexandrov inner product $h(\beta; X_i, \hat{\mu}_n)$. In the context of a Riemannian manifold, we can utilize the Riemannian logarithmic map $\text{Log}_{\hat{\mu}_n}$ and the identity $h(\beta; X_i, \hat{\mu}_n) = \langle \text{Log}_{\hat{\mu}_n}\beta, \text{Log}_{\hat{\mu}_n}X_i \rangle_{\hat{\mu}_n}$ for computation, where $\langle \cdot, \cdot \rangle_{\hat{\mu}_n}$ is the Riemannian metric at $\hat{\mu}_n$.

While there may be space-specific methods to compute the Alexandrov inner product for certain metric spaces (e.g., Riemannian manifolds in the above), for general metric spaces,

we may adopt the following method to compute the Alexandrov inner product $h(\beta; X_i, \hat{\mu}_n)$. The key is to computing the Alexandrov angle $\angle_{\hat{\mu}_n}(\beta, X_i)$ according to its definition in Equation (2), as follows. For a sufficiently small $r > 0$ and a sufficiently large integer $m > 0$, take two sequences $t_j = jr/m$ and $s_k = kr/m$ for $j, k = 1, \ldots, m$. Then we approximate the Alexandrov angle $\angle_{\hat{\mu}_n}(\beta, X_i)$ by

$$\angle_{\hat{\mu}_n}(\beta, X_i) \approx \sup_{j,k} \bar{\angle}_{\hat{\mu}_n}(\gamma^{\beta}_{\hat{\mu}_n}(t_j), \gamma^{X_i}_{\hat{\mu}_n}(s_k)),$$

where $\bar{\angle}$ is the comparison angle defined in Section 2 and can be computed straightforwardly. The approximation in the above becomes more accurate as $r$ decreases and $m$ increases. Once we obtain this approximate of the Alexandrov angle, we combine it with $d(\hat{\mu}_n, \beta)$ and $d(\hat{\mu}_n, X_i)$ to compute the Alexandrov inner product.

# B  Further Details on Data Application

## B.1  Preprocessing for the Data Application

In this section, we provide preprocessing details for data used in Section 7. According to the Desikan–Killiany atlas (Desikan et al., 2006), from the left hemisphere we choose $m = 8$ ROIs that may be related to motor and language processing based on the study of Barch et al. (2013). Specifically, we consider the superior temporal gyrus, three parts of inferior frontal gyrus (including the opercular part, triangular part and orbital part), frontal pole, precentral gyrus, lingual gyrus and lateral occipital cortex. To construct the functional connectivity $X_i$, denote by $V_{it}$ the $m$-dimensional vector of the BOLD signals from the $m$ ROIs of the $i$th subject at the $t$th time point for $i = 1, \ldots, n$ and $t = 1, \ldots, T$. For the motor task, there are 284 time points for a complete run. Removing the time points corresponding to the initiation and fixation blocks (Barch et al., 2013) results in $T = 210$

time points for the motor task. For the language processing task, there are no initiation or fixation blocks, and $T = 316$ for each complete run. For each subject $i$, we then compute the covariance matrix of $V_{it}$ with a moving local window of length $P$ for each time point $t$. Specifically, for $t = P+1, \ldots, T-P+1$, we have

$$S_{it} = \frac{1}{2P} \sum_{k=t-P}^{t+P-1} (V_{ik} - \bar{V}_{it})(V_{ik} - \bar{V}_{it})^\top \quad \text{with} \quad \bar{V}_{it} = \frac{1}{2P} \sum_{k=t-P}^{t+P-1} V_{ik}.$$

Finally, in analogy to that in Lin and Müller (2021), we aggregate the information for each subject $i$ across all time points to obtain a single covariate

$$X_i = \operatorname*{argmin}_{X \in \mathcal{S}_m^+} \sum_{t=P+1}^{T-P+1} d^2(X, S_{it}),$$

where $d$ is the Log-Cholesky metric (Lin, 2019)[1]. For the choice of $P$, Lin and Müller (2021) suggested $P \in [m, 3m]$. As the results are not sensitive to the choice of $P$ in this range, we simply set $P = 2m$ in our study. For better numerical stability, computation is performed after scaling each matrix by the constant $10^{-3}$.

## B.2 A Permutation Test for the Covariate Effect

For Euclidean spaces, no covariate effect is corresponding to $\beta^* = 0$, which is not meaningful in non-linear spaces. Instead, in our model, $\beta^* = \mu^*$ corresponds to no covariate effect, as in this case, the right-hand side of (6) is zero regardless of the value of $X$. Here, we may adopt a generic permutation test for $\beta^* = \mu^*$, as follows.

Given observations $(X_1, Y_2), \ldots, (X_n, Y_n)$, we produce the estimates $\hat{\beta}_n$ and $\hat{\mu}_n$, and compute a measure $G_0$ of goodness of fit (e.g., deviance). Then, given a large integer $B > 0$, for each $i = 1, \ldots, B$, we permute the covaraite randomly and reproduce the estimates and

---

[1] Results for other commonly used metrics on $\mathcal{S}_m^+$, such as the affine-invariant and Log-Euclidean metrics, are similar to those for the Log-Cholesky metric in our analysis.

the measure $G_i$ of goodness of fit based on the permuted data. As the empirical distribution of $\mathcal{G} = \{G_1, \ldots, G_B\}$ approximates the distribution of $G_0$ under the null hypothesis, the empirical $p$-value is the proportion of the event that $G_0$ is larger than those in $\mathcal{G}$ among the $B$ permutations. Finally, reject the null hypothesis that there is no covariate effect if the $p$-value is smaller than a pre-specified significance level (e.g., 5%). Note that each permutation of $X_1, \ldots, X_n$ leads to a different estimate of $\beta^*$ but the same empirical Fréchet mean $\hat{\mu}_n$. Therefore, the above permutation procedure tests only the significance of $\hat{\beta}_n$ but not the effect of $\hat{\mu}_n$, which is exactly what we intend to achieve.

For any estimates $\hat{\beta}$ and $\hat{\mu}$, the deviance corresponding to the proposed regression model (6) is

$$D = -2 \sum_{i=1}^{n} \left\{ Y_i h(\hat{\beta}; X_i, \hat{\mu}) - \log(1 + e^{h(\hat{\beta}; X_i, \hat{\mu})}) \right\}$$

where $h$ defined in (6). We adopt this deviance as the measure of goodness of fit and apply the above permutation test with $B = 1000$ to our data application. The deviances derived from the permuted data are represented by dots in Figure S1, where the deviance derived from the original data is indicated by the dashed line. None of the dots are below the dashed line, suggesting that the $p$-value is less than 0.001. Therefore, we reject the null hypothesis at the significance level 1% and conclude that functional connectivity has significantly different effect on motor skills and language processing.

## C  Odds Ratio and Model (6)

The odds ratio and its logarithmic version are commonly used for binary data analysis. Now we show that $d(\mu^*, \beta^*) \cos(\angle_{\mu^*}(X, \beta^*))$ in the model (6) can be interpreted as a log odds ratio. Specifically, given a point $X \in \mathcal{M}$, consider the geodesic $\gamma(t) := \gamma_{\mu^*}^X(t)$ emanating

9

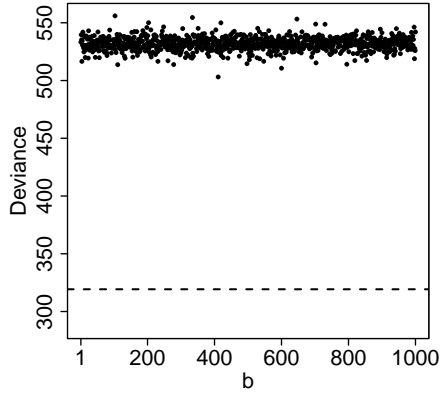

Figure S1: Deviances (dots) from $B = 1000$ permuted data and the deviance (dashed line) from the original data (the dashed line).

from $\mu^*$ to $X$. Then, for a one-unit change along the geodesic $\gamma(t)$, we have

$$R := \frac{P_{\gamma(t+1)}/(1 - P_{\gamma(t+1)})}{P_{\gamma(t)}/(1 - P_{\gamma(t)})} = \frac{\exp[d(\mu^*, \gamma(t+1))d(\mu^*, \beta^*)\cos(\angle_{\mu^*}(\gamma(t+1), \beta^*))]}{\exp[d(\mu^*, \gamma(t))d(\mu^*, \beta^*)\cos(\angle_{\mu^*}(\gamma(t), \beta^*))]}$$
$$= \frac{\exp[(t+1)d(\mu^*, \beta^*)\cos(\angle_{\mu^*}(X, \beta^*))]}{\exp[td(\mu^*, \beta^*)\cos(\angle_{\mu^*}(X, \beta^*))]}$$
$$= \exp[d(\mu^*, \beta^*)\cos(\angle_{\mu^*}(X, \beta^*))],$$

for all $t$, where we recall that $P_x = \mathbb{P}(Y = 1 \mid X = x)$ is the success probability for $x \in \mathcal{M}$. Therefore, the quantity $d(\mu^*, \beta^*)\cos(\angle_{\mu^*}(X, \beta^*))$, also viewed as the scalar projection of $\beta^*$ onto the geodesic $\gamma_{\mu^*}^X$, is the log odds ratio for a one-unit change along the geodesic $\gamma_{\mu^*}^X$.

## D Proofs of Propositions 3.1

*Proof of Proposition 3.1.* To ease the notation, we write $h(\beta) = h(\beta; X, \mu^*)$. We first observe

$$L(\beta) = \mathbb{E}\left[Yh(\beta) - \log(1 + e^{h(\beta)})\right]$$

$$= \mathbb{E}_X \left[ \frac{h(\beta) - \log(1 + e^{h(\beta)})}{1 + e^{-h(\beta^*)}} - \frac{\log(1 + e^{h(\beta)})}{1 + e^{h(\beta^*)}} \right]$$

$$= \mathbb{E}_X \left[ \frac{h(\beta)}{1 + e^{-h(\beta^*)}} - \log(1 + e^{h(\beta)}) \right],$$

where the expectation $\mathbb{E}_X$ is taken over $X$. Note that $L(\beta) = \mathbb{E}_X[M(h(\beta); h(\beta^*))]$ where

$$M(\theta; \theta^*) = \frac{\theta}{1 + e^{-\theta^*}} - \log(1 + e^\theta) \qquad (S4)$$

for any $\theta \in \mathbb{R}$ and a fixed $\theta^* \in \mathbb{R}$, and $\theta^*$ is the unique maximizer of $M(\theta; \theta^*)$ since $M$ is a strictly concave function in $\theta$ and $\frac{dM}{d\theta}\big|_{\theta=\theta^*} = 0$. Therefore, $L(\beta^*) = \mathbb{E}_X[M(h(\beta^*); h(\beta^*))] \geq \mathbb{E}_X[M(h(\beta); h(\beta^*))] = L(\beta)$, proving that $\beta^*$ is a maximizer of $L(\cdot)$ and is the only maximizer when $L(\cdot)$ has a unique maximizer. $\square$

# E  Proofs of Theorems 4.6 and 4.10

Theorems 4.6 and 4.10 follow directly from Theorem S1 below, which in turn follows from Theorem S2.

**Theorem S1.** *Let $k \geq 1$. Assume $\sigma_{X,m} := \mathbb{E}[d^m(\mu^*, X)] < \infty$ and $v_{X,n} := \mathbb{E}[d^m(\mu^*, \hat{\mu}_n)] < \infty$ for some $m \geq 2k$, and take $\tilde{\sigma}_{X,m} = \sigma_{X,m} + v_{X,n}$. Suppose Assumptions 4.1–4.4 hold with $\alpha_L > 1 + k/m$. Then, there exists a constant $C > 0$, such that*

$$\mathbb{E}\left[d^k(\beta^*, \hat{\beta}_n)\right] \leq C v_n^k,$$

*where*

- *if Assumption 4.5(a) holds, then*

$$v_n = (\tilde{\sigma}_{X,m}^{2/m}/\lambda_X^2)^{1/(2(\alpha_L - \alpha_U))} \left(\frac{D}{n}\right)^{1/(2(\alpha_L - \alpha_U))}.$$

- *if Assumption 4.5(b) holds, then*

$$v_n = (\tilde{\sigma}_{X,m}^{2/m}/\lambda_X^2)^{1/(2(\alpha_L-\alpha_U))} \cdot \begin{cases} n^{-1/(2(\alpha_L-\alpha_U))} & if\ D < 2\alpha_U, \\ (n/(\log n)^2)^{-1/(2(\alpha_L-\alpha_U))} & if\ D = 2\alpha_U, \\ n^{-\alpha_U/(D(\alpha_L-\alpha_U))} & if\ D > 2\alpha_U. \end{cases}$$

- *if Assumption 4.9(a) holds, then*

$$v_n = (\tilde{\sigma}_{X,m}^{2/m}/\lambda_X^2)^{1/(2(\alpha_L-\alpha_U+\gamma_0))} \left(\frac{D}{n}\right)^{1/(2(\alpha_L-\alpha_U+\gamma_0))},$$

where $\gamma_0 = W_0(1/2) \approx 0.35173$ and $W_0(\cdot)$ is the Lambert W function.

- *if Assumption 4.9(b) holds, then*

$$v_n = \begin{cases} (\tilde{\sigma}_{X,m}^{2/m}/\lambda_X^2)^{1/(2(\alpha_L-\alpha_U+D/2))} n^{-1/(2(\alpha_L-\alpha_U+D/2))} & if\ D < 2\alpha_U, \\ (\tilde{\sigma}_{X,m}^{2/m}/\lambda_X^2)^{1/(2\alpha_L)} (n/(\log n)^2)^{-1/(2\alpha_L)} & if\ D = 2\alpha_U, \\ (\tilde{\sigma}_{X,m}^{2/m}/\lambda_X^2)^{1/(2\alpha_L)} n^{-\alpha_U/(D\alpha_L)} & if\ D > 2\alpha_U. \end{cases}$$

*Remark* 1. By Assumption 4.3, the distribution of $X$ has a bounded support, then $\hat{\mu}_n$ falls into a neighborhood of $\mu^*$ almost surely. Therefore, $v_{X,n} = O(1) < \infty$ holds trivially.

*Proof.* By Theorem S2, one has

$$v_n^{-k} \mathbb{E}\left[d^k(\beta^*, \hat{\beta}_n)\right] = \int_0^\infty \mathbb{P}\left(v_n^{-k} d^k(\beta^*, \hat{\beta}_n) > t\right) dt$$
$$= \int_0^\infty \mathbb{P}\left(v_n^{-1} d(\beta^*, \hat{\beta}_n) > t^{\frac{1}{k}}\right) dt$$
$$\leq C \int_0^\infty t^{-\frac{m\xi}{2k}} dt,$$

where $\xi$ defined in Theorem S2, and we always have $m\xi/(2k) > 1$ since $\alpha_L > 1 + k/m$. Then the results are obtained by integrating the tail probabilities with the fact that

$$\int_0^\infty \min(1, bt^{-a}) dt = \frac{a}{a-1} b^{\frac{1}{a}},$$

for $a > 1$ and $b > 0$. □

To prepare for the proofs, recall

$$L_n(\beta) = \frac{1}{n}\sum_{i=1}^n \left\{Y_i h(\beta; X_i, \hat{\mu}_n) - \log(1 + e^{h(\beta; X_i, \hat{\mu}_n)})\right\},$$

$$L(\beta) = \mathbb{E}\left[Y h(\beta; X, \mu^*) - \log(1 + e^{h(\beta; X, \mu^*)})\right].$$

**Theorem S2.** *Under the same conditions of Theorem S1 with $k = 1$, we have*

$$\mathbb{P}\left((\tilde{\sigma}_{X,m}^{2/m}/\lambda_X^2 \cdot u_n)^{-1/\xi} d(\beta^*, \hat{\beta}_n) \geq t\right) \leq Ct^{-m\xi/2},$$

*where $C > 0$ is a constant, $\xi = 2(\alpha_L - \alpha)$, with $\alpha$ and $u_n$ defined as follows:*

- *When Assumption 4.5(a) holds, $\alpha = \alpha_U$, $u_n = D/n$.*

- *When Assumption 4.5(b) holds, $\alpha = \alpha_U$,*

$$u_n = \begin{cases} n^{-1}, & \text{for } D < 2\alpha_U, \\ \frac{(\log n)^2}{n}, & \text{for } D = 2\alpha_U, \\ n^{-2\alpha_U/D}, & \text{for } D > 2\alpha_U. \end{cases}$$

- *When Assumption 4.9(a) holds, $\alpha = \alpha_U - \gamma_0$, $u_n = D/n$, where $\gamma_0 = W_0(1/2) \approx 0.35173$ and $W_0(\cdot)$ is the Lambert W function.*

- *When Assumption 4.9(b) holds,*

$$\begin{cases} \alpha = \alpha_U - D/2, \ u_n = n^{-1}, & \text{for } D < 2\alpha_U, \\ \alpha = 0, \ u_n = \frac{(\log n)^2}{n}, & \text{for } D = 2\alpha_U, \\ \alpha = 0, \ u_n = n^{-2\alpha_U/D}, & \text{for } D > 2\alpha_U. \end{cases}$$

*Proof.* Define $M(\theta) = M(\theta; \theta^*)$, where $M(\theta; \theta^*)$ is defined in (S4). By Assumption 4.1, $X$ has a bounded support, and thus there exists a constant $B > 0$ such that

$$\left|\frac{d^2 M}{d\theta^2}\bigg|_{\theta=\theta^*}\right| = \frac{1}{2}\frac{1}{|e^{-\theta^*} + e^{\theta^*} + 2|} \geq B.$$

13

Therefore, for any $\theta$ in a neighborhood centered at $\theta^*$, by the Taylor expansion,

$$M(\theta) - M(\theta^*) \leq -B(\theta - \theta^*)^2. \tag{S5}$$

For any $\beta \in \mathcal{M}$, by setting $\theta = h(\beta; X, \mu^*)$ and $\theta^* = h(\beta^*; X, \mu^*)$ in (S5), with Assumption 4.4, we deduce

$$L(\beta) - L(\beta^*) = M(h(\beta; X, \mu^*)) - M(h(\beta^*; X, \mu^*)) \leq -C_1 \lambda_X d^{\alpha_L}(\beta, \beta^*), \tag{S6}$$

for some constant $C_1 > 0$.

Based on (S6), for $0 < a < b$, if $d(\beta^*, \hat{\beta}_n) \in [a, b]$, we have

$$C_1 \lambda_X a^{\alpha_L} \leq C_1 \lambda_X d^{\alpha_L}(\beta^*, \hat{\beta}_n) \leq L(\beta^*) - L(\hat{\beta}_n)$$

$$\leq L(\beta^*) - L(\hat{\beta}_n) + L_n(\hat{\beta}_n) - L_n(\beta^*)$$

$$\leq \sup_{d(\beta, \beta^*) \leq b} \Delta_n(\beta),$$

where

$$\Delta_n(\beta) := L(\beta^*) - L(\beta) + L_n(\beta) - L_n(\beta^*), \tag{S7}$$

and all constants in the above inequalities do not depend on $\hat{\beta}_n$.

For $s > 0$ and $m \geq 2$, setting $a_k = s2^k$, $k \in \mathbb{N}$, we have

$$\mathbb{P}\left(d(\beta^*, \hat{\beta}_n) \geq s\right) \leq \sum_{k=0}^{\infty} \mathbb{P}\left(d(\beta^*, \hat{\beta}_n) \in [a_k, a_{k+1}]\right)$$

$$\leq \sum_{k=0}^{\infty} \mathbb{P}\left(\sup_{d(\beta^*, \beta) \leq a_{k+1}} \Delta_n(\beta) \geq C_1 \lambda_X a_k^{\alpha_L}\right)$$

$$\leq C_2 \sum_{k=0}^{\infty} \frac{\mathbb{E}[\sup_{d(\beta^*, \beta) \leq a_{k+1}} \Delta_n(\beta)^m]}{\lambda_X^m a_k^{m\alpha_L}}, \tag{S8}$$

where we use Markov's inequality for the last inequality.

With $\tilde{\sigma}_{X,m} = \sigma_{X,m} + v_{X,n}$, by applying Claim S1 to $\mathbb{E}[\sup_{d(\beta^*,\beta) \leq a_{k+1}} \Delta_n(\beta)^m]$, we derive the desired upper bound under different metric entropy conditions:

14

- If Assumption 4.5(a) holds,

$$\mathbb{P}\left(d(\beta^*, \hat{\beta}_n) \geq s\right) \leq C_3 \sum_{k=0}^{\infty} \left(\frac{D}{n}\right)^{m/2} \frac{\tilde{\sigma}_{X,m} a_{k+1}^{m\alpha_U}}{\lambda_X^m a_k^{m\alpha_L}}$$

$$\leq C_4 \frac{\tilde{\sigma}_{X,m}}{\lambda_X^m} \left(\frac{D}{n}\right)^{m/2} s^{-m(\alpha_L - \alpha_U)} \sum_{k=0}^{\infty} 2^{-mk(\alpha_L - \alpha_U)}$$

$$\leq C_5 \frac{\tilde{\sigma}_{X,m}}{\lambda_X^m} \left(\frac{D}{n}\right)^{m/2} s^{-m(\alpha_L - \alpha_U)}.$$

For $t > 0$, taking $s = t\left(\tilde{\sigma}_{X,m}^{2/m}/\lambda_X^2 \cdot D/n\right)^{1/(2(\alpha_L - \alpha_U))}$ leads to

$$\mathbb{P}\left(\left(\tilde{\sigma}_{X,m}^{2/m}/\lambda_X^2 \cdot D/n\right)^{-1/(2(\alpha_L - \alpha_U))} d(\beta^*, \hat{\beta}_n) \geq t\right) \leq C_6 t^{-m(\alpha_L - \alpha_U)}.$$

- If Assumption 4.5(b) holds,

$$\mathbb{P}\left(d(\beta^*, \hat{\beta}_n) \geq s\right) \leq C_3 \sum_{k=0}^{\infty} \frac{\tilde{\sigma}_{X,m}}{\lambda_X^m} \frac{1}{a_k^{m\alpha_L}} \cdot \begin{cases} n^{-m/2} a_{k+1}^{m\alpha_U} & if\ D < 2\alpha_U, \\ \left(\frac{\log^2 n}{n}\right)^{m/2} a_{k+1}^{m\alpha_U} & if\ D = 2\alpha_U, \\ n^{-m\alpha_U/D} a_{k+1}^{m\alpha_U} & if\ D > 2\alpha_U, \end{cases}$$

$$\leq C_4 \frac{\tilde{\sigma}_{X,m}}{\lambda_X^m} \cdot \begin{cases} n^{-m/2} s^{-m(\alpha_L - \alpha_U)} & if\ D < 2\alpha_U, \\ \left(\frac{\log^2 n}{n}\right)^{m/2} s^{-m(\alpha_L - \alpha_U)} & if\ D = 2\alpha_U, \\ n^{-m\alpha_U/D} s^{-m(\alpha_L - \alpha_U)} & if\ D > 2\alpha_U. \end{cases}$$

The desired result then follows by taking

$$s = t(\tilde{\sigma}_{X,m}^{2/m}/\lambda_X^2)^{1/(2(\alpha_L - \alpha_U))} \cdot \begin{cases} n^{-1/(2(\alpha_L - \alpha_U))} & if\ D < 2\alpha_U, \\ (n/(\log n)^2)^{-1/(2(\alpha_L - \alpha_U))} & if\ D = 2\alpha_U, \\ n^{-\alpha_U/(D(\alpha_L - \alpha_U))} & if\ D > 2\alpha_U, \end{cases}$$

for $t > 0$.

- If Assumption 4.9(a) holds,

$$\mathbb{P}\left(d(\beta^*, \hat{\beta}_n) \geq s\right) \leq C_3 \sum_{k=0}^{\infty} \frac{\tilde{\sigma}_{X,m}}{\lambda_X^m} \left(\frac{D}{n}\right)^{m/2} \frac{a_{k+1}^{m(\alpha_U - \gamma_0)}}{a_k^{m\alpha_L}}$$

15

$$\leq C_4 \frac{\tilde{\sigma}_{X,m}}{\lambda_X^m} \left(\frac{D}{n}\right)^{m/2} s^{-m(\alpha_L - \alpha_U + \gamma_0)},$$

where $\gamma_0 = W_0(1/2) \approx 0.35173$ and $W_0(\cdot)$ is the Lambert W function. The desired result follows by taking $s = t \left(\tilde{\sigma}_{X,m}^{2/m}/\lambda_X^2 \cdot D/n\right)^{1/(2(\alpha_L - \alpha_U + \gamma_0))}$ for $t > 0$.

- If Assumption 4.9(b) holds,

$$\mathbb{P}\left(d(\beta^*, \hat{\beta}_n) \geq s\right) \leq C_3 \sum_{k=0}^{\infty} \frac{\tilde{\sigma}_{X,m}}{\lambda_X^m} \frac{1}{a_k^{m\alpha_L}} \cdot \begin{cases} n^{-m/2} a_{k+1}^{m(\alpha_U - D/2)} & \text{if } D < 2\alpha_U, \\ \left(\frac{\log^2 n}{n}\right)^{m/2} & \text{if } D = 2\alpha_U, \\ n^{-m\alpha_U/D} & \text{if } D > 2\alpha_U, \end{cases}$$

$$\leq C_4 \frac{\tilde{\sigma}_{X,m}}{\lambda_X^m} \cdot \begin{cases} n^{-m/2} s^{-m(\alpha_L - \alpha_U + D/2)} & \text{if } D < 2\alpha_U, \\ \left(\frac{\log^2 n}{n}\right)^{m/2} s^{-m\alpha_L} & \text{if } D = 2\alpha_U, \\ n^{-m\alpha_U/D} s^{-m\alpha_L} & \text{if } D > 2\alpha_U. \end{cases}$$

The desired result then follows by taking

$$s = t \cdot \begin{cases} (\tilde{\sigma}_{X,m}^{2/m}/\lambda_X^2)^{1/(2(\alpha_L - \alpha_U + D/2))} n^{-1/(2(\alpha_L - \alpha_U + D/2))} & \text{if } D < 2\alpha_U, \\ (\tilde{\sigma}_{X,m}^{2/m}/\lambda_X^2)^{1/(2\alpha_L)} (n/(\log n)^2)^{-1/(2\alpha_L)} & \text{if } D = 2\alpha_U, \\ (\tilde{\sigma}_{X,m}^{2/m}/\lambda_X^2)^{1/(2\alpha_L)} n^{-\alpha_U/(D\alpha_L)} & \text{if } D > 2\alpha_U, \end{cases}$$

for $t > 0$.

□

**Claim S1.** *Suppose Assumptions 4.1 to 4.3 hold, and let $\sigma_{X,m} = \mathbb{E}[d^m(\mu^*, X_1)]$ and $v_{X,n} = \mathbb{E}[d^m(\mu^*, \hat{\mu}_n)]$ with $m \geq 2$. For $\delta > 0$, we have*

$$\mathbb{E}\left[\sup_{d(\beta^*, \beta) \leq \delta} \Delta_n(\beta)^m\right] \leq C(\sigma_{X,m} + v_{X,n}) \eta_{n,\delta}^m,$$

*where $\Delta_n(\beta)$ is defined in (S7), $C > 0$ is a constant not depending on $n$ and $\delta$, and*

16

- *if Assumption 4.5(a) holds, then*

$$\eta_{n,\delta} = (D/n)^{1/2} \delta^{\alpha_U}.$$

- *if Assumption 4.5(b) holds, then*

$$\eta_{n,\delta} = \begin{cases} n^{-1/2}\delta^{\alpha_U} & if\ D < 2\alpha_U, \\ \frac{\log n}{\sqrt{n}}\delta^{\alpha_U} & if\ D = 2\alpha_U, \\ n^{-\alpha_U/D}\delta^{\alpha_U} & if\ D > 2\alpha_U. \end{cases}$$

- *if Assumption 4.9(a) holds, then*

$$\eta_{n,\delta} = (D/n)^{1/2} \delta^{\alpha_U - \gamma_0},$$

where $\gamma_0 = W_0(1/2) \approx 0.35173$ and $W_0(\cdot)$ is the Lambert W function.

- *if Assumption 4.9(b) holds, then*

$$\eta_{n,\delta} = \begin{cases} n^{-1/2}\delta^{\alpha_U - D/2} & if\ D < 2\alpha_U, \\ \frac{\log n}{\sqrt{n}} & if\ D = 2\alpha_U, \\ n^{-\alpha_U/D} & if\ D > 2\alpha_U. \end{cases}$$

*Proof.* Define

$$f(\beta; x, y, \mu^*) = yh(\beta; x, \mu^*) - \log(1 + e^{h(\beta;x,\mu^*)})$$

and

$$Z_i(\beta) = n^{-1}\left(\mathbb{E}[f(\beta^*; X_i, Y_i, \mu^*) - f(\beta; X_i, Y_i, \mu^*)] + f(\beta; X_i, Y_i, \hat{\mu}_n) - f(\beta^*; X_i, Y_i, \hat{\mu}_n)\right).$$

Then $\sup_{d(\beta^*,\beta)\leq\delta} \Delta_n(\beta) = \sup_{\beta \in B(\beta^*,\delta)} \sum_{i=1}^n Z_i(\beta)$ with $B(\beta^*, \delta)$ being the ball of radius $\delta$ centered at $\beta^*$. By the boundedness of the support of $X$ and Assumption 4.2, each $Z_i$ is integrable. Moreover, $Z_i(\cdot)$ is Hölder continuous (with $\alpha_U \in (0, 1]$) because

$$|Z_i(\beta_1) - Z_i(\beta_2)|$$

17

$$\leq \frac{1}{n}|\mathbb{E}[f(\beta_1; X_i, Y_i, \mu^*) - f(\beta_2; X_i, Y_i, \mu^*)]| + \frac{1}{n}|f(\beta_1; X_i, Y_i, \hat{\mu}_n) - f(\beta_2; X_i, Y_i, \hat{\mu}_n)|$$

$$\leq \frac{2}{n}\mathbb{E}|h(\beta_1; X_i, \mu^*) - h(\beta_2; X_i, \mu^*)| + \frac{2}{n}|h(\beta_1; X_i, \hat{\mu}_n) - h(\beta_2; X_i, \hat{\mu}_n)|$$

$$\leq \frac{2}{n}[\mathbb{E}d(\mu^*, X_i) + d(\hat{\mu}_n, X_i)]d^{\alpha_U}(\beta_1, \beta_2).$$

Given an independent copy $(Z_1', \ldots, Z_n')$ of $(Z_1, \ldots, Z_n)$, for each $i$, we then have

$$|Z_i(\beta_1) - Z_i(\beta_2) - Z_i'(\beta_1) + Z_i'(\beta_2)|$$
$$\leq \frac{4}{n}[\mathbb{E}d(\mu^*, X_i) + \mathbb{E}d(\mu^*, X_i') + d(\hat{\mu}_n, X_i) + d(\hat{\mu}_n, X_i')]d^{\alpha_U}(\beta_1, \beta_2).$$

Let $A_i = \frac{4}{n}[\mathbb{E}d(\mu^*, X_i) + \mathbb{E}d(\mu^*, X_i') + d(\hat{\mu}_n, X_i) + d(\hat{\mu}_n, X_i')]$ and $A = (A_1, \ldots, A_n)^\top$. To apply Theorem 6 in Schötz (2019), we first observe

$$\mathbb{E}[\|A\|_2^m] = \mathbb{E}\left[\left(\sum_{i=1}^n A_i^2\right)^{m/2}\right] \leq n^{m/2-1}\sum_{i=1}^n \mathbb{E}A_i^m$$

$$= \frac{C_1}{n^{m/2}}\mathbb{E}\left[(\mathbb{E}d(\mu^*, X_1) + \mathbb{E}d(\mu^*, X_1') + d(\hat{\mu}_n, X_1) + d(\hat{\mu}_n, X_1'))^m\right]$$

$$\leq C_1 2^m \frac{3^{m-1}}{n^{m/2}}\left(\mathbb{E}[d^m(\mu^*, X_1)] + \mathbb{E}[d^m(\mu^*, X_1')] + \mathbb{E}[d^m(\mu^*, \hat{\mu}_n)]\right)$$

$$= C_2 \frac{\sigma_{X,m} + v_{X,n}}{n^{m/2}},$$

where the two inequalities hold due to the basic inequality $|\sum_{k=1}^K x_k|^q \leq K^{q-1}\sum_{k=1}^K |x_k|^q$ for $q \geq 1$ and any real-valued sequence $\{x_k, k = 1, \ldots, K\}$. Then Theorem 6 in Schötz (2019) provides an upper bound for $\mathbb{E}[\sup_{d(\beta^*, \beta) \leq \delta} \Delta_n^m(\delta)]$:

$$\mathbb{E}[\sup_{d(\beta^*, \beta) \leq \delta} \Delta_n(\delta)^m] \leq 8C_3 \frac{\sigma_{X,m} + v_{X,n}}{n^{m/2}}[\text{entr}_n(B(\beta^*, \delta), d^{\alpha_U})]^m,$$

where $\text{entr}_n(Q, d)$ defined in (S1).

We complete the proof by bounding $\text{entr}_n(B(\beta^*, \delta), d^{\alpha_U})$ using Lemma S4, as follows, where $C_4 > 0$ is a constant depending only on $m$.

- If Assumption 4.5(a) holds, according to Lemma S4, we have

$$\mathbb{E}[\sup_{d(\beta^*, \beta) \leq \delta} \Delta_n(\delta)^m] \leq C_4(\sigma_{X,m} + v_{X,n})\left(\frac{D}{n}\right)^{m/2}\delta^{m\alpha_U}.$$

18

- If Assumption 4.5(b) holds, according to Lemma S4, we have

$$\mathbb{E}[\sup_{d(\beta^*,\beta)\leq \delta} \Delta_n(\delta)^m] \leq C_4(\sigma_{X,m} + v_{X,n}) \cdot \begin{cases} n^{-m/2}\delta^{m\alpha_U} & if \ D < 2\alpha_U, \\ \left(\frac{\log^2 n}{n}\right)^{m/2} \delta^{m\alpha_U} & if \ D = 2\alpha_U, \\ n^{-m\alpha_U/D}\delta^{m\alpha_U} & if \ D > 2\alpha_U. \end{cases}$$

- If Assumption 4.9(a) holds, according to Lemma S4, we have

$$\mathbb{E}[\sup_{d(\beta^*,\beta)\leq \delta} \Delta_n(\delta)^m] \leq C_4(\sigma_{X,m} + v_{X,n}) \left(\frac{D}{n}\right)^{m/2} \delta^{m(\alpha_U - \gamma_0)},$$

where $\gamma_0 = W_0(1/2) \approx 0.35173$ and $W_0(\cdot)$ is the Lambert W function.

- If Assumption 4.9(b) holds, according to Lemma S4, we have

$$\mathbb{E}[\sup_{d(\beta^*,\beta)\leq \delta} \Delta_n(\delta)^m] \leq C_4(\sigma_{X,m} + v_{X,n}) \cdot \begin{cases} n^{-m/2}\delta^{m(\alpha_U - D/2)} & if \ D < 2\alpha_U, \\ \left(\frac{\log^2 n}{n}\right)^{m/2} & if \ D = 2\alpha_U, \\ n^{-m\alpha_U/D} & if \ D > 2\alpha_U. \end{cases}$$

□

# F   Proofs of Theorems 4.7 and 4.11

In this section, we would consider a bounded subset $\mathcal{M}_0$ of $\mathcal{M}$, and then establish minimax lower bounds of $d^2(\beta, \hat{\beta}_n)$ over $\mathcal{M}$ via deriving the corresponding minimax lower bounds over $\mathcal{M}_0$ with the following relationship:

$$\inf_{\hat{\beta}_n} \sup_{\beta \in \mathcal{M}} \mathbb{E}_\beta[d^2(\beta, \hat{\beta}_n)] \geq \inf_{\hat{\beta}_n} \sup_{\beta \in \mathcal{M}_0} \mathbb{E}_\beta[d^2(\beta, \hat{\beta}_n)].$$

Let $\mathfrak{M}$ be the diameter of $\mathcal{M}_0$. Given a $\beta \in \mathcal{M}_0$, denote by $P_{XY}(\beta)$ a joint distribution of $X$ and $Y$ when the true parameter in the model (6) is $\beta$. Let $\mathcal{P}_{\mathcal{M}_0}$ be the set consisting

of $P_{XY}^n(\beta)$, the $n$-fold product distribution of $P_{XY}(\beta)$, for all $\beta \in \mathcal{M}_0$. Analogously, define the conditional distributions $P_X(\beta)$ and $P_X^n(\beta)$ of $Y$ given $X$.

*Proof of Theorem 4.7.* In light of the inequality $\mathbb{E}_\beta[d^2(\beta, \hat{\beta}_n)] \geq \{\mathbb{E}_\beta[d(\beta, \hat{\beta}_n)]\}^2$, it is sufficient to establish a minimax lower bound for $\mathbb{E}_\beta[d(\beta, \hat{\beta}_n)]$, and we accomplish this by using the local Fano method. To this end, we consider the idea from Lemma 3 in Yang and Barron (1999) (also see Theorem 6 in Massart and Nédélec (2006)) to construct a local packing first.

- Suppose $N(\epsilon, B(\beta^*, \delta), d) \asymp \left(\frac{\delta}{\epsilon}\right)^D$. That is, given $\delta > 0$, for all $0 < \epsilon \leq \delta$, there are $C_u \geq C_l > 0$, such that

$$\left(\frac{C_l \delta}{\epsilon}\right)^D \leq N(\epsilon, B(\beta^*, \delta), d) \leq \left(\frac{C_u \delta}{\epsilon}\right)^D.$$

Therefore, for the packing number $M(\epsilon, B(\beta^*, \delta), d)$, one has

$$\left(\frac{C_l \delta}{\epsilon}\right)^D \leq M(\epsilon, B(\beta^*, \delta), d) \leq \left(\frac{2C_u \delta}{\epsilon}\right)^D. \tag{S9}$$

For a $0 < \delta \leq \mathfrak{M}/4$ with $\mathfrak{M}$ being the diameter of $\mathcal{M}_0$, consider the ball $B(\beta^*, 2\delta) \subset \mathcal{M}_0$. Let $\mathcal{V}_\delta$ and $\mathcal{V}_{c\delta}$ be the largest $\delta$-packing set and the largest $c\delta$-packing set of $B(\beta^*, 2\delta)$, respectively, with a constant $c = C_l/(2eC_u) < 1$. It is clear that any point in $\mathcal{V}_{c\delta}$ must belong to some ball with radius $\delta$ centered at some point in $\mathcal{V}_\delta$. Let $\mathcal{V}$ be the intersection of $\mathcal{V}_{c\delta}$ with such a ball with maximal cardinality. Then for two distinct points $\beta_1, \beta_2 \in \mathcal{V}$, one has,

$$c\delta \leq d(\beta_1, \beta_2) \leq 2\delta, \tag{S10}$$

and

$$\log |\mathcal{V}| \geq \log M(c\delta, B(\beta^*, 2\delta), d) - \log M(\delta, B(\beta^*, 2\delta), d)$$

$$\geq D \log(2C_l\delta/(c\delta)) - D \log(4C_u\delta/\delta) = D, \tag{S11}$$

where the first inequality is from Lemma 3 in Yang and Barron (1999), and the second inequality is due to (S9).

- Suppose $\log N(\epsilon, B(\beta^*, \delta), d) \asymp \left(\frac{\delta}{\epsilon}\right)^D$. That is, given $\delta > 0$, for all $0 < \epsilon \leq \delta$, there are $C_u \geq C_l > 0$, such that

$$\left(\frac{C_l\delta}{\epsilon}\right)^D \leq \log M(\epsilon, B(\beta^*, \delta), d) \leq \left(\frac{2C_u\delta}{\epsilon}\right)^D.$$

Similar to the previous case, we can find a local packing $\mathcal{V}$ for $B(\beta^*, 2\delta)$ with $0 < \delta \leq \mathfrak{M}/4$ satisfying (S10) with $c = C_l/(2^{1/D+1}C_u) < 1$ and

$$\log |\mathcal{V}| \geq (4C_u)^D. \tag{S12}$$

With the local packing $\mathcal{V} \subset \mathcal{M}_0(\alpha_U)$, to apply the local Fano method, we need to derive an upper bound for the KL-divergence of each pair of distributions induced by elements in $\mathcal{V}$. For all $\beta_1, \beta_2 \in \mathcal{V}$,

$$\begin{aligned}
D_{\mathrm{kl}}(P_{XY}^n(\beta_1) \| P_{XY}^n(\beta_2)) &= \mathbb{E}_X[D_{\mathrm{kl}}(P_X^n(\beta_1) \| P_X^n(\beta_2))] \\
&= \mathbb{E}_X\left[\sum_{i=1}^n D_{\mathrm{kl}}(P_{X_i}(\beta_1) \| P_{X_i}(\beta_2))\right] \\
&\leq C_1 \mathbb{E}_X\left[\sum_{i=1}^n (h(\beta_1; X_i, \mu^*) - h(\beta_2; X_i, \mu^*))^2\right] \\
&\leq C_1 C_U^2 \mathbb{E}_X\left[\sum_{i=1}^n d^2(\mu^*, X_i)\right] d^{2\alpha_U}(\beta_1, \beta_2) \\
&= C_1 C_U^2 \sigma_X^2 n d^{2\alpha_U}(\beta_1, \beta_2) \leq C_2 C_U^2 \sigma_X^2 n \delta^{2\alpha_U}, \tag{S13}
\end{aligned}$$

where the expectation $\mathbb{E}_X$ is taken over $X_i, i = 1, \ldots, n$, the first inequality is due to Lemma S5, the second inequality is from Assumption 4.2, and the last inequality is from (S10).

21

Finally, by applying the Fano's method, setting

$$\delta = C_0 \left(\sigma_X^2\right)^{-1/(2\alpha_U)} \cdot \begin{cases} \left(\frac{D}{n}\right)^{1/(2\alpha_U)} & \text{if } N\left(\epsilon, B(\beta^*, \delta), d\right) \asymp \left(\frac{\delta}{\epsilon}\right)^D, \\ n^{-1/(2\alpha_U)} & \text{if } \log N\left(\epsilon, B(\beta^*, \delta), d\right) \asymp \left(\frac{\delta}{\epsilon}\right)^D, \end{cases}$$

for some suitable constant $C_0$, we obtain the desired results. □

Given a set $\mathcal{P}$ of probabilities, define the KL-covering number of $\mathcal{P}$ as

$$N_{\text{kl}}(\epsilon, \mathcal{P}) := \inf \left\{ N \in \mathbb{N} : \exists Q_i \in \mathcal{P}, i = 1, \ldots, N, \; \sup_{P \in \mathcal{P}} \min_i D_{\text{kl}}(P \| Q_i) \leq \epsilon^2 \right\},$$

where $N_{\text{kl}}(\epsilon, \mathcal{P}) = \infty$ if no such cover exists.

*Proof of Theorem 4.11.* In light of the inequality $\mathbb{E}_\beta[d^2(\beta, \hat{\beta}_n)] \geq \{\mathbb{E}_\beta[d(\beta, \hat{\beta}_n)]\}^2$, it is sufficient to establish a minimax lower bound for $\mathbb{E}_\beta[d(\beta, \hat{\beta}_n)]$.

- Suppose $N(\epsilon, \mathcal{M}, d) \asymp \left(\frac{C}{\epsilon}\right)^D$. The proof follows from the same argument in the proof of Theorem 4.7 via the local Fano method. In this case, for all $0 < \delta \leq \mathfrak{M}$, there are $C_u \geq C_l > 0$, such that

$$\left(\frac{C_l}{\delta}\right)^D \leq M(\delta, \mathcal{M}, d) \leq \left(\frac{2C_u}{\delta}\right)^D,$$

for the packing number $M(\delta, \mathcal{M}, d)$. Consequently, for $0 < \delta \leq \mathfrak{M}/2$, we can find a local packing $\mathcal{V}$ satisfying (S10) and (S11). By the same argument in the proof of Theorem 4.7, we conclude

$$\inf_{\hat{\beta}_n} \sup_{\beta \in \mathcal{M}(\alpha_U, C_U)} \mathbb{E}_\beta[d(\hat{\beta}_n, \beta)] \geq C_0 (\sigma_X^2)^{-1/(2\alpha_U)} \left(\frac{D}{n}\right)^{1/(2\alpha_U)},$$

for some constant $C_0 > 0$.

- Suppose $\log N(\epsilon, \mathcal{M}, d) \asymp \left(\frac{C}{\epsilon}\right)^D$. One may apply the same argument as before for this case, but the resulting lower bound is not tight. To obtain a tight lower bound,

22

we adapt the global Fano method in Yang and Barron (1999), in two steps. First, we choose $\epsilon_n > 0$ such that

$$\epsilon_n^2 \geq \log N_{\text{kl}}(\epsilon_n, \mathcal{P}_{\mathcal{M}}). \tag{S14}$$

Second, we choose the largest $\delta_n > 0$ such that

$$\log M(\delta_n, \mathcal{M}, d) \geq 4\epsilon_n^2 + 2 \log 2. \tag{S15}$$

With (S14) and (S15), the Fano's inequality can be applied; see Chapter 15.3.5 in Wainwright (2019) for details.

We complete the proof by finding $\epsilon_n$ and $\delta_n$ in the above. For this, we observe

$$\log N_{\text{kl}}(\epsilon, \mathcal{P}_{\mathcal{M}}) \leq N\left(\left(\frac{\epsilon}{C_1 \sqrt{n}\sigma_X}\right)^{1/\alpha_U}, \mathcal{M}, d\right) \leq C_2 \left(\frac{\sqrt{n}\sigma_X}{\epsilon}\right)^{D/\alpha_U},$$

for some constants $C_1, C_2 > 0$. $\epsilon_n^2 \asymp (\sigma_X^2 n)^{D/(2\alpha_U + D)}$ and $\delta_n \asymp (\epsilon_n^2)^{-1/D}$ fulfill (S14) as well as (S15) since $\log M(\delta_n, \mathcal{M}, d) \geq \log N(\delta_n, \mathcal{M}, d)$. Finally, this leads to the lower bound

$$\inf_{\hat{\beta}_n} \sup_{\beta \in \mathcal{M}(\alpha_U)} \mathbb{E}_\beta[d(\hat{\beta}_n, \beta)] \geq C_3 \delta_n = C_0 (\sigma_X^2)^{-1/(2(\alpha_U + D/2))} n^{-1/(2(\alpha_U + D/2))},$$

for some constants $C_0, C_3 > 0$.

$\square$

# G  Proofs of Theorems 5.1, 5.2 and 5.3

To prepare for the proofs, we state a consequence of Theorem S1 in the following corollary that provides the mean squared error of the estimated success probability.

**Corollary S1.** *For a complete Alexandrov space with curvature upper bounded by $\kappa$ and of diameter no larger than $D_\kappa/2$, if the Alexandrov inner product $h(\beta^*; x, \mu)$ is regular near*

23

$\mu^*$, then, under Assumption 4.3, there exists a constant $C > 0$ such that

$$\mathbb{E}\left[\left(P_X - \hat{P}_X\right)^2\right] \leq C \left(\mathbb{E}\left[d^2(\beta^*, \hat{\beta}_n)\right] + \mathbb{E}\left[d^2(\mu^*, \hat{\mu}_n)\right]\right),$$

If the assumptions of Theorem S1 with $k = 2$ also hold and $\mathbb{E}\left[d^2(\mu^*, \hat{\mu}_n)\right] \lesssim v_n^2$, then additionally,

$$\mathbb{E}\left[d^2(\beta^*, \hat{\beta}_n)\right] + \mathbb{E}\left[d^2(\mu^*, \hat{\mu}_n)\right] \leq C' v_n^2,$$

where $C' > 0$ is a constant and $v_n$ is defined in Theorem S1 with $\alpha_U = 1$.

*Remark* 2. In our setting where we assume $X$ has a bounded support, $d(\mu^*, \hat{\mu}_n)$ is bounded almost surely by a constant, and thus $\mathbb{E}\left[d^2(\mu^*, \hat{\mu}_n)\right] \lesssim v_n^2$ holds trivially. However, this condition may also hold in more general situations. According to Corollary 2 in Schötz (2019), $\mathbb{E}\left[d^2(\mu^*, \hat{\mu}_n)\right] \lesssim v_n^2$ holds under the quadruple inequality and some regularity conditions, with the value $\eta_{\beta,n}$ therein adjusted for different metric entropy conditions accordingly. The quadruple inequality holds for Hadamard spaces and bounded geodesic spaces (Schötz, 2019), and thus the assumption $\mathbb{E}\left[d^2(\mu^*, \hat{\mu}_n)\right] \lesssim v_n^2$ can be satisfied not only for some non-positively curved geodesic spaces, but also for some positively curved geodesic spaces which are bounded (Theorem 1.9, Petrunin and Tuschmann, 1999), without assuming the support of $X$ is bounded.

*Proof.* Recall $B(\mu^*, r)$ denotes the ball of radius $r$ centered at $\mu^*$. Also recall that, the regularity of the Alexandrov inner product $h(\beta^*; x, \mu)$ near $\mu^*$ implies that, there exist constants $C_{h1}, C_{h2} > 0$ and arbitrarily small but fixed constants $0 < r_1 \leq r_2$ such that, for all $\mu \in B(\mu^*, r_2)$,

- $|d(\beta^*, \mu) \cos(\angle_\mu(x, \beta^*)) - d(\beta^*, \mu^*) \cos(\angle_{\mu^*}(x, \beta^*))| \leq C_{h1} d(\mu, \mu^*)$ for all $x$ satisfying $d(x, \mu^*) = r_1$, and

- $|h(\beta^*; x, \mu) - h(\beta^*; x, \mu^*)| \leq C_{h2} d(\mu, \mu^*)$ for all $x \in B(\mu^*, r_2)$.

24

Note that Assumption 4.3 assumes a bounded support of $X$, which further implies that both $d(\mu^*, \hat{\mu}_n)$ and $d(\beta^*, \hat{\mu}_n)$ are bounded by a fixed constant almost surely.

Let $A$ be the event $\hat{\mu}_n \in B(\mu^*, r_1)$ and $\mathbb{I}_{\{A\}}$ the indicator function of $A$. We first observe

$$\mathbb{E}(P_X - \hat{P}_X)^2 \leq \mathbb{E}[(P_X - \hat{P}_X)^2 \mathbb{I}_{\{A\}}] + \mathbb{E}(\mathbb{I}_{\{A^c\}}). \tag{S16}$$

The term about $\mathbb{I}_{\{A^c\}}$ can be bounded by

$$\mathbb{E}(\mathbb{I}_{\{A^c\}}) \leq \mathbb{P}(d(\mu^*, \hat{\mu}_n) \geq r_1) \leq \frac{\mathbb{E}d^2(\mu^*, \hat{\mu}_n)}{r_1^2}. \tag{S17}$$

For the other term, we observe

$$\left(P_X - \hat{P}_X\right)^2 \mathbb{I}_{\{A\}}$$
$$= \left(\{1 + \exp(-h(\beta^*; X, \mu^*))\}^{-1} - \{1 + \exp(-h(\hat{\beta}_n; X, \hat{\mu}_n))\}^{-1}\right)^2 \mathbb{I}_{\{A\}}$$
$$\leq C_1 \left(h(\beta^*; X, \mu^*) - h(\hat{\beta}_n; X, \hat{\mu}_n)\right)^2 \mathbb{I}_{\{A\}}$$
$$\leq C_2 \left[(h(\beta^*; X, \mu^*) - h(\beta^*; X, \hat{\mu}_n))^2 \mathbb{I}_{\{A\}} + \left(h(\beta^*; X, \hat{\mu}_n) - h(\hat{\beta}_n; X, \hat{\mu}_n)\right)^2 \mathbb{I}_{\{A\}}\right]. \tag{S18}$$

According to Propositions S2 and S3, Assumption 4.2 holds with $\alpha_U = 1$. Then, for the second term on the right-hand side of (S18), we deduce

$$\mathbb{E}\left[\left(h(\beta^*; X, \hat{\mu}_n) - h(\hat{\beta}_n; X, \hat{\mu}_n)\right)^2 \mathbb{I}_{\{A\}}\right] \leq C_3 \mathbb{E}\left[d^2(\hat{\mu}_n, X) d^2(\beta^*, \hat{\beta}_n)\right]$$
$$\leq 2C_3 \mathbb{E}\left[(d^2(\mu^*, X) + d^2(\mu^*, \hat{\mu}_n)) d^2(\beta^*, \hat{\beta}_n)\right]$$
$$\leq C_4 \mathbb{E}\left[d^2(\beta^*, \hat{\beta}_n)\right] \leq C_5 v_n^2, \tag{S19}$$

where $\sigma_X^2 = \mathbb{E}[d^2(\mu^*, X)]$, and (only) the last inequality requires the assumptions and results of Theorem S1.

Next, we turn to bound the first term on the right-hand side of (S18). When $X < r_2$, the regularity of the Alexandrov inner product implies $|h(\beta^*; X, \mu^*) - h(\beta^*; X, \hat{\mu}_n)| \mathbb{I}_{\{A\}} \leq C_{h,1} d(\mu^*, \hat{\mu}_n)$. When $X \geq r_2 \geq r_1$, we claim that

$$|d(\beta^*, \mu^*) \cos(\angle_{\mu^*}(X, \beta^*)) - d(\beta^*, \hat{\mu}_n) \cos(\angle_{\hat{\mu}_n}(X, \beta^*))| \mathbb{I}_{\{A\}} \leq c d(\mu^*, \hat{\mu}_n) \tag{S20}$$

25

for some constant $c > 0$, and thus deduce

$$|h(\beta^*; X, \mu^*) - h(\beta^*; X, \hat{\mu}_n)| \, \mathbb{I}_{\{A\}}$$

$$\leq |d(X, \mu^*) d(\beta^*, \mu^*) \cos(\angle_{\mu^*}(X, \beta^*)) - d(X, \hat{\mu}_n) d(\beta^*, \hat{\mu}_n) \cos(\angle_{\hat{\mu}_n}(X, \beta^*))| \, \mathbb{I}_{\{A\}}$$

$$\leq |d(X, \mu^*) \left[ d(\beta^*, \mu^*) \cos(\angle_{\mu^*}(X, \beta^*)) - d(\beta^*, \hat{\mu}_n) \cos(\angle_{\hat{\mu}_n}(X, \beta^*)) \right]| \, \mathbb{I}_{\{A\}}$$

$$+ \left| [d(X, \mu^*) - d(X, \hat{\mu}_n)] d(\beta^*, \hat{\mu}_n) \cos(\angle_{\hat{\mu}_n}(X, \beta^*)) \right| \mathbb{I}_{\{A\}}$$

$$\leq C_6 \left[ d(X, \mu^*) d(\mu^*, \hat{\mu}_n) + d(\mu^*, \hat{\mu}_n) \right] \mathbb{I}_{\{A\}} \leq C_7 d(\mu^*, \hat{\mu}_n),$$

where in the second third inequality we relies on the fact that $d(\beta^*, \hat{\mu}_n)$ is bounded by a constant on the event $A$, and in the last inequality we utilize the boundedness of the support of $X$. Consequently,

$$\mathbb{E}\left[ (h(\beta^*; X, \mu^*) - h(\beta^*; X, \hat{\mu}_n))^2 \, \mathbb{I}_{\{A\}} \right] \leq C_8 \mathbb{E}\left[ d^2(\mu^*, \hat{\mu}_n) \right] \leq C_9 v_n^2. \quad (S21)$$

Combining (S16)–(S19) and (S21) proves the corollary.

It remains to establish the claim (S20). When $d(X, \mu^*) = r_1$, the claim follows from the assumed regularity of the Alexandrov inner product. Now consider the case $d(X, \mu^*) > r_1$. Let $\hat{x}$ be the point on the geodesic $\gamma^X_{\hat{\mu}_n}$ such that $d(\mu^*, \hat{x}) = r_1$ and $x^*$ the point on the geodesic $\gamma^X_{\mu^*}$ such that $d(\mu^*, x^*) = r_1$. We first observe

$$|d(\beta^*, \mu^*) \cos(\angle_{\mu^*}(X, \beta^*)) - d(\beta^*, \hat{\mu}_n) \cos(\angle_{\hat{\mu}_n}(X, \beta^*))| \, \mathbb{I}_{\{A\}}$$

$$= |d(\beta^*, \mu^*) \cos(\angle_{\mu^*}(x^*, \beta^*)) - d(\beta^*, \hat{\mu}_n) \cos(\angle_{\hat{\mu}_n}(\hat{x}, \beta^*))| \, \mathbb{I}_{\{A\}}$$

$$\leq |d(\beta^*, \mu^*) \cos(\angle_{\mu^*}(x^*, \beta^*)) - d(\beta^*, \mu^*) \cos(\angle_{\mu^*}(\hat{x}, \beta^*))| \, \mathbb{I}_{\{A\}} \quad (S22)$$

$$+ |d(\beta^*, \mu^*) \cos(\angle_{\mu^*}(\hat{x}, \beta^*)) - d(\beta^*, \hat{\mu}_n) \cos(\angle_{\hat{\mu}_n}(\hat{x}, \beta^*))| \, \mathbb{I}_{\{A\}}. \quad (S23)$$

The term of (S23) is bounded by $C_{h1} d(\hat{\mu}_n, \mu^*)$ due to the regularity near $\mu^*$ and $d(\mu^*, \hat{x}) = r_1$. For the other term, we observe that $d(\beta^*, \mu^*)$ is a constant and

$$|\cos(\angle_{\mu^*}(x^*, \beta^*)) - \cos(\angle_{\mu^*}(\hat{x}, \beta^*))| \leq |\angle_{\mu^*}(x^*, \beta^*) - \angle_{\mu^*}(\hat{x}, \beta^*)|$$

$$\leq \angle_{\mu^*}(x^*, \hat{x}) \leq \angle_{\mu^*}^{(\kappa)}(x^*, \hat{x})$$

$$\leq cd(x^*, \hat{x})$$

for a constant $c > 0$, with $\angle^{(\kappa)}$ denoting the Alexandrov angle in the model space $M_\kappa^2$, where the first inequality is due to the Lipschiz continuity of the cos function, the second is based on the triangle inequality of $\angle_{\mu^*}$, the third is due to the upper bound on the curvature and Proposition 1.7(4) of Bridson and Häfliger (1999), and the last is due to the smoothness of the model space $M_\kappa^2$ and $d(\mu^*, x^*) = d(\mu^*, \hat{x}) = r_1$. Lemma S3, with setting $u = x^*$, $v = \hat{x}$, $x = X$, $p = \mu^*$ and $q = \hat{\mu}_n$ therein, shows that $d(x^*, \hat{x}) \leq cd(\mu^*, \hat{\mu}_n)$ for a constant $c > 0$. This completes the proof of (S20). □

*Proof of Theorem 5.1.* According to Corollary 6.2 in Devroye et al. (1996), we can upper bound the excess risk via the difference between the corresponding success probabilities

$$\mathcal{E}(s^*, \hat{s}) \leq 2\sqrt{\mathbb{E}(P_X - \hat{P}_X)^2}.$$

Applying Corollary S1 completes the proof. □

*Proof of Theorem 5.2.* We first verify Assumptions 4.1–4.4.

- Assumption 4.1 holds for the assumed manifolds.

- Assumption 4.2 holds with $\alpha_U = 1$ due to Propositions S2 and S3.

- For Hadamard manifolds, the empirical Fréchet mean always exists, and the population Fréchet mean exists when $\mathbb{E}d^2(u, X) < \infty$ for some $u \in \mathcal{M}$. Therefore, with the assumed bounded support of $X$ (so that $\mathbb{E}d^2(u, X) < \infty$ for all $u$) and existence of $\hat{\beta}_n$, Assumption 4.3 holds for Hadamard manifolds. Now consider the other type of the manifold in the theorem. We first note that it is bounded and thus $X$ has a bounded

27

support. By the assumed condition $\inf_{p,q \in \mathcal{M}} \Lambda_{\min}(\text{Hess}_q d_p^2) > 0$ and Lemma S8, the function $d_p^2$ is strongly convex[2] with a common parameter $\eta > 0$ for all $p \in \mathcal{M}$. Then both $\mu^*$ and $\hat{\mu}_n$ exist according to Proposition 1.7 of Sturm (2003). Also, $\hat{\beta}_n$ exists according to Proposition S4 or Corollary S2. Consequently, Assumption 4.3 holds.

- Assumption 4.4 holds with $\alpha_L = 2$ since the smallest eigenvalue $\lambda_X$ of $\mathbb{E}[\text{Hess}_{\beta^*} h]$ is positive.

Also, the regularity of $h(\beta^*; x, \mu)$ near $\mu^*$ is satisfieid by Riemannian manifolds.

The assumption that $\mathbb{E}[d^2(\mu^*, \hat{\mu}_n)] \lesssim D/n$ holds for Hadamard manifolds, by applying Corollary 4 in Schötz (2019) with a bounded support of $X$. To establish $\mathbb{E}[d^2(\mu^*, \hat{\mu}_n)] \lesssim D/n$ for the other type of manifolds in the theorem, noting that positively curved geodesic space is always bounded (Theorem 1.9, Petrunin and Tuschmann, 1999) and the quadruple inequality holds for every bounded geodesic space, it is sufficient to check the Growth condition in Schötz (2019). That is to show, there is a constant $C_g > 0$, such that $\mathbb{E}[d^2(X, u)] - \mathbb{E}[d^2(X, \mu^*)] \geq C_g d^2(u, \mu^*)$ for all $u \in \mathcal{M}$. Let $f_X(\cdot) = d^2(\cdot, X)$ and $\ell = d(u, \mu^*)$. For a unit-speed geodesic $\gamma(t)$ with $\gamma(0) = \mu^*$ and $\gamma(\ell) = u$, we have

$$f_X(u) - f_X(\mu^*) = \int_0^\ell \langle \nabla f_X(\gamma(t)), \gamma'(t) \rangle_{\gamma(t)} dt.$$

By Taylor series expansion of $\nabla f_X(\gamma(t))$ at $\gamma(0) = \mu^*$ (similar to the first-order Taylor series expansion with the mean value theorem in the proof of Theorem 4 in Kendall and Le, 2011), we further have, for some constant $C > 0$,

$$f_X(u) - f_X(\mu^*) \geq \int_0^\ell \langle \tau_{\mu^*}^{\gamma(t)} \nabla f_X(\mu^*), \gamma'(t) \rangle_{\gamma(t)} dt + C d^2(u, \mu^*) \inf_{p,q \in \mathcal{M}} \Lambda_{\min}(\text{Hess}_q d_p^2),$$

---

[2] A function $f$ defined on $(\mathcal{M}, d)$ is strongly convex with parameter $\eta > 0$ if $f(\gamma_p^q(t)) \leq (1-t)f(p) + tf(q) - \eta t(1-t) d^2(p, q)$ for all $p, q \in \mathcal{M}$ and $t \in [0, 1]$, where $\gamma_p^q$ denotes the geodesic such that $\gamma(0) = p$ and $\gamma(1) = q$.

28

where $\tau_{\mu^*}^{\gamma(t)}$ denotes the parallel transport along the geodesic $\gamma(t)$ on the Riemannian manifold. Taking expectation of both sides of the above inequality and noting that $\nabla f_X(u) = -2\text{Log}_u X$ (Section 2 of Supplementary A, Pennec, 2006) for $u \in \mathcal{M}$ and $\mathbb{E}(\text{Log}_{\mu^*} X) = 0$, we verify the Growth condition since $\inf_{p,q\in\mathcal{M}} \Lambda_{\min}(\text{Hess}_q d_p^2) > 0$ is assumed.

Assumption 4.5(a) is satisfied for a $D$-dimensional Hadamard manifold of bounded sectional curvature in light of Lemma 4 of Sun et al. (2019). For a $D$-dimensional Riemannian manifold $\mathcal{M}$ with non-negative sectional curvature, denote $\mathcal{B} = B(\beta^*, \delta)$ for a given $\delta > 0$, and let $\|\cdot\|_q$ denote the norm on the tangent space $T_q\mathcal{B}$ for a fixed $q \in \mathcal{B}$. According to Proposition A.14 in Ahidar-Coutrix et al. (2020), with $\phi(y) = \text{Log}_q y$, where $\text{Log}_q$ is the Riemannian logarithmic map at $q$, we have $d(y_1, y_2) \leq \|\phi(y_1) - \phi(y_2)\|_q$. Consequently, since the metric space $(T_q\mathcal{B}, \|\cdot\|_q)$, which is isometric to $\mathbb{R}^D$, satisfies Assumption 4.5(a), the manifold $\mathcal{B}$, as a metric space, also satisfies the same assumption.

We have showed that all conditions of Corollary S1 are satisfied and Assumption 4.5(a) holds. We can now apply the corollary to conclude $\mathcal{E}(s^*, \hat{s}) \leq c\sqrt{D/n}$. □

*Proof of Theorem 5.3.* For any sufficiently small $t \geq 0$ to be chosen latter, given any $b \in H^D := \{0,1\}^D$, define $\eta_b = (\eta_{b1}, \ldots, \eta_{bD})$ by $\eta_{bj} = (1+t)/2$ if $b_j = 1$ and $\eta_{bj} = (1-t)/2$ if $b_j = 0$, for each $1 \leq j \leq D$.

By Lemma S7, for any $b \in H^D$, in a neighborhood $U$ of $\mu^*$, there exist a $\beta_b \in U$ and a set $\mathcal{X} = \{x_1, \ldots, x_D\} \subset U$ of $D$ points such that, for every $1 \leq j \leq D$,

$$|h(\beta_b; x_j, \mu^*)| = |\langle \text{Log}_{\mu^*}\beta_b, \text{Log}_{\mu^*}x_j\rangle_{\mu^*}| = \log\left(\frac{1+t}{1-t}\right) = \left|\log\left(\frac{\eta_{bj}}{1-\eta_{bj}}\right)\right|, \quad (S24)$$

where $\text{Log}_{\mu^*}$ is the Riemannian logarithmic map at $\mu^*$. In particular, $x_1, \ldots, x_D$ are chosen such that $h(\beta_b; x_j, \mu^*) = \log(\frac{\eta_{bj}}{1-\eta_{bj}})$ for $j = 1, \ldots, D$.

Let $\pi$ be the uniform probability distribution on $\mathcal{X}$, that is, $\pi(x_j) = 1/D$ for every $1 \leq j \leq D$. Let $(X, Y)$ be a random pair drawn from $\mathcal{X} \times \{0, 1\}$, and define $\mathbb{P}_b$ as the joint

29

distribution on $\mathcal{X} \times \{0, 1\}$ such that, under $\mathbb{P}_b$, $X$ has distribution $\pi$ and $Y$ given $X = x_j$ has a Bernoulli distribution with probability of success $\eta_{bj}$ for every $1 \leq j \leq D$. With this construction, $\mathbb{P}(Y = 1|X = x_j) = \eta_{bj} = 1/(1 + e^{-h(\beta_b; x_j, \mu^*)})$. That is, $(X, Y)$ follows the model (6) with true parameter $\beta_b$. Denote the corresponding Bayes classifier by $s_b^*$. Noting that $b_j = \mathbb{I}_{\{h(\beta_b; x_j, \mu^*) \geq 0\}}$ for every $1 \leq j \leq p$, we have $s_b^*(x_j) = b_j$ for every $1 \leq j \leq D$. Moreover, for any classifier $s : \mathcal{X} \to \{0, 1\}$,

$$\mathcal{E}(s_b^*, s) = \mathbb{E}_\pi[|2\mathbb{P}(Y = 1|X) - 1| \cdot |s_b^*(X) - s(X)|] \geq t\|s_b^* - s\|_1,$$

where $\mathbb{E}_\pi$ stands for the expectation with respect to $\pi$, and $\|\cdot\|_1$ denotes the $L_1(\pi)$-norm. Consequently,

$$\inf_{\hat{s}} \sup_{s^* \in \mathcal{C}} \mathcal{E}(s^*, \hat{s}) \geq \inf_{\hat{s}} \sup_{b \in H^D} \mathcal{E}(s_b^*, \hat{s}) \geq \inf_{\hat{s}} \sup_{b \in H^D} t \cdot \mathbb{E}_b[\|s_b^* - \hat{s}\|_1], \qquad \text{(S25)}$$

where $\mathbb{E}_b$ stands for the expectation with respect to the distribution induced by $b$.

Given an estimator $\hat{s}$, define $\hat{b} \in H^D$ such that

$$\|s_{\hat{b}}^* - \hat{s}\|_1 = \min_{b' \in H^p} \|s_{b'}^* - \hat{s}\|_1.$$

Then by the triangle inequality,

$$\|s_{\hat{b}}^* - s_b^*\|_1 \leq \|s_b^* - \hat{s}\|_1 + \|s_{\hat{b}}^* - \hat{s}\|_1 \leq 2\|s_b^* - \hat{s}\|_1,$$

which provides a further lower bound for (S25),

$$\inf_{\hat{s}} \sup_{s^* \in \mathcal{C}} \mathcal{E}(s^*, \hat{s}) \geq \frac{t}{2} \inf_{\hat{b} \in H^D} \sup_{b \in H^D} \mathbb{E}_b[\|s_{\hat{b}}^* - s_b^*\|_1].$$

According to Lemma 7 in Massart and Nédélec (2006), for every pair $b, b' \in H^D$, we have

$$\mathcal{H}^2(\mathbb{P}_b, \mathbb{P}_{b'}) = (1 - \sqrt{1 - t^2})\|s_b^* - s_{b'}^*\|_1 = D^{-1}(1 - \sqrt{1 - t^2})\left(\sum_{j=1}^{D} \mathbb{I}_{\{b_j \neq b_j'\}}\right),$$

where $\mathcal{H}^2(\cdot,\cdot)$ is the squared Hellinger distance between two distributions. Hence, applying the form of Assouad's lemma in Barron et al. (1999) (Lemma 7 therein), one has

$$\inf_{\hat{s}} \sup_{s^* \in \mathcal{C}} \mathcal{E}(s^*, \hat{s}) \geq \frac{D}{4} D^{-1} t \left[1 - \sqrt{2n\theta}\right],$$

where $\theta = D^{-1}(1 - \sqrt{1-t^2}) \leq t^2/D$. Finally, choosing $t = \sqrt{D/(4n)}$ gives the desired result. □

## H  Further Remarks on Assumptions

### H.1  Metric Spaces Satisfying Assumption 4.2

As shown in Lin and Müller (2021), Assumption 4.2 is satisfied by Hadamard spaces.

**Proposition S2** (Lemma S.1, Lin and Müller (2021)). *For a Hadamard space $\mathcal{M}$, for all $p, q, r, u \in \mathcal{M}$,*

$$|d(p,q)\cos(\angle_p(u,q)) - d(p,r)\cos(\angle_p(u,r))| \leq 5d(q,r).$$

It turns out that the above Lipschitz continuity holds for some subspaces of positively curved Alexandrov spaces. Recall that $M_\kappa^2$ denotes a reference space with curvature $\kappa$ with the distance function $d_\kappa$. For $p, q, r \in \mathcal{M}$, $\triangle(p,q,r)$ denotes the geodesic triangle formed by $p, q, r$ and $\triangle(\bar{p}, \bar{q}, \bar{r})$ is the corresponding comparison triangle in $M_\kappa^2$.

**Proposition S3.** *Suppose $(\mathcal{M}, d)$ is an Alexandrov space with a curvature upper bounded by $\kappa > 0$ and all geodesic triangles in $\mathcal{M}$ has perimeter no larger than $2D_\kappa$ with $D_\kappa = \pi/\sqrt{\kappa}$. For all $p, q, r, u \in \mathcal{M}$, if $d(p,q) \leq D_\kappa/2$, then we have*

$$|d(p,q)\cos(\angle_p(u,q)) - d(p,r)\cos(\angle_p(u,r))| \leq 5d(q,r).$$

*Proof.* For the case $d(q,r) > d(p,q)/2$, we have

$$|d(p,q)\cos(\angle_p(u,q)) - d(p,r)\cos(\angle_p(u,r))|$$
$$\leq d(p,q)|\cos(\angle_p(u,q))| + d(p,r)|\cos(\angle_p(u,r))|$$
$$\leq d(p,q) + d(p,r) \leq d(p,q) + d(p,q) + d(q,r)$$
$$\leq 5d(q,r). \tag{S26}$$

For the other case $d(q,r) \leq d(p,q)/2$, we first have

$$|d(p,q)\cos(\angle_p(u,q)) - d(p,r)\cos(\angle_p(u,r))| \tag{S27}$$
$$\leq |d(p,q)\cos(\angle_p(u,q)) - d(p,q)\cos(\angle_p(u,r))|$$
$$+ |d(p,q)\cos(\angle_p(u,r)) - d(p,r)\cos(\angle_p(u,r))|,$$

where the second term is bounded by

$$|d(p,q)\cos(\angle_p(u,r)) - d(p,r)\cos(\angle_p(u,r))| \leq |d(p,q) - d(p,r)| \leq d(q,r). \tag{S28}$$

For the first term,

$$|d(p,q)\cos(\angle_p(u,q)) - d(p,q)\cos(\angle_p(u,r))|$$
$$= d(p,q)|\cos(\angle_p(u,q)) - \cos(\angle_p(u,r))|$$
$$\leq d(p,q)|\angle_p(u,q) - \angle_p(u,r)|$$
$$\leq d(p,q)\angle_p(q,r),$$

where the first inequality is due to the Lipschitz continuity of $\cos(\cdot)$ and the last is due to the fact that the Alexandrov angle $\angle_p(\cdot,\cdot)$ defines a metric on the set of (germs of) geodesics emanating from $p$ (Holopainen, 2009, Theorem 2.17) and thus satisfies the triangle inequality. According to Proposition 1.7 of Bridson and Häfliger (1999), $\angle_p(q,r) \leq \angle_{\bar{p}}(\bar{q},\bar{r})$, where $\triangle(\bar{p},\bar{q},\bar{r})$ is the comparison triangle of $\triangle(p,q,r)$ in the reference space $M_\kappa^2$. For

an Alexandrov space with a curvature upper bounded by $\kappa > 0$, the reference space $M_\kappa^2$ is the sphere $\mathbb{S}^2 = \{(x, y, z) \in \mathbb{R}^3 : x^2 + y^2 + z^2 = 1\}$ with the angular distance function $d_\kappa(\bar{p}, \bar{q}) = \cos^{-1}(x_{\bar{p}} x_{\bar{q}} + y_{\bar{p}} y_{\bar{q}} + z_{\bar{p}} z_{\bar{q}})/\sqrt{\kappa}$. Since $d_\kappa(\bar{q}, \bar{r}) = d(q, r) \leq d(p, q)/2 = d_\kappa(\bar{p}, \bar{q})/2$, with Lemma S1, we have

$$d(p,q) \angle_p(q,r) \leq d_\kappa(\bar{p}, \bar{q}) \angle_{\bar{p}}(\bar{q}, \bar{r}) < 3 d_\kappa(\bar{q}, \bar{r}) = 3 d(q, r).$$

This combined with (S26)–(S28) completes the proof. □

## H.2 Existence of $\hat{\beta}_n$

**The case of bounded spaces.** The following proposition shows that $\hat{\beta}_n$ exists for totally bounded complete spaces satisfying Assumption 4.2, regardless of the distribution of $(X, Y)$. This contrasts with the case of unbounded spaces, where the existence of $\hat{\beta}_n$ requires additional distributional conditions on $(X, Y)$, even in the special case of Euclidean space (e.g., Albert and Anderson, 1984).

**Proposition S4.** *If $\mathcal{M}$ is complete and totally bounded, then $\hat{\beta}_n$ in (9) exists under Assumption 4.2.*

Since a (finite-dimensional) Riemannian manifold of positive curvature is bounded (and thus totally bounded) according to Theorem 1.9 of Petrunin and Tuschmann (1999), we have the following corollary.

**Corollary S2.** *If $\mathcal{M}$ is a complete Riemannian manifold of positive curvature, then $\hat{\beta}_n$ in (9) exists.*

*Proof of Proposition S4.* The Hölder continuity from Assumption 4.2 ensures that the function $h$ defined in (6), as a function of $\beta$ on $\mathcal{M}$, is continuous in $\beta$. Below we establish continuity of $L_n$.

To ease the notation, we write $h(\beta) = h(\beta; X, \mu^*)$. Define $M(\theta) = M(\theta; \theta^*)$, where $M(\theta; \theta^*)$ with $\theta^* = h(\beta^*)$ is defined in (S4). Then

$$|M'(\theta)| = \left|\frac{1}{1+e^{-\theta^*}} - \frac{1}{1+e^{-\theta}}\right| \leq 2.$$

With $\mathbb{E}d(\mu^*, X) < \infty$ and Assumption 4.2, we further deduce that

$$|L_n(\beta_1) - L_n(\beta_2)| = \left|\mathbb{E}_n[M(h(\beta_1)) - M(h(\beta_2))]\right|$$

$$\leq \sup_{\theta} |M'(\theta)| \cdot \mathbb{E}_n|h(\beta_1) - h(\beta_2)|$$

$$\leq 2C_U \mathbb{E}_n[d(\mu^*, X)d^{\alpha_U}(\beta_1, \beta_2)]$$

$$\leq C d^{\alpha_U}(\beta_1, \beta_2)$$

for some constant $C > 0$, where $\mathbb{E}_n$ represents the empirical average. This shows that the function $L_n$ in (9) is continuous in $\beta$. Finally, since $\mathcal{M}$ is totally bounded and complete and thus is compact, the extreme value theorem for topological spaces ensures existence of the maximizers $\beta^*$ and $\hat{\beta}_n$ in (8) and (9), respectively. $\square$

**The case of unbounded spaces.** For unbounded spaces, existence of $\hat{\beta}_n$ requires additional distributional conditions on $(X, Y)$, even in the special case of Euclidean space, such as the overlap condition (Albert and Anderson, 1984). To illustrate this type of distributional conditions outside of Euclidean realms, below we construct a condition in analogy to the overlap condition (Albert and Anderson, 1984) through the example of the SPD manifold $\mathcal{S}_m^+$ equipped with the Log-Euclidean metric. We first note that, in Euclidean spaces, the MLE of logistic regression exists if the data points $(X_i, Y_i) \in \{0,1\} \times \mathbb{R}^D$ satisfy the overlap condition (Albert and Anderson, 1984; Sur and Candès, 2019), in the sense that, there is at least one data point classified correctly (i.e., $(2Y_i - 1)X_i^\top b > 0$) and at least another one classified incorrectly (i.e., $(2Y_i - 1)X_i^\top b < 0$). For $\mathcal{S}_m^+$, given $X_i, \hat{\mu}_n, \beta \in \mathcal{S}_m^+$, let

34

$h_i(\beta) = \operatorname{tr}\left([\log(X_i) - \log(\hat{\mu}_n)]^\top [\log(\beta) - \log(\hat{\mu}_n)]\right)$, where $\log(A)$ is the matrix logarithm of a matrix $A$. Note that the tangent space $T_\mu \mathcal{S}_m^+ = \mathcal{S}_m$, a space consisting of symmetric matrices, is isometric to a Euclidean space. Therefore, the proposed MLE with covariates $X_i \in \mathcal{S}_m^+$ exists if the overlap condition holds, in the sense that, for every $\beta \neq \hat{\mu}_n$, there is at least one data point classified correctly (i.e., $(2Y_i - 1)h_i(\beta) > 0$) and at least another one classified incorrectly (i.e., $(2Y_i - 1)h_i(\beta) < 0$).

## H.3 Distributional Examples Satisfying Assumption 4.4

In Section 4, we mentioned that Assumption 4.4 is reduced to the basic condition that the covariance matrix of $X$ is positive-definite when $\mathcal{M}$ is Euclidean. This clearly indicates that Assumption 4.4 relates to the distribution of $X$. Hence, it is generally not possible to specify which metric spaces meet this assumption without considering the underlying distributions. Below we present two examples to illustrate Assumption 4.4 outside of Euclidean realms.

Let $\mathcal{M}$ be either (1) the unit $(D-1)$-sphere $\mathbb{S}^{D-1} \subset \mathbb{R}^D$ equipped with the Euclidean inner product as the Riemannian metric, or (2) the space $\mathcal{S}_m^+$ of $m \times m$ symmetric positive-definite matrices endowed with the Log-Euclidean metric (Arsigny et al., 2007) for any fixed positive integer $m$, where $\mathbb{S}^{D-1}$ is bounded while $\mathcal{S}_m^+$ is unbounded. Let $\operatorname{Log}_{\mu^*}$ be the Riemannian logarithmic map at $\mu^*$. As in Section 5 of Pennec (2006), we define the covariance of $X$ on $\mathcal{M}$ as the linear operator $\Sigma : T_{\mu^*}\mathcal{M} \to T_{\mu^*}\mathcal{M}$ by $\Sigma v = \mathbb{E}\{\langle \operatorname{Log}_{\mu^*} X, v \rangle \operatorname{Log}_{\mu^*} X\}$ for $v \in T_{\mu^*}\mathcal{M}$. Consider any distribution of $X$, e.g., the normal law with a full-rank concentration matrix on Riemannian manifolds (Section B, Pennec, 2006), such that the smallest eigenvalue $\lambda_X$ of $\Sigma$ is positive. Then Assumption 4.4 holds since

$$\mathbb{E}\left[|h(\beta; X, \mu^*) - h(\beta^*; X, \mu^*)|^2\right]$$
$$= \mathbb{E}[\langle \operatorname{Log}_{\mu^*} \beta - \operatorname{Log}_{\mu^*} \beta^*, \operatorname{Log}_{\mu^*} X \rangle^2]$$

$$= \langle \Sigma(\mathrm{Log}_{\mu^*}\beta - \mathrm{Log}_{\mu^*}\beta^*), (\mathrm{Log}_{\mu^*}\beta - \mathrm{Log}_{\mu^*}\beta^*)\rangle$$

$$\geq \lambda_X \|\mathrm{Log}_{\mu^*}\beta - \mathrm{Log}_{\mu^*}\beta^*\|_2^2 \geq \lambda_X d^2(\beta, \beta^*),$$

where the last inequality is due to the Rauch Comparison Theorem and the curvature of $\mathbb{S}^{D-1}$ and $\mathcal{S}_m^+$.

## H.4  Uniqueness of $\beta^*$

In this subsection, we first prove uniqueness of $\beta^*$ under Assumptions 4.3 and 4.4. Then, we provide two concrete examples.

**Proposition S5.** *If $X$ has a bounded support and $\mu^*$ exists and is unique, then $\beta^*$ is unique under Assumption 4.4.*

*Proof.* We adopt the strategy of proof by contradiciton, assuming $\beta$ is another maximizer. Consider the function $M(\theta)$ in the proof of Proposition S4. We first observe that $M''(\theta) < 0$ for all $\theta$ and is continuous. Taylor expansion asserts that

$$M(\theta^*) - M(\theta) = -M''(\bar{\theta})(\theta - \theta^*)^2, \tag{S29}$$

where $\bar{\theta}$ lies between $\theta$ and $\theta^*$. Now set $\theta = h(\beta; X, \mu^*)$ and $\theta^* = h(\beta^*; X, \mu^*)$ in (S29), and write

$$a = \min\left\{\inf_{x\in\mathrm{supp}(X)} h(\beta; x, \mu^*), \inf_{x\in\mathrm{supp}(X)} h(\beta^*; x, \mu^*)\right\}$$

and

$$b = \max\left\{\sup_{x\in\mathrm{supp}(X)} h(\beta; x, \mu^*), \sup_{x\in\mathrm{supp}(X)} h(\beta^*; x, \mu^*)\right\},$$

where $\mathrm{supp}(X)$ denotes the support of $X$. The bounded support of $X$ implies that both $a$ and $b$ are finite. With Assumption 4.4 and uniqueness of $\mu^*$, we deduce

$$L(\beta^*) - L(\beta) = \mathbb{E}\{M(h(\beta^*; X, \mu^*)) - M(h(\beta; X, \mu^*))\}$$

36

$$\geq \inf_{\bar{\theta}\in[a,b]}(-M''(\bar{\theta}))d^{\alpha_L}(\beta,\beta^*) > 0,$$

since the continuity and negativity of $M''$ imply

$$\inf_{\bar{\theta}\in[a,b]}(-M''(\bar{\theta})) > 0.$$

This contradicts $L(\beta^*) - L(\beta) = 0$ implied by the assumed maximizer $\beta$. □

The sphere example in H.3 also serves as an example for uniqueness of $\beta^*$, since it satisfies the assumptions of Proposition S5. Below we provide an example where the underlying space $\mathcal{M}$ is unbounded, more specifically, the SPD manifold $\mathcal{S}_m^+$ equipped with the Log-Euclidean metric (Arsigny et al., 2007).

With $h(\beta) = h(\beta; X, \mu^*)$, the population log-likelihood function is

$$L(\beta) = \mathbb{E}\left[Yh(\beta) - \log(1+e^{h(\beta)})\right] = \mathbb{E}_X\left[\frac{h(\beta)}{1+e^{-h(\beta^*)}} - \log(1+e^{h(\beta)})\right],$$

where the expectation $\mathbb{E}_X$ is taken over $X$. Let $M(\theta) = \frac{\theta}{1+e^{-\theta^*}} - \log(1+e^\theta)$ with $\theta = h(\beta)$ and $\theta^* = h(\beta^*)$, and observe $h(\beta) = \langle \text{Log}_{\mu^*}X, \text{Log}_{\mu^*}\beta\rangle_{\mu^*}$, where $\text{Log}_{\mu^*}$ and $\langle\cdot,\cdot\rangle_{\mu^*}$ are the Riemannian logarithmic map and the Riemannian metric at $\mu^*$ on $\mathcal{S}_m^+$, respectively. It is sufficient to show, for any geodesic $\gamma(t)$ in $\mathcal{S}_m^+$, $g(t) = M(h(\gamma(t)))$ is strictly concave in $t$, according to Proposition 1.7 in Sturm (2003).

For $\mathcal{S}_m^+$ equipped with the Log-Euclidean metric, by Corollary 3.9 in Arsigny et al. (2007),

$$h(\beta) = \langle \text{Log}_{\mu^*}X, \text{Log}_{\mu^*}\beta\rangle_{\mu^*} = \langle \log(X) - \log(\mu^*), \log(\beta) - \log(\mu^*)\rangle_{\text{Fr}}$$

$$= \text{tr}\left([\log(X) - \log(\mu^*)]^\top[\log(\beta) - \log(\mu^*)]\right),$$

where $\log(A)$ is the matrix logarithm of a matrix $A$. The geodesic connecting two matrices $S_1, S_2 \in \mathcal{S}_m^+$ is characterized as $\gamma(t) = \exp((1-t)\log(S_1) + t\log(S_2))$ for $t \in [0,1]$, where

37

$\exp(A)$ is the matrix exponential of a matrix $A$. Hence,

$$h(\gamma(t)) = \text{tr}([\log(X) - \log(\mu^*)]^\top [(1-t)\log(S_1) + t\log(S_2) - \log(\mu^*)]),$$

and $\frac{\partial h(\gamma(t))}{\partial t} = \text{tr}([\log(S_2) - \log(S_1)][\log(X) - \log(\mu^*)])$. Writing $W = [\log(S_2) - \log(S_1)][\log(X) - \log(\mu^*)]$, for $g(t) = M(h(\gamma(t)))$, one could find its first derivative

$$g'(t) = \text{tr}\left(\frac{\partial M}{\partial h} \frac{\partial h(\gamma(t))}{\partial t}\right) = \left(\frac{1}{1 + e^{-h(\beta^*)}} - \frac{e^{h(\gamma(t))}}{e^{h(\gamma(t))} + 1}\right) \text{tr}(W),$$

and its second derivative

$$g''(t) = -\frac{e^{h(\gamma(t))}}{(e^{h(\gamma(t))} + 1)^2} \text{tr}^2(W) < 0.$$

This shows $g(t)$ is strictly concave, as desired.

# I   Technical Lemmas

**Lemma S1.** *Let* $\mathbb{S}^2 = \{(x, y, z) \in \mathbb{R}^3 : x^2 + y^2 + z^2 = 1\}$ *be the sphere equipped with the angular distance function* $d(p, q) = \cos^{-1}(x_p x_q + y_p y_q + z_p z_q)/\sqrt{\kappa}$ *for* $\kappa > 0$. *For any* $p, q, r \in \mathbb{S}^2$, *if* $d(q, r) \leq d(p, q)/2$ *and* $d(p, q) \in [0, \pi/(2\sqrt{\kappa})]$, *then one must have*

$$d(p, q) \angle_p(q, r) < 3 d(q, r).$$

*Proof.* For a spherical triangle $\triangle(p, q, r)$ in $\mathbb{S}^2$, its perimeter is no larger than $2D_\kappa$ with $D_\kappa = \pi/\sqrt{\kappa}$, hence the length of any side is no larger than $D_\kappa$. By the spherical law of sines, $\sin(\angle_p(q, r))/\sin(\sqrt{\kappa}d(q, r)) = \sin(\angle_r(p, q))/\sin(\sqrt{\kappa}d(p, q))$. Hence,

$$\begin{aligned}
\sin(\angle_p(q, r)) &= \frac{\sin(\sqrt{\kappa}d(q, r))}{\sin(\sqrt{\kappa}d(p, q))} \sin(\angle_r(p, q)) \\
&\leq \frac{\sin(\sqrt{\kappa}d(q, r))}{\sin(\sqrt{\kappa}d(p, q))} \leq \frac{\sqrt{\kappa}d(q, r)}{\sin(\sqrt{\kappa}d(p, q))} \leq \frac{\pi}{2} \frac{\sqrt{\kappa}d(q, r)}{\sqrt{\kappa}d(p, q)} \leq \frac{\pi}{4} < 1,
\end{aligned}$$

where the second inequality is due to $\sin(x) \leq x$ for $x \geq 0$, and the third inequality is due to the convexity of $\sin(x)$ on $[0, \pi/2]$, as

$$\sin(x) \geq \sin(0) + (x-0)\frac{\sin(\pi/2) - \sin(0)}{\pi/2 - 0} = \frac{2}{\pi}x,$$

for $x \in [0, \pi/2]$.

We claim $\angle_p(q,r) \in [0, \pi/2]$ and defer its proof to the end. Then by the monotonicity of $\arcsin(\cdot)$ and the Taylor expansion of $\arcsin(\cdot)$ at $0$, we have

$$\angle_p(q,r) \leq \arcsin\left(\frac{\pi}{2}\frac{d(q,r)}{d(p,q)}\right) = \frac{1}{\sqrt{1-\theta^2}}\frac{\pi}{2}\frac{d(q,r)}{d(p,q)},$$

where $\theta \in [0, \frac{\pi}{2}\frac{d(q,r)}{d(p,q)}] \subset [0, \pi/4]$. Thus,

$$d(p,q)\angle_p(q,r) \leq \frac{1}{\sqrt{1-(\pi/4)^2}}\frac{\pi}{2}d(p,q)\frac{d(q,r)}{d(p,q)} < 3d(q,r).$$

It remains to establish the claim $\angle_p(q,r) \in [0, \pi/2]$. Consider the non-trivial case $\sqrt{\kappa}d(p,q) \in (0, \pi/2]$ and $\sqrt{\kappa}d(p,r) \in (0, \pi)$ (the length of any side is no larger than $D_\kappa = \pi/\sqrt{\kappa}$). By the spherical law of cosines,

$$\cos(\angle_p(q,r)) = \frac{\cos(\sqrt{\kappa}d(q,r)) - \cos(\sqrt{\kappa}d(p,q))\cos(\sqrt{\kappa}d(p,r))}{\sin(\sqrt{\kappa}d(p,q))\sin(\sqrt{\kappa}d(p,r))},$$

it is sufficient to show

$$\Delta := \cos(\sqrt{\kappa}d(q,r)) - \cos(\sqrt{\kappa}d(p,q))\cos(\sqrt{\kappa}d(p,r)) \geq 0.$$

It is seen that $\cos(\sqrt{\kappa}d(q,r)) \geq \cos(\sqrt{\kappa}d(p,q)) \geq 0$ due to the assumptions $d(q,r) \leq d(p,q)/2$ and $\sqrt{\kappa}d(p,q) < \pi/2$.

- For $\sqrt{\kappa}d(p,r) \geq \pi/2$, one has $\cos(\sqrt{\kappa}d(p,r)) \leq 0$, which implies $\Delta \geq 0$.

- For $\sqrt{\kappa}d(p,r) < \pi/2$, one has $0 < \cos(\sqrt{\kappa}d(p,r)) < 1$. Thus

  $$0 \leq \cos(\sqrt{\kappa}d(p,q))\cos(\sqrt{\kappa}d(p,r)) \leq \cos(\sqrt{\kappa}d(p,q)) \leq \cos(\sqrt{\kappa}d(q,r)),$$

  which also implies $\Delta \geq 0$.

This establishes $\angle_p(q,r) \in [0, \pi/2]$ and completes the proof. □

**Lemma S2.** *For points $p, q, x$ in $(M_\kappa^2, d)$, suppose the distances among $p, q, x$ are no larger than $D_\kappa/2$. Then, for $u \in \gamma_p^x$ and $v \in \gamma_q^x$,*

$$d(u,v) \leq 3\sqrt{2}d(p,q) + \sqrt{2}|d(p,u) - d(q,v)|.$$

*Proof.* Note that the conclusion trivially holds when $x = p$ or $x = q$ or $p = q$. Therefore, in the sequel we assume $x \neq p$, $x \neq q$ and $p \neq q$. We first consider the case of $\kappa \geq 0$; the case of $\kappa < 0$ will then follow due to the CAT(0) inequality.

Let $\gamma_p^q(t)$, for $t \in [0, \delta]$ with $\delta = d(p,q) > 0$, be the geodesic connecting $p$ to $q$. Parameterize $\gamma_p^q$ by the polar coordinate at $x$, so that $\gamma_p^q(t) = (\phi(t), \varphi(t))$ for $t \in [0, \delta]$. Let $\theta_1 = d(x,u)/d(x,p) = 1 - d(p,u)/d(x,p) \in [0,1]$ and $\theta_2 = d(x,v)/d(x,q) = 1 - d(q,v)/d(x,q) \in [0,1]$. Consider the curve $\chi(t) = (\theta(t)\phi(t), \varphi(t))$ for $\theta(t) = (1-t/\delta)\theta_1 + t\theta_2/\delta$. Clearly, $\chi$ connects $u$ to $v$, and $d(u,v) \leq \text{length}(\chi)$. To bound the length of $\chi$, we first define $g_\kappa(r) = r$ if $\kappa = 0$ and $g_\kappa(r) = (1/\sqrt{\kappa})\sin(\sqrt{\kappa}r)$ if $\kappa > 0$, for $r \in \mathbb{R}$. When $\kappa = 0$, we have $g_\kappa(\theta(t)\phi(t)) \leq g_\kappa(\phi(t))$ for all $t$ since $\theta(t) \in [0,1]$. When $\kappa > 0$, since $D_\kappa = \pi/\sqrt{\kappa}$, $\sup_t \phi(t) \leq D_\kappa/2$ and $\sup_t \sqrt{\kappa}\theta(t)\phi(t) \leq \sqrt{\kappa}D_\kappa/2 \leq \pi/2$, we still have $g_\kappa(\theta(t)\phi(t)) \leq g_\kappa(\phi(t))$ for all $t$. Based on this observation, together with $\sup_t \phi(t) \leq d(p,q) + \max\{d(x,p), d(x,q)\}$, we deduce that

$$d(u,v) \leq \text{length}(\chi) = \int_0^\delta |\dot\chi|(t)dt$$

$$= \int_0^\delta \sqrt{\{[\theta(t)\phi(t)]'\}^2 + g_\kappa^2(\theta(t)\phi(t))\{\varphi'(t)\}^2}\,dt$$

$$\leq \int_0^\delta \sqrt{2\{\theta'(t)\}^2\phi^2(t) + 2\theta^2(t)\{\phi'(t)\}^2 + g_\kappa^2(\phi(t))\{\varphi'(t)\}^2}\,dt$$

$$\leq \int_0^\delta \sqrt{2\{(\theta_2-\theta_1)/\delta\}^2\phi^2(t) + 2\{\phi'(t)\}^2 + 2g_\kappa^2(\phi(t))\{\varphi'(t)\}^2}\,dt$$

$$\leq \int_0^\delta \sqrt{2\{(\theta_2-\theta_1)/\delta\}^2\phi^2(t)}\,dt + \int_0^\delta \sqrt{2}\sqrt{\{\phi'(t)\}^2 + g_\kappa^2(\phi(t))\{\varphi'(t)\}^2}\,dt$$

40

$$\leq \sqrt{2}|\theta_2 - \theta_1|[d(p,q) + \max\{d(x,p), d(x,q)\}] + \sqrt{2}d(p,q)$$

$$\leq 2\sqrt{2}d(p,q) + \sqrt{2}|\theta_2 - \theta_1|\max\{d(x,p), d(x,q)\}. \tag{S30}$$

Let $r_1 = d(p,u)$ and $r_2 = d(q,v)$. When $d(x,p) > d(x,q)$, we have

$$|\theta_2 - \theta_1| = \left|\frac{r_1}{d(x,p)} - \frac{r_2}{d(x,q)}\right| = \left|\frac{r_1}{d(x,p)} - \frac{r_2}{d(x,p)} + \frac{r_2}{d(x,p)} - \frac{r_2}{d(x,q)}\right|$$

$$\leq \frac{|r_1 - r_2|}{d(x,p)} + \frac{r_2 d(p,q)}{d(x,q)d(x,p)} \leq \frac{|r_1 - r_2|}{d(x,p)} + \frac{d(p,q)}{d(x,p)} = \frac{|d(p,u) - d(q,v)| + d(p,q)}{d(x,p)}$$

and thus

$$|\theta_2 - \theta_1|\max\{d(x,p), d(x,q)\} \leq |d(p,u) - d(q,v)| + d(p,q).$$

When $d(x,p) \leq d(x,q)$, we have

$$|\theta_2 - \theta_1| = \left|\frac{r_1}{d(x,p)} - \frac{r_2}{d(x,q)}\right| = \left|\frac{r_1}{d(x,p)} - \frac{r_1}{d(x,q)} + \frac{r_1}{d(x,q)} - \frac{r_2}{d(x,q)}\right|$$

$$\leq \frac{r_1 d(p,q)}{d(x,q)d(x,p)} + \frac{|r_1 - r_2|}{d(x,q)} \leq \frac{d(p,q)}{d(x,q)} + \frac{|r_1 - r_2|}{d(x,q)} = \frac{|d(p,u) - d(q,v)| + d(p,q)}{d(x,q)}$$

and thus still

$$|\theta_2 - \theta_1|\max\{d(x,p), d(x,q)\} \leq |d(p,u) - d(q,v)| + d(p,q).$$

Together with (S30), this establishes the lemma for $\kappa \geq 0$.

Now consider the case of $\kappa < 0$. Let $d_0$ be the distance function on $M_0^2 = \mathbb{R}^2$ and $\bar{p}, \bar{q}, \bar{u}, \bar{v}$ the respectively comparison points of $p, q, u, v$ on $M_0^2$. Then

$$d(u,v) \leq d_0(\bar{u}, \bar{v}) \leq 3\sqrt{2}d_0(\bar{p}, \bar{q}) + \sqrt{2}|d_0(\bar{p}, \bar{u}) - d_0(\bar{q}, \bar{v})|$$

$$= 3\sqrt{2}d(p,q) + \sqrt{2}|d(p,u) - d(q,v)|,$$

where the first inequality is due to the CAT(0) inequality, the second follows from the lemma in the case of $\kappa \geq 0$, and the last equality is based on $d_0(\bar{p}, \bar{q}) = d(p,q)$, $d_0(\bar{p}, \bar{u}) = d(p,u)$ and $d_0(\bar{q}, \bar{v}) = d(q,v)$. This completes the proof. □

41

**Lemma S3.** *For a complete Alexandrov space with curvature upper bounded by $\kappa$ and of diameter no larger than $D_\kappa/2$, suppose $u \in \gamma_p^x$ and $v \in \gamma_q^x$ such that $d(p,v) = d(p,u)$. Then $d(u,v) \leq 4\sqrt{2}d(p,q)$.*

*Proof.* Let $d_\kappa$ be the distance on the model space $M_\kappa^2$. Let $\triangle(\bar{x},\bar{p},\bar{q})$ be a comparison triangle of $\triangle(x,p,q)$ in $M_\kappa^2$, and $\bar{u},\bar{v}$ the corresponding comparison points of $u$ and $v$. Since $d(p,q) = d_\kappa(\bar{p},\bar{q})$, $d(p,u) = d_\kappa(\bar{p},\bar{u})$ and $d(q,v) = d_\kappa(\bar{q},\bar{v})$, we have

$$|d_\kappa(\bar{p},\bar{u}) - d_\kappa(\bar{q},\bar{v})| = |d(p,u) - d(q,v)| = |d(p,v) - d(q,v)| \leq d(p,q).$$

This implies

$$d(u,v) \leq d_\kappa(\bar{u},\bar{v}) \leq 3\sqrt{2}d_\kappa(\bar{p},\bar{q}) + \sqrt{2}|d_\kappa(\bar{p},\bar{u}) - d_\kappa(\bar{q},\bar{v})| \leq 4\sqrt{2}d(p,q),$$

where the first inequality is due to the CAT($\kappa$) inequality and the second follows from Lemma S2. □

Recall $\mathsf{entr}_n(Q,d)$ is defined in (S1) for a pseudo-metric space $(Q,d)$. Also recall $B(x,\delta) \subset \mathcal{R}$ represents the ball of radius $\delta$ centered at $x$.

**Lemma S4.** *Suppose $(\mathcal{R},d)$ is a pseudo-metric space whose diameter is $R > 0$. Let $\alpha \in (0,1]$.*

*(a) Assume that, for all $0 < r < R$, there exist constants $C, D > 0$ such that*

$$N(r, B(x^*, \delta), d) \leq C\left(\frac{\delta}{r}\right)^D.$$

*Then $\mathsf{entr}_n(B(x^*,\delta), d^\alpha) \leq C\frac{\Gamma(3/2)D^{1/2}}{\alpha^{1/2}}\delta^\alpha$, where $\Gamma(\cdot)$ is the Gamma function.*

*(b) Assume that, for all $0 < r < R$, there exist constants $C, D > 0$ such that*

$$\log N(r, B(x^*, \delta), d) \leq C\left(\frac{\delta}{r}\right)^D.$$

(i) If $D < 2\alpha$ then $\mathsf{entr}_n(B(x^*, \delta), d^\alpha) \leq C\frac{2\alpha}{2\alpha-D}\delta^\alpha$.

(ii) If $D = 2\alpha$ then $\mathsf{entr}_n(B(x^*, \delta), d^\alpha) \leq C\delta^\alpha \log n$ for all sufficiently large $n$.

(iii) If $D > 2\alpha$ then $\mathsf{entr}_n(B(x^*, \delta), d^\alpha) \leq C'\delta^\alpha n^{-\alpha/D+1/2}$ with a constant $C' > 0$ depending only on $D$, $\alpha$ and $C$.

(c) Assume that, for all $0 < r < R$, there exist constants $C, D > 0$ such that

$$N(r, \mathcal{R}, d) \leq C\left(\frac{R}{r}\right)^D.$$

Then $\mathsf{entr}_n(B(x^*, \delta), d^\alpha) \leq C\frac{\alpha}{1-\gamma_0}D^{1/2}R^{\gamma_0}\delta^{\alpha-\gamma_0}$, where $\gamma_0 = W_0(1/2) \approx 0.35173$ and $W_0(\cdot)$ is the Lambert W function.

(d) Assume that, for all $0 < r < R$, there exist constants $C, D > 0$ such that

$$\log N(r, \mathcal{R}, d) \leq C\left(\frac{R}{r}\right)^D.$$

(i) If $D < 2\alpha$ then $\mathsf{entr}_n(B(x^*, \delta), d^\alpha) \leq C\frac{2\alpha}{2\alpha-D}R^{D/2}\delta^{\alpha-D/2}$.

(ii) If $D = 2\alpha$ then $\mathsf{entr}_n(B(x^*, \delta), d^\alpha) \leq CR^\alpha \log n$ for all sufficiently large $n$.

(iii) If $D > 2\alpha$ then $\mathsf{entr}_n(B(x^*, \delta), d^\alpha) \leq C'n^{-\alpha/D+1/2}$ with a constant $C' > 0$ depending only on $D$, $\alpha$, $C$ and $R$.

*Proof.* We provide proofs for (b) and (c); proofs for (a) and (d) are similar. Recall $\mathsf{entr}_n(Q, b) = \inf_{\epsilon>0}\left(\epsilon\sqrt{n} + \int_\epsilon^\infty \sqrt{\log N(r, Q, b)}dr\right)$. Then

$$\mathsf{entr}_n(B(x^*, \delta), d^\alpha) = \inf_{\epsilon>0}\left(\epsilon\sqrt{n} + \int_\epsilon^\infty \sqrt{\log N(r, B(x^*, \delta), d^\alpha)}dr\right)$$
$$\leq \inf_{\epsilon>0}\left(\epsilon\sqrt{n} + \int_\epsilon^\infty \sqrt{\log N\left(r^{\frac{1}{\alpha}}, B(x^*, \delta), d\right)}dr\right)$$
$$= \inf_{\epsilon>0}\left(\epsilon\sqrt{n} + \alpha\int_{\epsilon^{1/\alpha}}^\delta s^{\alpha-1}\sqrt{\log N(s, B(x^*, \delta), d)}ds\right).$$

- Consider the case (b).

43

(i) If $D < 2\alpha$ and thus $-\frac{D}{2} + \alpha > 0$, then

$$\text{entr}_n(B(x^*, \delta), d^\alpha) \leq C\alpha \int_0^\delta s^{\alpha-1} \left(\frac{\delta}{s}\right)^{D/2} ds$$

$$= C\alpha \delta^{\frac{D}{2}} \int_0^\delta s^{-\frac{D}{2}+\alpha-1} ds$$

$$= C\alpha \delta^{\frac{D}{2}} \frac{\delta^{-\frac{D}{2}+\alpha}}{\alpha - \frac{D}{2}} = C \frac{2\alpha}{2\alpha - D} \delta^\alpha.$$

(ii) If $D = 2\alpha$ and thus $-\frac{D}{2} + \alpha = 0$, then

$$\text{entr}_n(B(x^*, \delta), d^\alpha) \leq \inf_{\epsilon > 0} \left(\epsilon \sqrt{n} + C\alpha \delta^{\frac{D}{2}} \int_{\epsilon^{1/\alpha}}^\delta s^{-1} ds\right)$$

$$= \inf_{\epsilon > 0} \left(\epsilon \sqrt{n} + C\alpha \delta^{\frac{D}{2}} \left(\log \delta - \alpha^{-1} \log \epsilon\right)\right).$$

Let $f(\epsilon) = \sqrt{n}\epsilon - c \log \epsilon$ with $c = C\delta^{\frac{D}{2}} = C\delta^\alpha$. Its second derivative $f''(\epsilon) = c/\epsilon^2$ is always positive. Therefore, by solving the equation $f'(\epsilon) = \sqrt{n} - c/\epsilon = 0$, we find the minimum of $f(\epsilon)$ at $\epsilon_0 = c/\sqrt{n}$. Consequently,

$$\text{entr}_n(B(x^*, \delta), d^\alpha) \leq c\left(\log \delta + 1 - \log c + \log \sqrt{n}\right) \leq c \log n = C\delta^\alpha \log n,$$

where the second inequality holds for all sufficiently large $n$.

(iii) If $D > 2\alpha$ and thus $\frac{D}{2} - \alpha > 0$, then

$$\text{entr}_n(B(x^*, \delta), d^\alpha) \leq \inf_{\epsilon > 0} \left(\epsilon \sqrt{n} + C\alpha \delta^{\frac{D}{2}} \int_{\epsilon^{1/\alpha}}^\delta s^{-\frac{D}{2}+\alpha-1} ds\right)$$

$$= \inf_{\epsilon > 0} \left(C \frac{\alpha \delta^\alpha}{\alpha - \frac{D}{2}} + \epsilon \sqrt{n} + C \frac{\delta^{\frac{D}{2}}}{\frac{D}{2\alpha} - 1} \epsilon^{-\frac{D}{2\alpha}+1}\right).$$

Let $f(\epsilon) = \sqrt{n}\epsilon + c\epsilon^{-\frac{D}{2\alpha}+1}$ with $c = C\delta^{\frac{D}{2}}/(\frac{D}{2\alpha} - 1)$. Its second derivative $f''(\epsilon) = C\delta^{\frac{D}{2}} \frac{D}{2\alpha} \epsilon^{-1-\frac{D}{2\alpha}}$ is positive for $\epsilon > 0$. Therefore, by solving the equation $f'(\epsilon) = \sqrt{n} - c(\frac{D}{2\alpha} - 1)\epsilon^{-\frac{D}{2\alpha}} = 0$, we find the minimum of $f(\epsilon)$ at $\epsilon_0 = C^{\frac{2\alpha}{D}} \frac{\delta^\alpha}{n^{\alpha/D}}$ with

$$f(\epsilon_0) = C'\delta^\alpha n^{-\frac{\alpha}{D}+\frac{1}{2}},$$

44

for some constant $C' > 0$ depending only on $D$, $\alpha$ and $C$. Since $\frac{\alpha\delta^\alpha}{\alpha-\frac{D}{2}} < 0$, we deduce

$$\mathsf{entr}_n(B(x^*, \delta), d^\alpha) \leq f(\epsilon_0) = C'\delta^\alpha n^{-\frac{\alpha}{D}+\frac{1}{2}}.$$

- Consider the case (c). We first observe

$$\mathsf{entr}_n(B(x^*, \delta), d^\alpha) \leq \alpha \int_0^\delta s^{\alpha-1} \sqrt{\log N(s, B(x^*, \delta), d)} ds$$

$$\leq C\alpha D^{1/2} \int_0^\delta s^{\alpha-1} \sqrt{\log \frac{R}{s}} ds$$

$$= C\alpha D^{1/2} R^\alpha \int_0^{\delta/R} t^{\alpha-1} \sqrt{\log \frac{1}{t}} dt$$

$$\leq C\alpha D^{1/2} R\delta^{\alpha-1} \int_0^{\delta/R} \sqrt{\log \frac{1}{t}} dt$$

$$= C\alpha D^{1/2} R\delta^{\alpha-1} \int_{\log \frac{R}{\delta}}^\infty x^{\frac{1}{2}} e^{-x} dx. \tag{S31}$$

To further upper bound (S31), we claim there is a $\gamma_0 \in (0, 1)$ such that $x^{1/2} e^{-\gamma_0 x} \leq 1$ for $x \geq 1$. With this $\gamma_0$, the integral on the right-hand side of (S31) can be bounded, as follows.

(i) If $\log \frac{R}{\delta} \geq 1$, then

$$\int_{\log \frac{R}{\delta}}^\infty x^{\frac{1}{2}} e^{-x} dx = \int_{\log \frac{R}{\delta}}^\infty x^{\frac{1}{2}} e^{-\gamma_0 x} e^{-(1-\gamma_0)x} dx$$

$$\leq \int_{\log \frac{R}{\delta}}^\infty e^{-(1-\gamma_0)x} dx = \frac{1}{(1-\gamma_0)R^{1-\gamma_0}} \delta^{1-\gamma_0}.$$

(ii) If $\log \frac{R}{\delta} < 1$, then

$$\int_{\log \frac{R}{\delta}}^\infty x^{\frac{1}{2}} e^{-x} dx = \int_{\log \frac{R}{\delta}}^1 x^{\frac{1}{2}} e^{-x} dx + \int_1^\infty x^{\frac{1}{2}} e^{-x} dx$$

$$\leq \int_{\log \frac{R}{\delta}}^1 e^{-(1-\gamma_0)x} dx + \int_1^\infty e^{-(1-\gamma_0)x} dx$$

$$\leq \int_{\log \frac{R}{\delta}}^\infty e^{-(1-\gamma_0)x} dx = \frac{1}{(1-\gamma_0)R^{1-\gamma_0}} \delta^{1-\gamma_0}.$$

45

Combining the above bounds with (S31), we conclude

$$\operatorname{entr}_n(B(x^*, \delta), d^\alpha) \leq C \frac{\alpha}{1 - \gamma_0} D^{1/2} R^{\gamma_0} \delta^{\alpha - \gamma_0}.$$

It remains to find a $\gamma_0 \in (0, 1)$ such that $x^{1/2} e^{-\gamma_0 x} \leq 1$ for $x \geq 1$. Consider the function $f(x) = e^{\gamma_0 x} - x^{1/2}$ for $x \geq 1$. It is then sufficient to find some $\gamma_0 \in (0, 1)$ such that $f(x) \geq 0$ on $[1, \infty)$. Observe that $f(1) > 0$, and the second derivative of $f(x)$ is $f''(x) = \gamma_0^2 e^{\gamma_0 x} + x^{-3/2}/4$, which is always positive for $x \geq 1$. Therefore, one sufficient condition for $f(x) \geq 0$ on $[1, \infty)$ is $f'(1) = 0$. Solving

$$f'(1) = \gamma_0 e^{\gamma_0} - 1/2 = 0$$

for $\gamma_0$, we find that $\gamma_0 = W_0(1/2) \approx 0.35173$, where $W_0(\cdot)$ is the Lambert W function.

$\square$

**Lemma S5.** *Given two real numbers $a, b \in [-M, M]$ for some $M > 0$, let $p_a = 1/(1 + e^{-a})$ and $p_b = 1/(1 + e^{-b})$. Then, we have*

$$D_{\mathrm{kl}}(p_a \| p_b) \leq C(a - b)^2,$$

*for some constant $C > 0$.*

*Proof.* By the definition and the relation between the KL-divergence and the $\chi^2$-divergence (Tsybakov, 2009, Section 2.4), we have

$$D_{\mathrm{kl}}(p_a \| p_b) \leq D_{\chi^2}(p_a \| p_b) = \frac{p_a^2}{p_b} + \frac{(1 - p_a)^2}{1 - p_b} - 1 = \frac{(p_a - p_b)^2}{p_b(1 - p_b)}.$$

Since $a, b$ are bounded, $p_b(1 - p_b)$ is bounded from zero, and by the mean-value theorem, we also have $|p_a - p_b| \leq C_1 |a - b|$ for some constant $C_1 > 0$. Combining these facts yields the conclusion. $\square$

**Lemma S6.** *Given two real numbers $a, b \in \mathbb{R}$, let $p_a = 1/(1+e^a)$ and $p_b = 1/(1+e^b)$. Then, we have*

$$D_{\mathrm{kl}}(p_a \| p_b) + D_{\mathrm{kl}}(p_b \| p_a) \leq (a-b)^2.$$

*Proof.* We first observe

$$D_{\mathrm{kl}}(p_a \| p_b) + D_{\mathrm{kl}}(p_b \| p_a) = p_a \log \frac{p_a}{p_b} + (1-p_a) \log \frac{1-p_a}{1-p_b} + p_b \log \frac{p_b}{p_a} + (1-p_b) \log \frac{1-p_b}{1-p_a}$$

$$= (p_a - p_b) \log \frac{p_a}{p_b} + (p_b - p_a) \log \frac{1-p_a}{1-p_b}$$

$$= (p_a - p_b) \log \left( \frac{p_a}{1-p_a} \frac{1-p_b}{p_b} \right).$$

Note that $p_a/(1-p_a) = e^{-a}$ and $(1-p_b)/p_b = e^b$. Then,

$$D_{\mathrm{kl}}(p_a \| p_b) + D_{\mathrm{kl}}(p_b \| p_a) = \left( \frac{1}{1+e^a} - \frac{1}{1+e^b} \right) \log\left(e^{b-a}\right) = (b-a)\left( \frac{1}{1+e^a} - \frac{1}{1+e^b} \right). \tag{S32}$$

Assume without loss of generality that $b \geq a$. Noting that $e^x \geq 1+x$ by convexity, we have

$$\frac{1}{1+e^a} - \frac{1}{1+e^b} = \frac{e^b - e^a}{(1+e^a)(1+e^b)} \leq \frac{e^b - e^a}{e^b} = 1 - e^{a-b} \leq 1 - (1+(a-b)) = b-a.$$

Combining the above with (S32) yields the desired result. □

For $\mu \in \mathcal{M}$, recall that $\langle \cdot, \cdot \rangle_\mu$ is the Riemannian metric on the tangent space $T_\mu \mathcal{M}$, and $\mathrm{Log}_\mu$ denotes the logarithmic manifold maps at $\mu$.

**Lemma S7.** *Let $\mathcal{M}$ be a Riemannian manifold of dimension $D$. For any sufficiently small $t > 0$, there is a set $\{\beta, x_1, \ldots, x_D\} \subset \mathcal{M}$ of $D+1$ distinct points belonging to a neighborhood of $\mu$ such that*

$$\left| \langle \mathrm{Log}_\mu x_j, \mathrm{Log}_\mu \beta \rangle_\mu \right| = \log\left( \frac{1+t}{1-t} \right) \quad \text{for every } 1 \leq j \leq D.$$

*In addition, for any $b = (b_1, \ldots, b_D) \in \{-1, +1\}^D$, the points $x_1, \ldots, x_D$ can be chosen so that $\mathrm{sign}(\langle \mathrm{Log}_\mu x_j, \mathrm{Log}_\mu \beta \rangle_\mu) = b_j$ for $j = 1, \ldots, D$.*

*Proof.* First, we prove the statement for $\mathcal{M} = \mathbb{R}^D$. Let $\{\mathbf{e}_1, \ldots, \mathbf{e}_D\}$ be the standard basis of $\mathbb{R}^D$, and represent $\mu = (\mu_1, \ldots, \mu_D)^\top$ by its coordinate. The desired $\beta$ and $\{x_1, \ldots, x_D\}$ can be constructed as $\beta = \mu + r \cdot \mathbf{1}_D$ and $x_j = \mu + b_j T/r \cdot \mathbf{e}_j$, where $T = \log(\frac{1+t}{1-t})$, $r > 0$ is a parameter to be chosen later, and $\mathbf{1}_D$ is a $D$-dimensional vector of ones. Consider the ball $B(\mu, R)$ with radius $R > 0$ centered at $\mu$. Choosing $r = 2T/R$, one can see that for sufficiently small $t$ (note $T > 0$ and decreases as $t$ decreasing), we have $\|\beta - \mu\|_E < R$ and $\|x_j - \mu\|_E < R$ for all $1 \leq j \leq D$. In other words, we find a set $\{\beta, x_1, \ldots, x_D\}$ of $D+1$ points belonging to a ball centered at $\mu$ such that, for every $1 \leq j \leq D$,

$$\langle \mathrm{Log}_\mu x_j, \mathrm{Log}_\mu \beta \rangle_\mu = (\beta - \mu)^\top (x_j - \mu) = b_j T.$$

For a general $D$-dimensional Riemannian manifold $\mathcal{M}$, as the tangent space $T_\mu \mathcal{M}$ is isometric to $\mathbb{R}^D$, the above arguments show that there exists a set $\{\beta^E, x_1^E, \ldots, x_D^E\} \subset T_\mu \mathcal{M}$ such that $\langle \beta^E, x_j^E \rangle_\mu = b_j T$. As $\mathrm{Exp}_\mu$ is a local diffeomorphism, for sufficiently small $t > 0$, the set $\{\beta, x_1, \ldots, x_D\}$ with $\beta = \mathrm{Exp}_\mu \beta^E$ and $x_j = \mathrm{Exp}_\mu x_j^E$ contains distinct points and has the stated properties in the lemma. $\square$

**Lemma S8.** *Let $\mathcal{A} \subset \mathcal{M}$ be a geodesically convex set of a smooth Riemannian manifold $\mathcal{M}$ and $f : \mathcal{A} \to \mathbb{R}$ be a smooth function. If $\inf_{p \in \mathcal{A}} \Lambda_{\min}(\mathrm{Hess}_p f) \geq \eta > 0$, then $f$ is strongly convex with a parameter $\eta/2$.*

*Proof.* For any $p, q \in \mathcal{M}$, let $\gamma(t)$ be the constant-speed geodesic connecting them such that $\gamma(0) = p$ and $\gamma(1) = q$. Our goal is to establish $(1-t)f(p) + tf(q) - (\eta/2)t(1-t)d^2(p,q) \geq f(\gamma(t))$.

Define $\phi(t) = f(\gamma(t))$. Taylor series expansion of $\phi$ at $t$ implies

$$\phi(0) - \phi(t) = \phi'(t)(-t) + \frac{1}{2}\phi''(t_1)t^2 \quad \text{for some } t_1 \in [0, t]$$

$$\phi(1) - \phi(t) = \phi'(t)(1-t) + \frac{1}{2}\phi''(t_2)(1-t)^2 \quad \text{for some } t_2 \in [t, 1],$$

48

which further imply

$$(1-t)[\phi(0) - \phi(t)] + t[\phi(1) - \phi(t)] = \frac{1}{2}\phi''(t_1)(1-t)t^2 + \frac{1}{2}\phi''(t_2)(1-t)^2 t. \qquad (S33)$$

Note that, for all $t \in [0,1]$

$$\phi''(t) = \nabla_{\gamma'(t)}(df)(\gamma') = \langle \nabla_{\gamma'(t)} \mathrm{grad} f, \gamma'(t) \rangle_{\gamma(t)} = \langle (\mathrm{Hess}_{\gamma(t)} f)\gamma'(t), \gamma'(t) \rangle_{\gamma(t)}$$

$$\geq \eta \|\gamma'(t)\|^2 = \eta d^2(p,q).$$

Applying this result to the right-hand side of (S33) leads to $(1-t)[\phi(0) - \phi(t)] + t[\phi(1) - \phi(t)] \geq \eta t(1-t)d^2(p,q)/2$. The proof is completed by observing that $(1-t)[\phi(0) - \phi(t)] + t[\phi(1) - \phi(t)] = (1-t)f(p) + tf(q) - f(\gamma(t))$. □

## J  Supplement to Numeric Studies

We first describe the competing classifiers referenced in Sections 6 and 7, and then present additional simulation results.

**SPDNet.** This classifier is proposed in Huang and Van Gool (2017). It adopts a deep neural network architecture and preserves the SPD structure via BiRe blocks with non-linear maps on SPD matrices. In practice, one needs to determine the number of BiRe blocks and the shape of the blocks. As in Huang and Van Gool (2017), we tried various SPDNets with different BiRe blocks, and found that the SPDNet (SPDNet-3) containing one BiRe block with a transformation matrix of the shape $3 \times 3$ yields the best performance in the simulation studies among all SPDNets, and the SPDNet (SPDNet-8) containing one BiRe block with a transformation matrix of the shape $8 \times 8$ yields the best performance in the data application. Therefore, we adopt SPDNet-3 in Section 6 for simulation studies and SPDNet-8 in Section 7 for the data application.

***k*-NN.** We adopt the *k*-NN classifier (Chaudhuri and Dasgupta, 2014) with the Log-Cholesky distance as the distance metric. The hyperparameter $k$ is determined by the five-fold cross validation on the training data.

**Kernel SVM (kSVM).** We adopt the kernel SVM proposed in Jayasumana et al. (2013), which is a kernel method on the Riemannian manifold of SPD matrices. For a positive integer $m$, define the function $K: \mathcal{S}_m^+ \times \mathcal{S}_m^+ \to \mathbb{R}$ by

$$K(p, q) = \exp\left(-\frac{d^2(p, q)}{2\sigma^2}\right), \tag{S34}$$

where $d$ is the Log-Cholesky distance on $\mathcal{S}_m^+$. Then, according to Theorem 4.1 in Jayasumana et al. (2013), $K$ is a positive definite kernel for all $\sigma > 0$. Consequently, one can develop a kernel SVM classifier on the SPD matrices with this kernel $K$. The tuning parameter $\sigma$ is determined by the five-fold cross validation on training data.

**Kernel density classifier (KDC).** The kernel density classifier in Section 6.6.2 of Hastie et al. (2009) can be applied to classify data sampled from Riemannian manifolds as long as a density estimate of the predictor $X$ is available. To obtain a kernel density classifier on SPD matrices, we employ the kernel density estimator proposed by Pelletier (2005) for Riemannian manifolds and select the bandwidth via the five-fold cross validation on training data.

Below are additional results of the simulation studies, where

- Table S1 summarizes the results for the estimators $\hat{\mu}_n$ and $\hat{\beta}_n$ under different settings of Case 1; the results are similar to those presented in Table 1 in the main paper.

- Tables S2–S4 summarize the results in classification under different settings and cases; the results are similar to those in Table 2.

The numeric studies testify that there is no universally best classifier. In practice, given a dataset, we can use cross validation to benchmark the proposed model against other classifiers. If the proposed model exhibits a performance comparable to that of the best classifier, then we would recommend the proposed model since it also offers unparalleled interpretability; see Section 7 for a demonstration.

Table S1: Empirical estimation performance of the proposed method under the other settings of Case 1. The standard deviations based on the 500 independent Monte Carlo replicates are given in the parentheses. The value of $\sigma_X^2$ is approximated numerically.

| $\mu^*$ | $\beta^*$ | $\sigma_X^2$ | $n$ | $d(\mu^*, \hat{\mu}_n)$ | $d(\beta^*, \hat{\beta}_n)$ | RMSE |
|---|---|---|---|---|---|---|
| $I_3$ | $3I_3$ | 2.248 | 100 | 0.145(0.047) | 1.128(0.391) | 0.142(0.057) |
| | | | 500 | 0.063(0.020) | 0.475(0.140) | 0.025(0.008) |
| | | 4.496 | 100 | 0.205(0.066) | 0.871(0.316) | 0.151(0.060) |
| | | | 500 | 0.089(0.028) | 0.356(0.108) | 0.026(0.008) |
| $\Sigma_{AR}$ | $3I_3$ | 3.637 | 100 | 0.179(0.063) | 0.901(0.461) | 0.169(0.076) |
| | | | 500 | 0.079(0.028) | 0.325(0.112) | 0.029(0.009) |
| | | 7.273 | 100 | 0.245(0.087) | 0.815(0.455) | 0.201(0.099) |
| | | | 500 | 0.112(0.039) | 0.300(0.108) | 0.033(0.011) |
| | $3\Sigma_{AR}$ | 3.637 | 100 | 0.180(0.064) | 1.170(0.421) | 0.155(0.062) |
| | | | 500 | 0.079(0.028) | 0.521(0.162) | 0.026(0.008) |
| | | 7.273 | 100 | 0.249(0.087) | 0.976(0.372) | 0.168(0.071) |
| | | | 500 | 0.112(0.039) | 0.439(0.148) | 0.029(0.009) |

Table S2: Empirical classification performance of the classifiers when $\mu^* = I_3$ in Case 1. The standard deviations based on the 500 independent Monte Carlo replicates are given in the parentheses.

| $\beta^*$ | $\sigma_X^2$ | $n$ | Method | Accuracy | AUC | Sensitivity | Specificity | EER |
|---|---|---|---|---|---|---|---|---|
| $3I_3$ | 2.248 | 100 | SPDNet | 0.566(0.114) | 0.591(0.138) | 0.562(0.197) | 0.568(0.202) | 0.034(0.121) |
| | | | $k$-NN | 0.530(0.115) | 0.555(0.094) | 0.513(0.264) | 0.539(0.268) | 0.070(0.147) |
| | | | kSVM | 0.535(0.114) | 0.546(0.086) | 0.517(0.352) | 0.538(0.345) | 0.065(0.144) |
| | | | KDC | 0.522(0.115) | 0.537(0.077) | 0.586(0.375) | 0.452(0.377) | 0.079(0.153) |
| | | | Proposed | 0.568(0.110) | 0.624(0.111) | 0.556(0.165) | 0.583(0.166) | 0.032(0.103) |
| | | 500 | SPDNet | 0.600(0.053) | 0.638(0.059) | 0.609(0.103) | 0.590(0.105) | 0.009(0.038) |
| | | | $k$-NN | 0.558(0.052) | 0.560(0.046) | 0.562(0.131) | 0.553(0.132) | 0.051(0.055) |
| | | | kSVM | 0.581(0.056) | 0.583(0.052) | 0.590(0.162) | 0.572(0.161) | 0.027(0.049) |
| | | | KDC | 0.553(0.061) | 0.556(0.050) | 0.586(0.273) | 0.521(0.281) | 0.056(0.068) |
| | | | Proposed | 0.602(0.052) | 0.638(0.060) | 0.601(0.070) | 0.602(0.076) | 0.007(0.036) |
| | 4.496 | 100 | SPDNet | 0.605(0.112) | 0.650(0.134) | 0.605(0.179) | 0.604(0.185) | 0.033(0.098) |
| | | | $k$-NN | 0.557(0.120) | 0.576(0.102) | 0.550(0.251) | 0.560(0.253) | 0.081(0.142) |
| | | | kSVM | 0.557(0.119) | 0.565(0.096) | 0.550(0.313) | 0.561(0.312) | 0.081(0.147) |
| | | | KDC | 0.545(0.114) | 0.553(0.089) | 0.579(0.318) | 0.500(0.335) | 0.093(0.150) |
| | | | Proposed | 0.610(0.110) | 0.668(0.114) | 0.595(0.167) | 0.629(0.159) | 0.028(0.091) |
| | | 500 | SPDNet | 0.639(0.052) | 0.691(0.056) | 0.651(0.087) | 0.626(0.087) | 0.007(0.032) |
| | | | $k$-NN | 0.600(0.052) | 0.601(0.048) | 0.602(0.122) | 0.598(0.119) | 0.046(0.051) |
| | | | kSVM | 0.627(0.054) | 0.627(0.053) | 0.630(0.116) | 0.624(0.114) | 0.019(0.042) |
| | | | KDC | 0.607(0.055) | 0.606(0.052) | 0.611(0.182) | 0.601(0.174) | 0.040(0.052) |
| | | | Proposed | 0.641(0.052) | 0.692(0.057) | 0.642(0.069) | 0.639(0.074) | 0.005(0.030) |
| $3\Sigma_{AR}$ | 4.496 | 100 | SPDNet | 0.656(0.109) | 0.719(0.127) | 0.646(0.174) | 0.664(0.175) | 0.044(0.101) |
| | | | $k$-NN | 0.631(0.111) | 0.633(0.105) | 0.614(0.212) | 0.641(0.219) | 0.069(0.122) |
| | | | kSVM | 0.642(0.112) | 0.644(0.105) | 0.635(0.231) | 0.647(0.230) | 0.058(0.110) |
| | | | KDC | 0.621(0.112) | 0.621(0.105) | 0.613(0.239) | 0.620(0.252) | 0.079(0.123) |
| | | | Proposed | 0.678(0.106) | 0.746(0.119) | 0.668(0.163) | 0.687(0.157) | 0.022(0.089) |
| | | 500 | SPDNet | 0.679(0.046) | 0.745(0.050) | 0.669(0.077) | 0.690(0.077) | 0.030(0.036) |
| | | | $k$-NN | 0.673(0.048) | 0.674(0.047) | 0.677(0.099) | 0.670(0.094) | 0.037(0.042) |
| | | | kSVM | 0.697(0.046) | 0.697(0.046) | 0.700(0.087) | 0.694(0.086) | 0.013(0.033) |
| | | | KDC | 0.686(0.047) | 0.687(0.046) | 0.692(0.119) | 0.681(0.118) | 0.024(0.039) |
| | | | Proposed | 0.703(0.047) | 0.776(0.046) | 0.705(0.071) | 0.703(0.070) | 0.006(0.028) |

Table S3: Empirical classification performance of the classifiers when $\mu^* = \Sigma_{AR}$ in Case 1. The standard deviations based on the 500 independent Monte Carlo replicates are given in the parentheses.

| $\beta^*$ | $\sigma_X^2$ | $n$ | Method | Accurary | AUC | Sensitivity | Specificity | EER |
|---|---|---|---|---|---|---|---|---|
| $3I_3$ | 3.637 | 100 | SPDNet | 0.647(0.118) | 0.711(0.128) | 0.671(0.172) | 0.625(0.179) | 0.024(0.093) |
| | | | $k$-NN | 0.606(0.114) | 0.617(0.100) | 0.602(0.217) | 0.609(0.216) | 0.065(0.118) |
| | | | kSVM | 0.612(0.122) | 0.618(0.106) | 0.599(0.259) | 0.626(0.245) | 0.059(0.116) |
| | | | KDC | 0.592(0.120) | 0.599(0.100) | 0.602(0.263) | 0.578(0.279) | 0.079(0.130) |
| | | | Proposed | 0.649(0.109) | 0.711(0.122) | 0.638(0.160) | 0.663(0.162) | 0.022(0.089) |
| | | 500 | SPDNet | 0.668(0.047) | 0.737(0.052) | 0.676(0.078) | 0.660(0.078) | 0.009(0.034) |
| | | | $k$-NN | 0.646(0.051) | 0.645(0.051) | 0.648(0.097) | 0.642(0.096) | 0.032(0.048) |
| | | | kSVM | 0.660(0.049) | 0.660(0.048) | 0.665(0.099) | 0.655(0.095) | 0.017(0.039) |
| | | | KDC | 0.650(0.051) | 0.649(0.050) | 0.656(0.131) | 0.643(0.125) | 0.027(0.046) |
| | | | Proposed | 0.672(0.048) | 0.742(0.052) | 0.673(0.069) | 0.671(0.073) | 0.005(0.028) |
| | 7.273 | 100 | SPDNet | 0.695(0.106) | 0.770(0.114) | 0.716(0.154) | 0.676(0.161) | 0.026(0.083) |
| | | | $k$-NN | 0.649(0.112) | 0.653(0.104) | 0.650(0.198) | 0.645(0.203) | 0.072(0.116) |
| | | | kSVM | 0.669(0.116) | 0.671(0.112) | 0.665(0.210) | 0.673(0.216) | 0.052(0.102) |
| | | | KDC | 0.641(0.118) | 0.651(0.108) | 0.633(0.217) | 0.647(0.230) | 0.080(0.118) |
| | | | Proposed | 0.696(0.107) | 0.775(0.108) | 0.680(0.146) | 0.716(0.154) | 0.025(0.082) |
| | | 500 | SPDNet | 0.713(0.044) | 0.792(0.046) | 0.721(0.072) | 0.706(0.070) | 0.012(0.031) |
| | | | $k$-NN | 0.693(0.047) | 0.693(0.047) | 0.694(0.088) | 0.693(0.085) | 0.032(0.041) |
| | | | kSVM | 0.710(0.044) | 0.709(0.044) | 0.714(0.079) | 0.705(0.081) | 0.015(0.033) |
| | | | KDC | 0.698(0.046) | 0.698(0.046) | 0.700(0.093) | 0.695(0.098) | 0.027(0.040) |
| | | | Proposed | 0.721(0.045) | 0.802(0.045) | 0.722(0.065) | 0.720(0.070) | 0.004(0.024) |
| $3\Sigma_{AR}$ | 3.637 | 100 | SPDNet | 0.555(0.114) | 0.582(0.136) | 0.509(0.202) | 0.602(0.203) | 0.081(0.136) |
| | | | $k$-NN | 0.531(0.113) | 0.557(0.097) | 0.522(0.262) | 0.536(0.259) | 0.106(0.143) |
| | | | kSVM | 0.543(0.113) | 0.554(0.089) | 0.535(0.332) | 0.538(0.328) | 0.094(0.144) |
| | | | KDC | 0.522(0.113) | 0.539(0.082) | 0.581(0.347) | 0.447(0.352) | 0.115(0.149) |
| | | | Proposed | 0.604(0.113) | 0.668(0.118) | 0.598(0.166) | 0.612(0.172) | 0.033(0.097) |
| | | 500 | SPDNet | 0.598(0.052) | 0.637(0.058) | 0.630(0.099) | 0.565(0.109) | 0.047(0.051) |
| | | | $k$-NN | 0.565(0.052) | 0.567(0.048) | 0.575(0.131) | 0.555(0.127) | 0.080(0.059) |
| | | | kSVM | 0.610(0.055) | 0.609(0.054) | 0.618(0.122) | 0.600(0.112) | 0.035(0.048) |
| | | | KDC | 0.560(0.058) | 0.560(0.051) | 0.582(0.241) | 0.534(0.240) | 0.085(0.064) |
| | | | Proposed | 0.638(0.049) | 0.688(0.055) | 0.635(0.070) | 0.641(0.070) | 0.007(0.030) |
| | 7.273 | 100 | SPDNet | 0.589(0.112) | 0.627(0.134) | 0.538(0.188) | 0.635(0.188) | 0.091(0.125) |
| | | | $k$-NN | 0.556(0.111) | 0.572(0.098) | 0.547(0.237) | 0.568(0.242) | 0.124(0.134) |
| | | | kSVM | 0.578(0.109) | 0.580(0.097) | 0.575(0.299) | 0.571(0.301) | 0.102(0.129) |
| | | | KDC | 0.543(0.111) | 0.558(0.088) | 0.594(0.297) | 0.482(0.312) | 0.137(0.138) |
| | | | Proposed | 0.661(0.109) | 0.722(0.121) | 0.652(0.163) | 0.670(0.164) | 0.019(0.086) |
| | | 500 | SPDNet | 0.623(0.051) | 0.671(0.055) | 0.667(0.090) | 0.578(0.098) | 0.069(0.051) |
| | | | $k$-NN | 0.603(0.050) | 0.603(0.048) | 0.611(0.124) | 0.595(0.120) | 0.088(0.054) |
| | | | kSVM | 0.657(0.052) | 0.656(0.050) | 0.663(0.109) | 0.649(0.102) | 0.035(0.042) |
| | | | KDC | 0.614(0.052) | 0.614(0.049) | 0.626(0.150) | 0.602(0.147) | 0.077(0.055) |
| | | | Proposed | 0.685(0.047) | 0.749(0.049) | 0.685(0.069) | 0.686(0.069) | 0.006(0.025) |

Table S4: Empirical classification performance of classifiers in Case 2 and Case 3 when $\sigma_X^2 = 4.496$. The standard deviations based on the 500 independent Monte Carlo replicates are given in the parentheses.

| Model | $n$ | Method | Accurary | AUC | Sensitivity | Specificity | EER |
|---|---|---|---|---|---|---|---|
| Case 2 | 100 | SPDNet | 0.627(0.114) | 0.673(0.128) | 0.621(0.193) | 0.635(0.187) | 0.147(0.132) |
| | | $k$-NN | 0.606(0.114) | 0.616(0.100) | 0.598(0.230) | 0.611(0.231) | 0.168(0.134) |
| | | kSVM | 0.604(0.129) | 0.612(0.111) | 0.600(0.280) | 0.615(0.287) | 0.170(0.144) |
| | | KDC | 0.582(0.123) | 0.591(0.104) | 0.605(0.285) | 0.560(0.311) | 0.192(0.147) |
| | | Proposed | 0.658(0.110) | 0.700(0.116) | 0.651(0.167) | 0.665(0.169) | 0.115(0.127) |
| | 500 | SPDNet | 0.671(0.050) | 0.705(0.054) | 0.667(0.082) | 0.675(0.085) | 0.109(0.052) |
| | | $k$-NN | 0.652(0.050) | 0.652(0.050) | 0.658(0.095) | 0.647(0.097) | 0.127(0.056) |
| | | kSVM | 0.695(0.049) | 0.695(0.049) | 0.699(0.093) | 0.690(0.088) | 0.085(0.053) |
| | | KDC | 0.668(0.052) | 0.668(0.051) | 0.674(0.124) | 0.661(0.127) | 0.112(0.057) |
| | | Proposed | 0.718(0.047) | 0.729(0.053) | 0.720(0.065) | 0.716(0.072) | 0.062(0.048) |
| Case 3 | 100 | SPDNet | 0.761(0.097) | 0.796(0.119) | 0.512(0.232) | 0.871(0.094) | 0.057(0.095) |
| | | $k$-NN | 0.726(0.106) | 0.621(0.116) | 0.895(0.108) | 0.346(0.248) | 0.092(0.112) |
| | | kSVM | 0.722(0.108) | 0.586(0.112) | 0.943(0.078) | 0.230(0.252) | 0.096(0.109) |
| | | KDC | 0.710(0.109) | 0.572(0.103) | 0.934(0.108) | 0.209(0.248) | 0.108(0.112) |
| | | Proposed | 0.676(0.105) | 0.783(0.113) | 0.626(0.135) | 0.787(0.188) | 0.142(0.115) |
| | 500 | SPDNet | 0.776(0.040) | 0.816(0.045) | 0.514(0.096) | 0.890(0.042) | 0.047(0.039) |
| | | $k$-NN | 0.763(0.045) | 0.663(0.055) | 0.915(0.046) | 0.410(0.120) | 0.060(0.043) |
| | | kSVM | 0.781(0.043) | 0.693(0.055) | 0.915(0.042) | 0.471(0.117) | 0.043(0.039) |
| | | KDC | 0.737(0.045) | 0.583(0.051) | 0.973(0.048) | 0.193(0.127) | 0.086(0.048) |
| | | Proposed | 0.688(0.051) | 0.804(0.049) | 0.633(0.065) | 0.814(0.073) | 0.135(0.054) |

			
	\end{document}